\documentclass[stslayout, noinfoline]{imsart}
\usepackage{geometry}
\usepackage{graphicx}
\usepackage{amssymb, amsmath}
\usepackage{natbib}
\usepackage{epstopdf}
\usepackage{subfigure}
\usepackage{algorithm}
\usepackage[noend]{algpseudocode}
\usepackage{tcolorbox}
\usepackage{tikz}
\usepackage[colorlinks,citecolor=blue,urlcolor=blue]{hyperref}

\usetikzlibrary{intersections, backgrounds, calc}

\definecolor{light}{RGB}{220, 188, 188}
\definecolor{mid}{RGB}{185, 124, 124}
\definecolor{dark}{RGB}{143, 39, 39}
\definecolor{highlight}{RGB}{0, 255, 0}
\definecolor{gray10}{gray}{0.1}
\definecolor{gray20}{gray}{0.2}
\definecolor{gray30}{gray}{0.3}
\definecolor{gray40}{gray}{0.4}
\definecolor{gray60}{gray}{0.6}
\definecolor{gray70}{gray}{0.7}
\definecolor{gray80}{gray}{0.8}
\definecolor{gray90}{gray}{0.9}
\definecolor{gray90}{gray}{0.9}
\definecolor{gray91}{gray}{0.91}
\definecolor{gray92}{gray}{0.92}
\definecolor{gray93}{gray}{0.93}
\definecolor{gray94}{gray}{0.94}
\definecolor{gray95}{gray}{0.95}
\definecolor{comment}{gray}{0.50}

\begin{document}

\begin{frontmatter}

\title{A Geometric Theory of Higher-Order Automatic Differentiation}
\runtitle{Geometric Automatic Differentiation}

\begin{aug}
  \author{Michael Betancourt%
  \ead[label=e1]{betanalpha@gmail.com}}
  
  \runauthor{Betancourt}

  \address{Michael Betancourt is the principle research scientist
           at Symplectomorphic, LLC. \printead{e1}.}

\end{aug}

\begin{abstract}
First-order automatic differentiation is a ubiquitous tool 
across statistics, machine learning, and computer science.  
Higher-order implementations of automatic differentiation, 
however, have yet to realize the same utility.  In this 
paper I derive a comprehensive, differential geometric 
treatment of automatic differentiation that naturally 
identifies the higher-order differential operators amenable 
to automatic differentiation as well as explicit procedures 
that provide a scaffolding for high-performance implementations.
\end{abstract}

\end{frontmatter}

\newpage

\setcounter{tocdepth}{2}
\tableofcontents

\pagebreak

Automatic differentiation is a powerful and increasingly pervasive 
tool for numerically evaluating the derivatives of a function 
implemented as a computer program; see \cite{Margossian:2018, 
GriewankEtAl:2008, BuckerEtAl:2006} for thorough reviews and
\cite{autodifforg:2018} for a extensive list of automatic 
differentiation software.  

Most popular implementations of automatic differentiation focus 
on, and optimize for, evaluating the first-order differential 
operators required in methods such as gradient descent 
\citep{Bishop:2006}, Langevin Monte Carlo \citep{XifaraEtAl:2014}
and Hamiltonian Monte Carlo \citep{Betancourt:2017a}.  Some 
implementations also offer support for the evaluation of 
higher-order differential operators that arise in more 
sophisticated methods such as Newton-Raphson optimization 
\citep{Bishop:2006}, gradient-based Laplace Approximations 
\citep{KristensenEtAl:2016}, and certain Riemannian Langevin 
and Hamiltonian Monte Carlo implementations \citep{Betancourt:2013b}.  

Most higher-order automatic differentiation implementations, 
however, exploit the recursive application of first-order 
methods \citep{GriewankEtAl:2008}.  While practically 
convenient, this recursive strategy can confound the nature of 
the underlying differential operators and obstruct the 
optimization of their implementations.  In this paper I derive 
a comprehensive, differential geometric treatment of higher-order 
automatic differentiation that naturally identifies valid 
differential operators as well as explicit rules for their 
propagating through a computer program.  This treatment not 
only identifies novel automatic differentiation techniques but
also illuminates the challenges for efficient software 
implementations.  

I begin with a conceptual review of first-order automatic 
differentiation and its relationship to differential 
geometry before introducing the jet spaces that generalize 
the underlying theory to higher-orders.  Next I show how the 
natural transformations of jets explicitly realize higher-order 
differential operators before finally considering the 
consequences of the theory on the efficient implementation 
of these methods in practice.

\section{First-Order Automatic Differentiation}

First-order automatic differentiation has proven a powerful
tool in practice, but the mechanisms that admit the method
are often overlooked.  Before considering higher-order methods 
let's first review first-order methods, starting with the 
chain rule over composite functions and the opportunities 
for algorithmic differentiation.  We'll then place these 
techniques in a geometric perspective which will guide 
principled generalizations.

\subsection{Composite Functions and the Chain Rule}

The first-order differential structure of a function from
the real numbers into the real numbers,
$F_{1} : \mathbb{R}^{D_0} \rightarrow \mathbb{R}^{D_1}$,
in the neighborhood of a point, $x \in \mathbb{R}^{D_0}$ 
is completely quantified by the \emph{Jacobian matrix} 
of first-order partial derivatives evaluated at that point,
\begin{equation*}
(\mathcal{J}_{F_{1}})^{i}_{j} (x) = \frac{ \partial (F_{1})^{i} }{ \partial x^{j} }(x).
\end{equation*}

Given another function 
$F_{2} : \mathbb{R}^{D_1} \rightarrow \mathbb{R}^{D_2}$
we can construct the \emph{composition}
\begin{alignat*}{6}
F = F_{2} \circ F_{1} :\; &\mathbb{R}^{D_0}& &\rightarrow& \; &\mathbb{R}^{D_2}&
\\
&x& &\mapsto& &F(x) = F_{2}(F_{1}(x))&.
\end{alignat*}
The Jacobian matrix for this composition is given by the infamous 
chain rule,
\begin{align*}
(\mathcal{J}_{F})^{i}_{j} (x)
&= \sum_{k = 0}^{D_{1}} 
\frac{ \partial (F_{2})^{i} }{\partial y^{k}} (F_{1}(x))
\frac{ \partial (F_{1})^{k} }{\partial x^{j}} (x)
\\
&= \sum_{k = 1}^{D_{1}} 
(\mathcal{J}_{F_{2}})^{i}_{k} (F_{1}(x))
\, \cdot \,
(\mathcal{J}_{F_{1}})^{k}_{j} (x).
\end{align*}

The structure of the chain rule tells us that the Jacobian matrix 
of the composite function is given by the matrix product of the 
Jacobian matrix of the component functions,
\begin{equation*}
\mathcal{J}(x) = \mathcal{J}_{F_{2}}(F_{1}(x)) \cdot \mathcal{J}_{F_{1}}(x).
\end{equation*}
Applying the chain rule iteratively demonstrates that this 
product form holds for the composition of any number of
component functions.  In particular, given the component functions
\begin{alignat*}{6}
F_{n} :\; &\mathbb{R}^{D_{n - 1}}& &\rightarrow& \; &\mathbb{R}^{D_{n}}&
\\
&x_{n - 1}& &\mapsto& &x_{n} = F_{n}(x_{n - 1})&.
\end{alignat*}
we can construct the composite function
\begin{equation*}
F = F_{N} \circ F_{N - 1} \circ \cdots \circ F_{2} \circ F_{1},
\end{equation*}
and its Jacobian matrix
\begin{equation*}
\mathcal{J}_{F}(x) = 
\mathcal{J}_{F_{N}}(x_{N - 1}) \cdot 
\mathcal{J}_{F_{N - 1}}(x_{N - 2}) \cdot
\ldots \cdot 
\mathcal{J}_{F_{2}}(x_{1}) \cdot
\mathcal{J}_{F_{1}}(x_{0}).
\end{equation*} 

The iterative nature of the chain rule implies that we can
build up the Jacobian matrix evaluated at a given input for 
any composite function as we apply each of the component 
functions to that input.  In doing so we use only calculations 
local to each component function and avoid having to deal 
with the global structure of the composite function directly.

\subsection{Automatic Differentiation}

Automatic differentiation exploits the structure of the chain
rule to evaluate differential operators of a function as the
function itself is being evaluated as a computer program.
A differential operator is a mapping that takes an input
function to a new function that captures well-defined 
differential information about the input function.  For
example, first-order differential operators of many-to-one
real-valued functions include directional derivatives and
gradients.  At higher-orders well-posed differential 
operators are more subtle and tend to deviate from our
expectations; for example the two-dimensional array of
second-order partial derivatives commonly called the 
Hessian is not a well-posed differential operator.

The Jacobian matrix can be used to construct first-order
differential operators that map vectors from the input space 
to vectors in the output space, and vice versa.  For example,
let's first consider the action of the Jacobian matrix on a 
vector $v_{0}$ in the input space $\mathbb{R}^{D_{0}}$,
\begin{equation*}
v = \mathcal{J}_{F} \cdot v_{0},
\end{equation*}
or in terms of the components of the vector,
\begin{equation*}
v^{i} 
= \sum_{j = 1}^{D_{0}} (\mathcal{J}_{F})^{i}_{j}(x_{0}) \cdot (v_{0})^{j}
= \sum_{j = 1}^{D_{0}} \frac{ \partial F^{i} }{\partial x^{j}} (x_{0}) \cdot (v_{0})^{j}.
\end{equation*}
We can consider the input vector $v_{0}$ as a small perturbation 
of the input point with the action of the Jacobian producing a
vector $v$ that approximately quantifies how the output of the 
composite function changes under that perturbation,
\begin{equation*}
v^{i} \approx F^{i}(x + v_{0}) - F^{i}(x).
\end{equation*}

Because of the product structure of the composite Jacobian 
matrix this action can be calculated sequentially, at each 
iteration multiplying one of the component Jacobian matrices 
by an intermediate vector,
\begin{align*}
\mathcal{J}_{F}(x) \cdot v_{0}
&=
\mathcal{J}_{F_{N}}(x_{N - 1}) 
\cdot \ldots \cdot 
\mathcal{J}_{F_{2}}(x_{1}) 
\cdot 
\;\, \mathcal{J}_{F_{1}}(x_{0}) \cdot v_{0}
\\
&=
\mathcal{J}_{F_{N}}(x_{N - 1}) 
\cdot \ldots \cdot 
\mathcal{J}_{F_{2}}(x_{1})  
\cdot 
(\mathcal{J}_{F_{1}}(x_{0}) \cdot v_{0})
\\
&=
\mathcal{J}_{F_{N}}(x_{N - 1}) 
\cdot \ldots \cdot 
\mathcal{J}_{F_{2}}(x_{1}) \cdot \quad\quad\;\; v_{1}
\\
&=
\mathcal{J}_{F_{N}}(x_{N - 1}) 
\cdot \ldots \cdot 
(\mathcal{J}_{F_{2}}(x_{1})  \cdot v_{1})
\\
&=
\mathcal{J}_{F_{N}}(x_{N - 1}) 
\cdot \ldots \cdot 
\quad\quad\quad\; v_{2}
\\
&\ldots
\\
&=
\mathcal{J}_{N}(x_{N - 1}) \cdot v_{N - 1}
\\
&=
v_{N}.
\end{align*} 
In words, if we can evaluate the component Jacobian matrices  
then we can evaluate the action of the composite 
Jacobian matrix on an input vector by propagating that input 
vector through each of the component Jacobian matrices in turn.  
Different choices of $v_{0}$ extract different projections of 
the composite Jacobian matrix which isolate different details
about the differential behavior of the composite function around
the input point.

When the component functions are defined by subexpressions in 
a computer program the propagation of an input vector implements 
\emph{forward mode automatic differentiation}.  In this context
the intermediate vectors are denoted \emph{tangents} or 
\emph{perturbations}.

The same methodology can be used to compute the action of the
transposed Jacobian matrix on a vector $\alpha_{0}$ in the 
output space, $\mathbb{R}^{D_{N}}$,
\begin{equation*}
\alpha = \mathcal{J}_{F}^{T}(x_{0}) \cdot \alpha_{0},
\end{equation*}
or in terms of the components of the vector,
\begin{equation*}
(\alpha)_{i} 
= \sum_{j = 1}^{D_{N}} (\mathcal{J}_{F})^{j}_{i}(x_{0}) \cdot  (\alpha_{0})_{j} 
= \sum_{j = 1}^{D_{N}} \frac{ \partial F^{j} }{\partial x^{i}} (x_{0}) \cdot (\alpha_{0})_{j}.
\end{equation*}  
Intuitively the output vector $\alpha_{0}$ quantifies a
perturbation in the output of the composite function 
while the action of the transposed Jacobian yields a 
vector $\alpha$ that quantifies the change in the input 
space needed to achieve that specific perturbation.

The transpose of the Jacobian matrix of a composite 
function is given by
\begin{equation*}
\mathcal{J}_{F}^{T} = 
\mathcal{J}_{1}^{T}(x) 
\cdot \ldots \cdot 
\mathcal{J}_{N}^{T}(x_{N - 1}),
\end{equation*}
and its action on $\alpha_{0}$ can be computed sequentially as
\begin{align*}
\mathcal{J}^{T}_{F}(x_{0}) \cdot \alpha_{0}
&=
\mathcal{J}_{F_{1}}^{T}(x_{0}) 
\cdot \ldots \cdot 
\mathcal{J}_{F_{N - 1}}^{T} (x_{N - 2}) \cdot 
\;\, \mathcal{J}_{F_{N}}^{T}(x_{N - 1}) \cdot \alpha_{0}
\\
&=
\mathcal{J}_{F_{1}}^{T}(x_{0}) 
\cdot \ldots \cdot 
\mathcal{J}_{F_{N - 1}}^{T} (x_{N - 2}) \cdot 
(\mathcal{J}_{F_{N}}^{T}(x_{N - 1}) \cdot \alpha_{0})
\\
&=
\mathcal{J}_{F_{1}}^{T}(x_{0}) 
\cdot \ldots \cdot 
\mathcal{J}_{F_{N - 1}}^{T}(x_{N - 2}) \cdot 
\quad\quad\quad\quad\;\; \alpha_{1}
\\
&=
\mathcal{J}_{F_{1}}^{T}(x_{0}) 
\cdot \ldots \cdot 
( \mathcal{J}_{F_{N - 1}}^{T}(x_{N - 2}) 
\cdot \alpha_{1})
\\
&=
\mathcal{J}_{F_{1}}^{T}(x_{0}) 
\cdot \ldots \cdot 
\quad\quad\quad\quad\;\; \alpha_{2}
\\
&\ldots
\\
&=
\mathcal{J}_{F_{1}}^{T}(x) \cdot \alpha_{N - 1}
\\
&=
\alpha_{N}.
\end{align*} 

As before, if we can explicitly evaluate the component Jacobian matrices 
then we can evaluate the action of the transposed
composite Jacobian matrix on an output vector by propagating said
vector through each of the component Jacobian matrices, only now
in the reversed order of the composition.  The action of the
transpose Jacobian allows for different components of the 
composite Jacobian matrix to be projected out with each sweep 
backwards through the component functions.

When the component functions are defined by subexpressions in a 
computer program the backwards propagation of an output vector 
implements \emph{reverse mode automatic differentiation}.  In 
this case the intermediate vectors are denoted \emph{adjoints} 
or \emph{sensitivities}.

Higher-order automatic differentiation is significantly more 
complicated due to the way that higher-order partial derivatives
mix with lower-order partial derivatives.  Most implementations 
of higher-order automatic differentiation recursively apply 
first-order methods, implicitly propagating generalized tangents 
and adjoints back and forth across the composite function while 
automatically differentiating the analytic first-order partial 
derivatives to obtain the necessary higher-order partial derivatives 
at the same time.  Although this approach removes the burden of 
specifying higher-order partial derivatives and the more complex
update rules from users, it limits the clarity of those updates
as well as the potential to optimize software implementations.

\subsection{The Geometric Perspective of Automatic Differentiation}

Using the Jacobian to propagate vectors forwards and
backwards along a composite function isn't unreasonable
once its been introduced, and it certainly leads to useful
algorithms in practice.  There isn't much initial motivation,
however, as to why this technique would be initially worth 
exploring, at least the way it its typically presented.  
Fortunately that motivation isn't absent entirely, it's just 
hiding in the field of differential geometry.

Differential geometry is the study of \emph{smooth manifolds},
spaces that admit well-defined derivatives which generalize 
the tools of calculus beyond the real numbers \citep{Lee:2013}.  
This generalization is useful for expanding the scope of 
derivatives and integrals, but it also strips away the structure 
of the real numbers irrelevant to differentiation which then 
facilitates the identification of well-defined differential 
calculations.

In differential geometry tangents are more formally identified 
as \emph{tangent vectors}, elements of the \emph{tangent space},
$T_{x} X$, at the input point, $x \in X$, where the composite 
function is being evaluated.  These tangent vectors specify
directions and magnitudes in a small neighborhood around $p$.  
The propagation of a tangent vector through a function,
\begin{alignat*}{6}
F :\; &X& &\rightarrow& \; &\;\;Y&
\\
&x& &\mapsto& &F(x)&,
\end{alignat*}
is the natural \emph{pushforward} operation which maps the 
input tangent space to the output tangent space,
\begin{alignat*}{6}
F_{*} :\; &T_{x} X& &\rightarrow& \; &T_{F(x)} Y&
\\
&\;\;\;v& &\mapsto& &\;\;\;\;v_{*}&.
\end{alignat*}
The pushforward tangent vector quantifies the best linear 
approximation of the function along the direction of the 
tangent vector.  The best linear approximation of a 
composite function can be computed as a series of best 
linear approximations of the intermediate composite 
functions by applying their respective pushforward 
operations to the initial tangent vector.

Similarly, adjoints are more formally identified as
\emph{cotangent vectors}, elements of the \emph{cotangent 
space} $T^{*}_{x} X$ at the point $x \in X$ where the 
composite function is being evaluated.  The propagation 
of a cotangent vector backwards through a function is 
the natural \emph{pullback} operation which maps the 
output cotangent space to the input cotangent space,
\begin{alignat*}{6}
F^{*} :\; &T^{*}_{F(x)} Y& &\rightarrow& \; &T^{*}_{x} X&
\\
&\;\;\;\alpha& &\mapsto& &\;\;\;\;\alpha^{*}&.
\end{alignat*}

Consequently the basis of automatic differentiation follows 
immediately from the objects and  transformations that arise 
naturally in differential geometry \citep{Pusch:1996}.  
The utility of this more formal perspective, besides the 
immediate generalization of automatic differentiation to 
any smooth manifold, is that it identifies the well-posed 
differential operators amenable to automatic differentiation.  
In particular, the extension of the geometric picture to 
higher-order derivatives clearly lays out the well-defined 
higher-order differential operations and their implementations, 
providing a principled foundation for building high-performance 
higher-order automatic differentiation algorithms.

\section{Jet Setting}

Tangent and cotangent vectors are the simplest instances of
\emph{jets} which approximate functions between smooth
manifolds through their differential structure.  More general 
jet spaces, especially the velocity and covelocity spaces, 
define higher-order approximations of functions and their 
transformations that lie at the heart of higher-order
differential geometry.

In this section I will assume a familiarity with differential
geometry and its contemporary notation -- for a review I
highly recommend \cite{Lee:2013}.  For a more formal review
of the theory of jet spaces see \cite{KolarEtAl:1993}.

\subsection{Jets and Some of Their Properties}

Tangent vectors, cotangent vectors, and jets are all equivalence
classes of particular functions that look identical up to a certain 
differential order.  A tangent vector at a point $x \in X$, for 
example, is an equivalence class of curves in a manifold,
\begin{equation*}
c : \mathbb{R} \rightarrow X
\end{equation*}
that intersect at $x$ and share the same first derivative at that 
intersection.  Similarly, a cotangent vector at $x \in X$ is an 
equivalence class of real-valued functions that vanish at $x$,
\begin{equation*}
f : X \rightarrow \mathbb{R}, f(x) = 0,
\end{equation*}
and share the same first derivative at $x$.  Jets generalize 
these concepts to equivalence classes of functions between any 
two smooth manifolds, $F : X \rightarrow Y$ that share the same 
Taylor expansion truncated to a given order.

More formally consider a point in the input space $x \in X$ 
and a surrounding neighborhood $x \in U \subset X$.  Let 
$\phi_{X} : U \rightarrow \mathbb{R}^{D_{X}}$ be a chart over 
$U$ centered on $x$ with the $D_{X}$ coordinate functions 
$x^{i}_{\phi} : U \rightarrow \mathbb{R}$ satisfying
$x^{i}_{\phi}(x) = 0$.  Finally let 
$\phi_{Y} : F(U) \rightarrow \mathbb{R}^{D_{Y}}$ be a chart 
over the image neighborhood $F(x) \in F(U) \subset Y$ with 
the $D_{Y}$ corresponding coordinate functions
$y^{i}_{\phi} : U \rightarrow \mathbb{R}$.

Composing a function $F: X \rightarrow Y$ with the chart on $U$ 
and coordinate functions in $F(U)$ yields the real-valued function
\begin{equation*}
y^{i}_{\phi} \circ F \circ \phi^{-1}_{X}: 
\mathbb{R}^{D_{X}} \rightarrow \mathbb{R}^{D_{Y}}
\end{equation*}
from which we can define the \emph{higher-order Jacobian arrays},
\begin{equation*}
(\mathcal{J}_{F})^{j}_{i_{1} \ldots i_{D_{X}}} (x)
\equiv
\frac{ \partial (y^{j}_{\phi} \circ F \circ \phi^{-1}_{X}) }
{ \partial x^{i_{1}} \ldots \partial x^{i_{D_{X}}} }(x).
\end{equation*}
Here the index $j$ ranges from $1$ to $D_{Y}$ and the
indices $i_{k}$ each range from $1$ to $D_{X}$.  These Jacobian 
arrays are \emph{not} tensors but rather transform in much
more subtle ways, as we will see below.

Using these higher-order Jacobian arrays we can define the 
Taylor expansion of $F$ at $x$ as
\begin{align*}
y^{l}_{\phi} \circ F(x')
&= \quad\quad\,
y^{l}_{\phi} \circ F(x) 
\\
& \quad\quad\, +
(\mathcal{J}_{F})^{l}_{i}(x) \cdot x^{i}_{\phi} (x')
\\
& \quad +
\frac{1}{2!} \,
(\mathcal{J}_{F})^{l}_{ij}(x)
\cdot x^{i}_{\phi} (x') \cdot x^{j}_{\phi} (x')
\\
& \quad +
\frac{1}{3!} \,
(\mathcal{J}_{F})^{l}_{ijk}(x)
\cdot x^{i}_{\phi} (x') \cdot x^{j}_{\phi} (x') \cdot x^{k}_{\phi} (x')
\\
& \quad + \ldots.
\end{align*}
An \emph{$R$-truncated Taylor expansion} in the neighborhood $U$
then follows as a Taylor expansion truncated to include only the 
first $R$ terms.  For example, the $2$-truncated Taylor expansion
would be
\begin{align*}
T_{2} (y^{l}_{\phi} \circ F(x'))
&= \quad\quad\,
y^{l}_{\phi} \circ F(x) 
\\
& \quad\quad\, +
(\mathcal{J}_{F})^{l}_{i}(x) \cdot x^{i}_{\phi} (x')
\\
& \quad +
\frac{1}{2!} \,
(\mathcal{J}_{F})^{l}_{ij}(x)
\cdot x^{i}_{\phi} (x') \cdot x^{j}_{\phi} (x').
\end{align*}

An equivalence class of functions from $X$ to $Y$ that share the 
same $R$-truncated Taylor expansion at $x \in X$ is known as an 
\emph{$R$-jet}, $j^{R}_{x}$, from the \emph{source} $x$ to the 
\emph{target} $F(x)$.  The jet corresponding to a particular 
function $F : X \rightarrow Y$ is denoted $j^{R}_{x} (F)$. 
While the specific value of each term in a Taylor expansion depends 
on the chosen charts, these equivalence classes, and hence the
jets themselves, do not and hence characterize the inherent
geometric structure of the maps.

The \emph{jet space} of all $R$-jets between $X$ and $Y$ 
regardless of their sources or targets is denoted $J^{R}(X, Y)$.  
Likewise the restriction of the jet space to the source $x$ 
is denoted $J^{R}_{x}(X, Y)$ the restriction to the target 
$y$ is denoted $J^{R}(X, Y)_{y}$, and the restriction to the
source $x$ and the target $y$ at the same time is denoted 
\begin{equation*}
J^{R}_{x}(X, Y)_{y} = J^{R}_{x}(X, Y) \cap J^{R}(X, Y)_{y}.
\end{equation*}

In order for two functions to share the same $R$-truncated Taylor 
expansion they must also share the same $(R - 1)$-truncated Taylor
expansion, the same $(R - 2)$-truncated Taylor expansion, and so
on.  Consequently there is a natural projection from the $R$-th
order jet space to any lower-order jet space,
\begin{equation*}
\pi^{R}_{R - n} : J^{R}(X, Y) \rightarrow J^{R - n} (X, Y)
\end{equation*}
for any $1 \le n \le R$.  This projective structure is particularly
useful for specifying coordinate representations of jets as in 
order to preserve this projective structure the coordinates will 
necessarily decompose into first-order coordinates, second-order 
coordinates, and so on.

One of the most important properties of jets is that they inherit 
a compositional structure from the compositional structure of 
their underlying functions.  In particular we can define the 
composition of a jet $j^{R}_{x} (F_{1}) \in J^{R}_{x}(X, Y)$ and 
a jet $j^{R}_{y} (F_{2}) \in J^{R}_{y}(Y, Z)$ as the jet
\begin{equation*}
j^{R}_{y}(F_{2}) \circ j^{R}_{x}(F_{1}) 
= 
j^{R}_{x} (F_{2} \circ F_{1})
\in J^{R}_{x}(X, Z).
\end{equation*}
This compositional structure provides the basis for propagating 
higher-order jets, and the higher-order differential information 
that they encode, through a composite function.  These propagations, 
however, will prove to be much more complex than those for tangent 
vectors and cotangent vectors because the coordinates of 
higher-order jets transform in more intricate ways.

Within the charts $\phi_{X}$ and $\phi_{Y}$ an $R$-jet can be
specified by coordinates equal to the coefficients of the 
$R$-truncated Taylor expansion.  For example, a $2$-jet can be 
specified by the output values, $y^{l}_{\phi} \circ F(x)$, 
the components of the gradient, $(\mathcal{J}_{F})^{l}_{i}(x)$, 
and the components of the Hessian, $(\mathcal{J}_{F})^{l}_{ij}(x)$.  
Like the jets themselves, these coordinates transform in a complex 
manner.  These coordinates are isomorphic to the coefficients of 
a rank $R$ polynomial which allows for a more algebraic treatment 
of jets and their transformations as convenient.  In particular, 
in these coordinates the compositional structure of jets reduces 
to the compositional structure of the quotient of a polynomial 
ring by its $(R + 1)$st-order ideal.

Any jet space encodes differential information of functions 
between two manifolds, but two jet spaces in particular naturally 
generalize the geometric perspective of automatic differentiation 
to higher-orders: velocity spaces and covelocity spaces.

\subsection{Velocity Spaces}

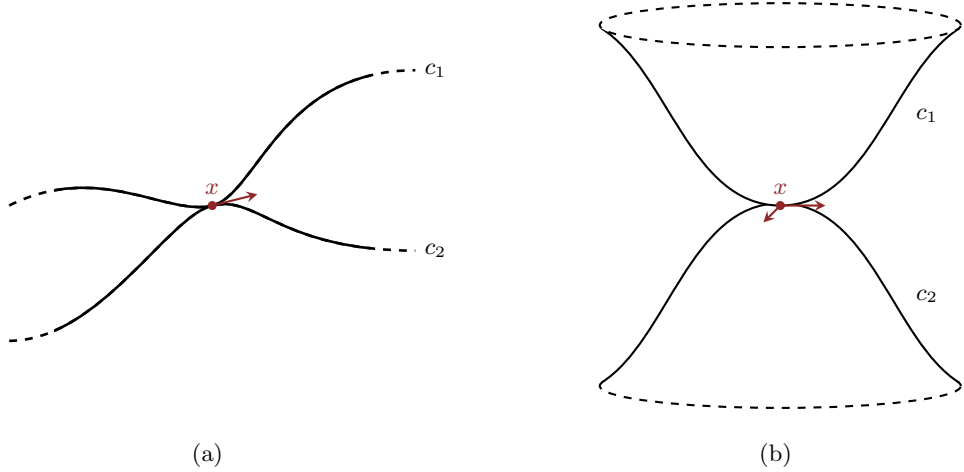
\begin{figure*}
\centering
\subfigure[]{
\begin{tikzpicture}[scale=0.3, thick]
  \draw [rounded corners=2pt, color=white] (-12, -10) rectangle (12, 10);
  
  \draw [line width=1, dashed] (-9, 0) .. controls (-5, 2) and (-2, -0.5) .. (0, 0)
  								 .. controls (2, 0.5) and (3, 6) .. (9, 6)
  node[right] { $c_{1}$ };

  \draw [line width=1, dashed] (-9, -6) .. controls (-5, -6) and (-2, -0.5) .. (0, 0)
  							      .. controls (2, 0.5) and (3, -2) .. (9, -2)
  node[right] { $c_{2}$ };

  \begin{scope}
  	\clip (-7, -7) rectangle (7, 7);
  	\draw [line width=1] (-9, 0) .. controls (-5, 2) and (-2, -0.5) .. (0, 0)
       							   .. controls (2, 0.5) and (3, 6) .. (9, 6);

    \draw [line width=1] (-9, -6) .. controls (-5, -6) and (-2, -0.5) .. (0, 0)
  							        .. controls (2, 0.5) and (3, -2) .. (9, -2);
  \end{scope}

  \node[color=dark] at (0, 0.75) { $x$ }; 
  \fill [fill=dark] (0, 0) circle (0.2);
  \draw[color=dark, ->, >=stealth] (0, 0) -- (2, 0.5);
\end{tikzpicture}
}
% Two-dimensional
\subfigure[]{
\begin{tikzpicture}[scale=0.3, thick,
declare function={ fr(\t) = 3 * (1 - \t) * (1 - \t) * \t * 4
                           + 3 * (1 - \t) * \t * \t * 5 + \t * \t * \t * 8;},
declare function={ fh(\t) =  3 * (1 - \t) * \t * \t * 6 + \t * \t * \t * 8;} 
]
  \draw [rounded corners=2pt, color=white] (-12, -10) rectangle (12, 10);

  \node[] at (6.5, 4) { $c_{1}$ };
  \node[] at (6.5, -4) { $c_{2}$ };

  \draw[dashed] (0, -8) circle [x radius=8, y radius={8 * sin(7)}];

  \filldraw[draw=black, fill=white] (-7.92, -7.85) .. controls (-5, -6) and (-4, 0.1) 
    .. (0, 0.1) .. controls (4, 0.1) and (5, -6) .. (7.92, -7.85);
    
  \filldraw[draw=black, fill=white] (-7.92, 7.85) .. controls (-5, 6) and (-4, 0) 
    .. (0, 0) .. controls (4, 0) and (5, 6) .. (7.92, 7.85);

  \filldraw[draw=black, fill=white, dashed] (0, 8) circle [x radius=8, y radius={8 * sin(7)}];

  \node[color=dark] at (0, 0.75) { $x$ }; 
  \fill [fill=dark] (0, 0) circle (0.2);
  \draw[color=dark, ->, >=stealth] (0, 0) -- (2, 0);
  \draw[color=dark, ->, >=stealth] (0, 0) -- (-0.75, -0.75);
\end{tikzpicture}
}
\caption{
Velocities are equivalence classes of surfaces immersed into
a manifold $X$ that behave the same within a neighborhood of a
given point $x \in X$. (a) $(1, 1)$-velocities at $x$ are 
equivalence classes of one-dimensional curves with the same
first-order behavior around $x$.  Here $c_{1}$ and $c_{2}$ 
fall into the same equivalence class.  (b) $(2, 1)$-velocities
at $x$ are equivalence classes of two-dimensional surfaces
with the same first-order behavior at $x$.  Here $c_{1}$ and
$c_{2}$ once again fall into the same equivalence class.
Higher-order velocities correspond to equivalence classes
of higher-order differential behavior around $x$.
}
\label{fig:velocities} 
\end{figure*}

Tangent vectors are formally defined as equivalence classes
of one-dimensional curves $c : \mathbb{R} \rightarrow X$ centered 
on a point $x \in X$.  Consequently a natural candidate for 
their generalization are the equivalence classes of higher-dimensional 
\emph{surfaces} immersed in $X$, $c^{k} : \mathbb{R}^{k} \rightarrow X$, or 
more formally the jet spaces  $J^{R}_{0} (\mathbb{R}^{k}, X)_{x}$
(Figure \ref{fig:velocities}).  Jets belonging to these spaces are 
denoted \emph{$(k, R)$-velocities at $x$}, with $k$ the dimension 
of the velocity and $R$ its order.

The term velocity here refers to a general notion of how 
quickly a surface is traversed around $x$ and not just the 
first-order speed for which the term velocity is used in 
physics.   For example, a $(1, 1)$-velocity corresponds to 
the curves sharing the same physical velocity at $x$, but 
a $(1, 2)$-velocity corresponds to curves sharing the same 
physical velocity and \emph{physical acceleration}.  
Higher-dimensional velocities correspond to equivalence 
classes of higher-dimensional surfaces within $X$ and so
require consideration of changes along multiple directions.

In local coordinates the Taylor expansion of a one-dimensional 
curve can be written as
\begin{equation*}
x^{i} (t_{1}) = 
\sum_{r = 1}^{R} \frac{t_{1}^{r}}{r!} 
\frac{\partial^{r} c^{i} }{ \partial t^{r}_{1}}.
\end{equation*}
Denoting
\begin{equation*}
(\delta^{r} v)^{i}
\equiv
\frac{\partial^{r} c^{i} }{ \partial t^{r}_{1}}
\end{equation*}
we can then specify an $(1, R)$-velocity locally with the 
coordinates
\begin{equation*}
\left(
\frac{\partial c^{i} }{ \partial t_{1}}, \ldots, 
\frac{\partial^{r} c^{i} }{ \partial t^{r}_{1}}
\right)
=
( (\delta v^{i}), \ldots, (\delta^{r} v^{i}) ).
\end{equation*}
This $\delta^{r}$ notation serves as a convenient bookkeeping
aid which allows any calculation to be quickly doubled-checked: 
any term involved in a calculation of the $R$-th order coordinates
of an $R$-jet will require $R$ $\delta$s.

Keep in mind that for $r > 1$ the $\delta^{r} v^{i}$ are simply
the elements of a one-dimensional array and do \emph{not} 
correspond to the components of a vector.  As we will see the 
higher-order cooordinates of a velocity transform much more 
intricately than tangent vectors.

Higher-dimensional velocities probe the local structure of $X$
in multiple directions at the same time.  In coordinates a 
$k$-dimensional velocity is specified with perturbations along 
each of the $k$ axes along with the corresponding cross terms.  
For example, a $(2, 2)$-velocity 
in $J^{2}_{0}(\mathbb{R}^{2}, X)_{0}$ features the aligned
coordinates
\begin{align*}
(\delta v)^{i} = \frac{\partial c^{i} }{ \partial t_{1}}, \,
(\delta^{2} v)^{i} = \frac{\partial^{2} c^{i} }{ \partial t^{2}_{1}},
\\
(\delta u)^{i} = \frac{\partial c^{i} }{ \partial t_{2}}, \,
(\delta^{2} u)^{i} = \frac{\partial^{2} c^{i} }{ \partial t^{2}_{2}},
\end{align*}
and the crossed coordinates,
\begin{equation*}
(\delta v \delta u)^{i} = \frac{\partial^{2} c^{i} }{ \partial t_{1} \partial t_{2}}.
\end{equation*}

Similarly, a $(3, 3)$-velocity in $J^{3}_{0}(\mathbb{R}^{3}, X)$ 
features the aligned coordinates
\begin{align*}
(\delta v)^{i} = \frac{\partial c^{i} }{ \partial t_{1}}, \,
(\delta^{2} v)^{i} &= \frac{\partial^{2} c^{i} }{ \partial t_{1}^{2}}, \,
(\delta^{3} v)^{i} = \frac{\partial^{3} c^{i} }{ \partial t_{1}^{3}},
\\
(\delta u)^{i} = \frac{\partial c^{i} }{ \partial t_{2}}, \,
(\delta^{2} u)^{i} &= \frac{\partial^{2} c^{i} }{ \partial t_{2}^{2}}, \,
(\delta^{3} u)^{i} = \frac{\partial^{3} c^{i} }{ \partial t_{2}^{3}},
\\
(\delta w)^{i} = \frac{\partial c^{i} }{ \partial t_{3}}, \,
(\delta^{2} w)^{i} &= \frac{\partial^{2} c^{i} }{ \partial t_{3}^{2}}, \,
(\delta^{3} w)^{i} = \frac{\partial^{3} c^{i} }{ \partial t_{3}^{3}},
\end{align*}
and the crossed coordinates,
\begin{align*}
(\delta v \delta u)^{i} = \frac{\partial^{2} c^{i} }{ \partial t_{1} \partial t_{2}}, \,
(\delta u \delta w)^{i} &= \frac{\partial^{2} c^{i} }{ \partial t_{2} \partial t_{3}}, \,
(\delta v \delta w)^{i} = \frac{\partial^{2} c^{i} }{ \partial t_{1} \partial t_{3}},
\\
(\delta^{2} v \delta u)^{i} = \frac{\partial^{3} c^{i} }{ \partial t_{1}^{2} \partial t_{2}}, \,
(\delta v \delta^{2} u)^{i} &= \frac{\partial^{3} c^{i} }{ \partial t_{1} \partial t_{2}^{2}}, \,
(\delta^{2} u \delta w)^{i} = \frac{\partial^{3} c^{i} }{ \partial t_{2}^{2} \partial t_{3}},
\\
(\delta u \delta^{2} w)^{i} = \frac{\partial^{3} c^{i} }{ \partial t_{2} \partial t_{3}^{2}}, \,
(\delta^{2} v \delta w)^{i} &= \frac{\partial^{3} c^{i} }{ \partial t_{1}^{2} \partial t_{3}}, \,
(\delta v \delta^{2} w)^{i} = \frac{\partial^{3} c^{i} }{ \partial t_{1} \partial t_{3}^{2}},
\\
(\delta v \delta u \delta w)^{i} 
&= 
\frac{\partial^{3} c^{i} }{ \partial t_{1} \partial t_{2} \partial t_{3}}.
\end{align*}

The coordinates with $R$ $\delta$s specify the $R$-th order 
structure of the velocity, and exhibit the natural projective
structure of jets.  For example, a $(2, 2)$-velocity
specified with the coordinates 
\begin{equation*}
\delta v \delta u = ( (\delta v)^{i}, (\delta u)^{i}, 
(\delta^{2} v)^{i}, (\delta v \delta u)^{i}, (\delta^{2} u)^{i} )
\end{equation*}
projects to a $(2, 1)$-velocity specified with the coordinates
\begin{equation*}
\pi^{2}_{1} (\delta v \delta u) = ( (\delta v)^{i}, (\delta u)^{i} ).
\end{equation*}
Because of the product structure of the real numbers, the
$(2, 1)$-velocity can further be projected to the 
$(1, 1)$-velocity with coordinates $(\delta v)^{i}$
or the $(1, 1)$-velocity with coordinates $(\delta u)^{i}$.

\subsection{Covelocity Spaces}

\begin{figure*}
\centering
\subfigure[]{
\begin{tikzpicture}[scale=0.3, thick,
declare function={ f1_1(\x) = 0 + 0.333333 * \x;},
declare function={ f1_2(\x) = 0 + 0.333333 * \x + 0.5 * 1.97531 * \x * \x;},
declare function={ f2_1(\x) = 0 + 0.333333 * \x;},
declare function={ f2_2(\x) = 0 + 0.333333 * \x + 0.5 * -0.987654 * \x * \x;}
]

  \pgfmathsetmacro{\dx}{0}
  \pgfmathsetmacro{\dy}{0}
  
  \draw [rounded corners=2pt, color=white] (-11 + \dx, -5 + \dy) rectangle +(22, 10);

  \draw [color=gray91]   (-9 + \dx, 1 + \dy) .. controls (-5 + \dx, 5 + \dy) and (-1.5 + \dx, -0.5 + \dy) 
    				  .. (0 + \dx, 0 + \dy) .. controls (1.5 + \dx, 0.5 + \dy) and (3 + \dx, 6 + \dy) 
   					  .. (9 + \dx, 3 + \dy);

  \draw [color=gray92]    (-9 + \dx, 3 + \dy) .. controls (-5 + \dx, -2 + \dy) and (-2 + \dx, -1 + \dy) 
                       .. (0 + \dx, 0 + \dy) .. controls (2 + \dx, 1 + \dy) and (3 + \dx, 3 + \dy) 
                       .. (9 + \dx, 1 + \dy);
								 
  \draw [color=gray93]    (-9 + \dx, -3 + \dy) .. controls (-5 + \dx, -5 + \dy) and (-1.5 + \dx, -0.5 + \dy) 
                       .. (0 + \dx, 0 + \dy) .. controls (1.5 + \dx, 0.5 + \dy) and (3 + \dx, 1 + \dy)
                       .. (9 + \dx, -2 + \dy);
								 
  \draw [color=gray94]    (-9 + \dx, 0 + \dy) .. controls (-5 + \dx, 5 + \dy) and (-2 + \dx, 2 + \dy) 
                       .. (0 + \dx, 0 + \dy) .. controls (2 + \dx, -2 + \dy) and (3 + \dx, -3 + \dy) 
                       .. (9 + \dx, -4 + \dy);
								 
  \draw [color=gray95]    (-9 + \dx, -1 + \dy) .. controls (-5 + \dx, -3 + \dy) and (-2 + \dx, -1.5 + \dy) 
                       .. (0 + \dx, 0 + \dy) .. controls (2 + \dx, 1.5 + \dy) and (3 + \dx, 4 + \dy) 
                       .. (9 + \dx, 2 + \dy);		
								 
  \draw [line width=1] (-9 + \dx, 0 + \dy) -- +(18, 0)
  node[right] { $X$ };
								 
  \draw [color=mid, line width=1] 
       (-9 + \dx, 1 + \dy) .. controls (-5 + \dx, 5 + \dy) and (-1.5 + \dx, -0.5 + \dy) 
    .. (0 + \dx, 0 + \dy) .. controls (1.5 + \dx, 0.5 + \dy) and (3 + \dx, 6 + \dy) 
    .. (9 + \dx, 3 + \dy);
  \node[color=mid] at (-10 + \dx, 1 + \dy) { $f_{1}$ };

  \fill [fill=dark] (0 + \dx, 0 + \dy) circle (0.15);
  \node[color=dark] at (-0 + \dx, -0.75 + \dy) { $x$ };
  
  \node[color=mid] at (3.5 + \dx, 0.75 + \dy) { $f_{1}(x) = 0$ };
  
\pgfmathsetmacro{\dx}{24}
  \pgfmathsetmacro{\dy}{0}
  
  \draw [rounded corners=2pt, color=white] (-11 + \dx, -5 + \dy) rectangle +(22, 10);

  \draw [color=gray91]   (-9 + \dx, 1 + \dy) .. controls (-5 + \dx, 5 + \dy) and (-1.5 + \dx, -0.5 + \dy) 
    				  .. (0 + \dx, 0 + \dy) .. controls (1.5 + \dx, 0.5 + \dy) and (3 + \dx, 6 + \dy) 
   					  .. (9 + \dx, 3 + \dy);

  \draw [color=gray92]    (-9 + \dx, 3 + \dy) .. controls (-5 + \dx, -2 + \dy) and (-2 + \dx, -1 + \dy) 
                       .. (0 + \dx, 0 + \dy) .. controls (2 + \dx, 1 + \dy) and (3 + \dx, 3 + \dy) 
                       .. (9 + \dx, 1 + \dy);
								 
  \draw [color=gray93]    (-9 + \dx, -3 + \dy) .. controls (-5 + \dx, -5 + \dy) and (-1.5 + \dx, -0.5 + \dy) 
                       .. (0 + \dx, 0 + \dy) .. controls (1.5 + \dx, 0.5 + \dy) and (3 + \dx, 1 + \dy)
                       .. (9 + \dx, -2 + \dy);
								 
  \draw [color=gray94]    (-9 + \dx, 0 + \dy) .. controls (-5 + \dx, 5 + \dy) and (-2 + \dx, 2 + \dy) 
                       .. (0 + \dx, 0 + \dy) .. controls (2 + \dx, -2 + \dy) and (3 + \dx, -3 + \dy) 
                       .. (9 + \dx, -4 + \dy);
								 
  \draw [color=gray95]    (-9 + \dx, -1 + \dy) .. controls (-5 + \dx, -3 + \dy) and (-2 + \dx, -1.5 + \dy) 
                       .. (0 + \dx, 0 + \dy) .. controls (2 + \dx, 1.5 + \dy) and (3 + \dx, 4 + \dy) 
                       .. (9 + \dx, 2 + \dy);		
								 
  \draw [line width=1] (-9 + \dx, 0 + \dy) -- +(18, 0)
  node[right] { $X$ };
								 
  \draw [color=mid, line width=1] 
       (-9 + \dx, -3 + \dy) .. controls (-5 + \dx, -5 + \dy) and (-1.5 + \dx, -0.5 + \dy) 
    .. (0 + \dx, 0 + \dy) .. controls (1.5 + \dx, 0.5 + \dy) and (3 + \dx, 1 + \dy)
    .. (9 + \dx, -2 + \dy);
  \node[color=mid] at (-10 + \dx, -3 + \dy) { $f_{2}$ };

  \fill [fill=dark] (0 + \dx, 0 + \dy) circle (0.15);
  \node[color=dark] at (-0 + \dx, -0.75 + \dy) { $x$ };
  
  \node[color=mid] at (-1 + \dx, 1 + \dy) { $f_{2}(x) = 0$ };

\end{tikzpicture}
}
\subfigure[]{
\begin{tikzpicture}[scale=0.3, thick,
declare function={ f1_1(\x) = 0 + 0.333333 * \x;},
declare function={ f1_2(\x) = 0 + 0.333333 * \x + 0.5 * 1.97531 * \x * \x;},
declare function={ f2_1(\x) = 0 + 0.333333 * \x;},
declare function={ f2_2(\x) = 0 + 0.333333 * \x + 0.5 * -0.987654 * \x * \x;}
]

\pgfmathsetmacro{\dx}{0}
  \pgfmathsetmacro{\dy}{-12}
  
  \draw [rounded corners=2pt, color=white] (-11 + \dx, -5 + \dy) rectangle +(22, 10);

  \draw [color=gray91]   (-9 + \dx, 1 + \dy) .. controls (-5 + \dx, 5 + \dy) and (-1.5 + \dx, -0.5 + \dy) 
    				  .. (0 + \dx, 0 + \dy) .. controls (1.5 + \dx, 0.5 + \dy) and (3 + \dx, 6 + \dy) 
   					  .. (9 + \dx, 3 + \dy);

  \draw [color=gray92]    (-9 + \dx, 3 + \dy) .. controls (-5 + \dx, -2 + \dy) and (-2 + \dx, -1 + \dy) 
                       .. (0 + \dx, 0 + \dy) .. controls (2 + \dx, 1 + \dy) and (3 + \dx, 3 + \dy) 
                       .. (9 + \dx, 1 + \dy);
								 
  \draw [color=gray93]    (-9 + \dx, -3 + \dy) .. controls (-5 + \dx, -5 + \dy) and (-1.5 + \dx, -0.5 + \dy) 
                       .. (0 + \dx, 0 + \dy) .. controls (1.5 + \dx, 0.5 + \dy) and (3 + \dx, 1 + \dy)
                       .. (9 + \dx, -2 + \dy);
								 
  \draw [color=gray94]    (-9 + \dx, 0 + \dy) .. controls (-5 + \dx, 5 + \dy) and (-2 + \dx, 2 + \dy) 
                       .. (0 + \dx, 0 + \dy) .. controls (2 + \dx, -2 + \dy) and (3 + \dx, -3 + \dy) 
                       .. (9 + \dx, -4 + \dy);
								 
  \draw [color=gray95]    (-9 + \dx, -1 + \dy) .. controls (-5 + \dx, -3 + \dy) and (-2 + \dx, -1.5 + \dy) 
                       .. (0 + \dx, 0 + \dy) .. controls (2 + \dx, 1.5 + \dy) and (3 + \dx, 4 + \dy) 
                       .. (9 + \dx, 2 + \dy);				
								 
  \fill [color=light, opacity=0.25] (-1 + \dx, -5 + \dy) rectangle +(2, 10);
  \draw [dashed, color=light, opacity=0.25] (-1 + \dx, -5 + \dy) rectangle +(2, 10);

  \draw [line width=1] (-9 + \dx, 0 + \dy) -- +(18, 0)
  node[right] { $X$ };
								 
  \draw [color=mid, line width=1] 
       (-9 + \dx, 1 + \dy) .. controls (-5 + \dx, 5 + \dy) and (-1.5 + \dx, -0.5 + \dy) 
    .. (0 + \dx, 0 + \dy) .. controls (1.5 + \dx, 0.5 + \dy) and (3 + \dx, 6 + \dy) 
    .. (9 + \dx, 3 + \dy);
  \node[color=mid] at (-10 + \dx, 1 + \dy) { $f_{1}$ };
  
  \fill [fill=dark] (0 + \dx, 0 + \dy) circle (0.15);
  \node[color=dark] at (-0 + \dx, -0.75 + \dy) { $x$ };
  
  \draw[domain=-1:1, smooth, samples=20, variable=\x, line width=1, color=dark] 
    plot ({\x + \dx}, {f1_1(\x) + \dy});

  \pgfmathsetmacro{\dx}{24}
  \pgfmathsetmacro{\dy}{-12}
  
  \draw [rounded corners=2pt, color=white] (-11 + \dx, -5 + \dy) rectangle +(22, 10);

  \draw [color=gray91]   (-9 + \dx, 1 + \dy) .. controls (-5 + \dx, 5 + \dy) and (-1.5 + \dx, -0.5 + \dy) 
    				  .. (0 + \dx, 0 + \dy) .. controls (1.5 + \dx, 0.5 + \dy) and (3 + \dx, 6 + \dy) 
   					  .. (9 + \dx, 3 + \dy);

  \draw [color=gray92]    (-9 + \dx, 3 + \dy) .. controls (-5 + \dx, -2 + \dy) and (-2 + \dx, -1 + \dy) 
                       .. (0 + \dx, 0 + \dy) .. controls (2 + \dx, 1 + \dy) and (3 + \dx, 3 + \dy) 
                       .. (9 + \dx, 1 + \dy);
								 
  \draw [color=gray93]    (-9 + \dx, -3 + \dy) .. controls (-5 + \dx, -5 + \dy) and (-1.5 + \dx, -0.5 + \dy) 
                       .. (0 + \dx, 0 + \dy) .. controls (1.5 + \dx, 0.5 + \dy) and (3 + \dx, 1 + \dy)
                       .. (9 + \dx, -2 + \dy);
								 
  \draw [color=gray94]    (-9 + \dx, 0 + \dy) .. controls (-5 + \dx, 5 + \dy) and (-2 + \dx, 2 + \dy) 
                       .. (0 + \dx, 0 + \dy) .. controls (2 + \dx, -2 + \dy) and (3 + \dx, -3 + \dy) 
                       .. (9 + \dx, -4 + \dy);
								 
  \draw [color=gray95]    (-9 + \dx, -1 + \dy) .. controls (-5 + \dx, -3 + \dy) and (-2 + \dx, -1.5 + \dy) 
                       .. (0 + \dx, 0 + \dy) .. controls (2 + \dx, 1.5 + \dy) and (3 + \dx, 4 + \dy) 
                       .. (9 + \dx, 2 + \dy);	
                       
  \fill [color=light, opacity=0.25] (-1 + \dx, -5 + \dy) rectangle +(2, 10);
  \draw [dashed, color=light, opacity=0.25] (-1 + \dx, -5 + \dy) rectangle +(2, 10);

  \draw [line width=1] (-9 + \dx, 0 + \dy) -- +(18, 0)
  node[right] { $X$ };
								 
  \draw [color=mid, line width=1] 
       (-9 + \dx, -3 + \dy) .. controls (-5 + \dx, -5 + \dy) and (-1.5 + \dx, -0.5 + \dy) 
    .. (0 + \dx, 0 + \dy) .. controls (1.5 + \dx, 0.5 + \dy) and (3 + \dx, 1 + \dy)
    .. (9 + \dx, -2 + \dy);
  \node[color=mid] at (-10 + \dx, -3 + \dy) { $f_{2}$ };
  
  \fill [fill=dark] (0 + \dx, 0 + \dy) circle (0.15);
  \node[color=dark] at (-0 + \dx, -0.75 + \dy) { $x$ };
  
  \draw[domain=-1:1, smooth, samples=20, variable=\x, line width=1, color=dark] 
    plot ({\x + \dx}, {f2_1(\x) + \dy});

\end{tikzpicture}
}
\subfigure[]{
\begin{tikzpicture}[scale=0.3, thick,
declare function={ f1_1(\x) = 0 + 0.333333 * \x;},
declare function={ f1_2(\x) = 0 + 0.333333 * \x + 0.5 * 1.97531 * \x * \x;},
declare function={ f2_1(\x) = 0 + 0.333333 * \x;},
declare function={ f2_2(\x) = 0 + 0.333333 * \x + 0.5 * -0.987654 * \x * \x;}
]

  \pgfmathsetmacro{\dx}{0}
  \pgfmathsetmacro{\dy}{-24}
  
  \draw [rounded corners=2pt, color=white] (-11 + \dx, -5 + \dy) rectangle +(22, 10);

  \draw [color=gray91]   (-9 + \dx, 1 + \dy) .. controls (-5 + \dx, 5 + \dy) and (-1.5 + \dx, -0.5 + \dy) 
    				  .. (0 + \dx, 0 + \dy) .. controls (1.5 + \dx, 0.5 + \dy) and (3 + \dx, 6 + \dy) 
   					  .. (9 + \dx, 3 + \dy);

  \draw [color=gray92]    (-9 + \dx, 3 + \dy) .. controls (-5 + \dx, -2 + \dy) and (-2 + \dx, -1 + \dy) 
                       .. (0 + \dx, 0 + \dy) .. controls (2 + \dx, 1 + \dy) and (3 + \dx, 3 + \dy) 
                       .. (9 + \dx, 1 + \dy);
								 
  \draw [color=gray93]    (-9 + \dx, -3 + \dy) .. controls (-5 + \dx, -5 + \dy) and (-1.5 + \dx, -0.5 + \dy) 
                       .. (0 + \dx, 0 + \dy) .. controls (1.5 + \dx, 0.5 + \dy) and (3 + \dx, 1 + \dy)
                       .. (9 + \dx, -2 + \dy);
								 
  \draw [color=gray94]    (-9 + \dx, 0 + \dy) .. controls (-5 + \dx, 5 + \dy) and (-2 + \dx, 2 + \dy) 
                       .. (0 + \dx, 0 + \dy) .. controls (2 + \dx, -2 + \dy) and (3 + \dx, -3 + \dy) 
                       .. (9 + \dx, -4 + \dy);
								 
  \draw [color=gray95]    (-9 + \dx, -1 + \dy) .. controls (-5 + \dx, -3 + \dy) and (-2 + \dx, -1.5 + \dy) 
                       .. (0 + \dx, 0 + \dy) .. controls (2 + \dx, 1.5 + \dy) and (3 + \dx, 4 + \dy) 
                       .. (9 + \dx, 2 + \dy);	
								 
  \fill [color=light, opacity=0.25] (-1 + \dx, -5 + \dy) rectangle +(2, 10);
  \draw [dashed, color=light, opacity=0.25] (-1 + \dx, -5 + \dy) rectangle +(2, 10);

  \draw [line width=1] (-9 + \dx, 0 + \dy) -- +(18, 0)
  node[right] { $X$ };	
  
  \draw [color=mid, line width=1] 
       (-9 + \dx, 1 + \dy) .. controls (-5 + \dx, 5 + \dy) and (-1.5 + \dx, -0.5 + \dy) 
    .. (0 + \dx, 0 + \dy) .. controls (1.5 + \dx, 0.5 + \dy) and (3 + \dx, 6 + \dy) 
    .. (9 + \dx, 3 + \dy);			 
  \node[color=mid] at (-10 + \dx, 1 + \dy) { $f_{1}$ };
  
  \fill [fill=dark] (0 + \dx, 0 + \dy) circle (0.15);
  \node[color=dark] at (-0 + \dx, -0.75 + \dy) { $x$ };
  
  \draw[domain=-1:1, smooth, samples=20, variable=\x, line width=1, color=dark] 
    plot ({\x + \dx}, {f1_2(\x) + \dy});

  \pgfmathsetmacro{\dx}{24}
  \pgfmathsetmacro{\dy}{-24}
  
  \draw [rounded corners=2pt, color=white] (-11 + \dx, -5 + \dy) rectangle +(22, 10);

  \draw [color=gray91]   (-9 + \dx, 1 + \dy) .. controls (-5 + \dx, 5 + \dy) and (-1.5 + \dx, -0.5 + \dy) 
    				  .. (0 + \dx, 0 + \dy) .. controls (1.5 + \dx, 0.5 + \dy) and (3 + \dx, 6 + \dy) 
   					  .. (9 + \dx, 3 + \dy);

  \draw [color=gray92]    (-9 + \dx, 3 + \dy) .. controls (-5 + \dx, -2 + \dy) and (-2 + \dx, -1 + \dy) 
                       .. (0 + \dx, 0 + \dy) .. controls (2 + \dx, 1 + \dy) and (3 + \dx, 3 + \dy) 
                       .. (9 + \dx, 1 + \dy);
								 
  \draw [color=gray93]    (-9 + \dx, -3 + \dy) .. controls (-5 + \dx, -5 + \dy) and (-1.5 + \dx, -0.5 + \dy) 
                       .. (0 + \dx, 0 + \dy) .. controls (1.5 + \dx, 0.5 + \dy) and (3 + \dx, 1 + \dy)
                       .. (9 + \dx, -2 + \dy);
								 
  \draw [color=gray94]    (-9 + \dx, 0 + \dy) .. controls (-5 + \dx, 5 + \dy) and (-2 + \dx, 2 + \dy) 
                       .. (0 + \dx, 0 + \dy) .. controls (2 + \dx, -2 + \dy) and (3 + \dx, -3 + \dy) 
                       .. (9 + \dx, -4 + \dy);
								 
  \draw [color=gray95]    (-9 + \dx, -1 + \dy) .. controls (-5 + \dx, -3 + \dy) and (-2 + \dx, -1.5 + \dy) 
                       .. (0 + \dx, 0 + \dy) .. controls (2 + \dx, 1.5 + \dy) and (3 + \dx, 4 + \dy) 
                       .. (9 + \dx, 2 + \dy);		
								 
  \fill [color=light, opacity=0.25] (-1 + \dx, -5 + \dy) rectangle +(2, 10);
  \draw [dashed, color=light, opacity=0.25] (-1 + \dx, -5 + \dy) rectangle +(2, 10);

  \draw [line width=1] (-9 + \dx, 0 + \dy) -- +(18, 0)
  node[right] { $X$ };
								 
  \draw [color=mid, line width=1] 
       (-9 + \dx, -3 + \dy) .. controls (-5 + \dx, -5 + \dy) and (-1.5 + \dx, -0.5 + \dy) 
    .. (0 + \dx, 0 + \dy) .. controls (1.5 + \dx, 0.5 + \dy) and (3 + \dx, 1 + \dy)
    .. (9 + \dx, -2 + \dy);
  \node[color=mid] at (-10 + \dx, -3 + \dy) { $f_{2}$ };

  \fill [fill=dark] (0 + \dx, 0 + \dy) circle (0.15);
  \node[color=dark] at (-0 + \dx, -0.75 + \dy) { $x$ };
  
  \draw[domain=-1:1, smooth, samples=20, variable=\x, line width=1, color=dark] 
    plot ({\x + \dx}, {f2_2(\x) + \dy});

\end{tikzpicture}
}
\caption{
Covelocities at a point $x \in X$ are equivalence classes of real-valued 
functions that vanish at $x$ with the same differential behavior around 
$x$ up to a given order.  (a) One-dimensional covelocities at $x$ are 
equivalence classes of scalar real-valued functions.  (b) Because $f_{1}$
and $f_{2}$ have the same linear behavior around $x$ they belong to the
same $(1, 1)$-covelocity.  (c) The two functions feature different 
second-order behavior, however, and hence belong to \emph{different}
$(1, 2)$-covelocities.
}
\label{fig:covelocities} 
\end{figure*}
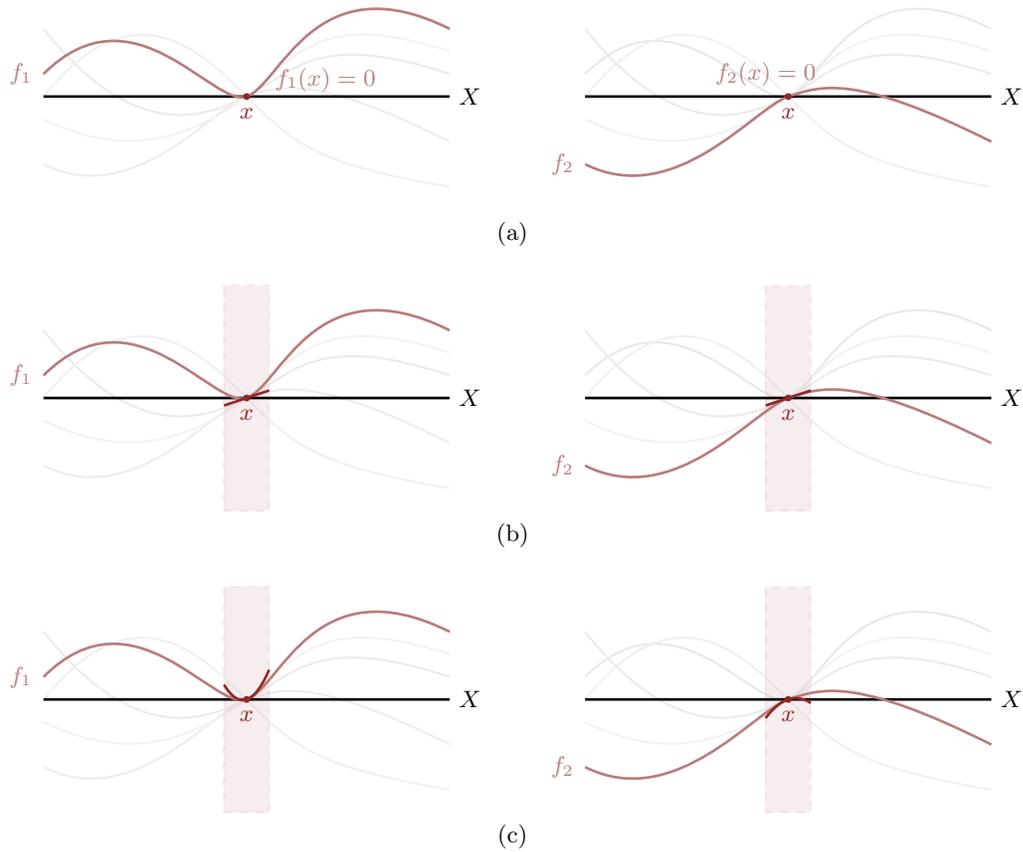

Cotangent vectors are equivalence classes of real-valued
functions $f : X \rightarrow \mathbb{R}$  that vanish at 
$x \in X$.  Consequently a natural jet spaces to consider 
for generalizing the behavior of cotangent vectors are 
$J^{R}_{x}(X, \mathbb{R}^{k})_{0}$ (Figure \ref{fig:covelocities}).  
Jets in these spaces are denoted \emph{$(k, R)$-covelocties} 
where once again $k$ refers to the dimension of a covelocity 
and $R$ its order.

In local coordinates the Taylor expansion of a function 
$f : X \rightarrow \mathbb{R}$ that vanishes at the input 
point is completely specified by the corresponding Jacobian 
arrays.  Consequently we can define coordinates for 
$(1, R)$-covelocities as collections of real numbers with 
the same symmetries as the Jacobian arrays. For example, 
a $(1, 1)$-covelocity is specified with the coordinates
\begin{equation*}
( (\mathfrak{d} \alpha)_{i}).
\end{equation*}
Similarly a $(1, 2)$-covelocity is specified with the 
coordinates
\begin{equation*}
( (\mathfrak{d} \alpha)_{i}, (\mathfrak{d}^{2} \alpha)_{ij}),
\end{equation*}
where $(\mathfrak{d}^{2} \alpha)_{ij} = (\mathfrak{d}^{2} \alpha)_{ji}$,
and a $(1, 3)$-covelocity is specified with the coordinates
\begin{equation*}
( (\mathfrak{d} \alpha)_{i}, (\mathfrak{d}^{2} \alpha)_{ij}, (\mathfrak{d}^{3} \alpha)_{ijk}),
\end{equation*}
where 
$(\mathfrak{d}^{3} \alpha)_{ijk} = (\mathfrak{d}^{3} \alpha)_{jik} 
= (\mathfrak{d}^{3} \alpha)_{kji}$.  In each case the indices run 
from 1 to $D_{X}$.

The coordinates of a covelocity belonging to a specific 
function $f : X \rightarrow \mathbb{R}$ are given by the
corresponding Jacobian arrays.  For example, a (1, 3)-covelocity
for the function $f$ at $x$ is specified by the coordinates
\begin{alignat*}{3}
(\mathfrak{d} \alpha (f) )_{i}
&=
(\mathcal{J}_{f})^{l}_{i} (x)
&=& \,
\frac{ \partial (y^{l}_{\phi} \circ f) }{ \partial x^{i} } (x)
\\
(\mathfrak{d}^{2} \alpha (f) )_{ij}
&=
(\mathcal{J}_{f})^{l}_{ij} (x)
&=& \,
\frac{ \partial (y^{l}_{\phi} \circ f) }
{ \partial x^{i} \partial x^{j} } (x)
\\
(\mathfrak{d}^{3} \alpha (f) )_{ijk}
&=
(\mathcal{J}_{F})^{l}_{ijk} (x) \,
&=& \,
\frac{ \partial (y^{j}_{\phi} \circ f) }
{ \partial x^{i} \partial x^{j} \partial x^{k} } (x).
\end{alignat*}

Higher-dimensional covelocity spaces inherit the product structure
of $\mathbb{R}^{k}$,
\begin{equation*}
J^{R}(\mathbb{R}^{k}, X)_{0} = \otimes_{k' = 1}^{k} J^{R}(\mathbb{R}, X)_{0}.
\end{equation*}
Consequently the properties of higher-dimensional covelocity
spaces simply replicate the properties of the one-dimensional 
covelocity space.  Because of this in this paper we will limit 
our consideration to one-dimensional covelocity spaces.

As with velocities, the coordinates representations of 
covelocities manifest the projective structure of jets,
with the components with $R$ $\mathfrak{d}$s specifying the 
$R$-th order structure of the covelocity.  For example, a 
$(1, 2)$-covelocity specified with the coordinates 
\begin{equation*}
\mathfrak{d}^{2} \alpha = ( (\mathfrak{d} \alpha)_{i}, (\mathfrak{d}^{2} \alpha)_{ij} )
\end{equation*}
projects to a $(1, 1)$-covelocity specified with the coordinates
\begin{equation*}
\pi^{2}_{1} (\mathfrak{d}^{2} \alpha) = ( (\mathfrak{d} \alpha)_{i} ).
\end{equation*}

\section{Higher-Order Automatic Differentiation} \label{sec:higher_order}

The natural compositional structure of jets immediately defines 
natural pushforwards of velocities and pullbacks of covelocities.  
These transformations then allow us to sequentially propagate these 
objects, and the differential structure they encode, through composite 
functions.  This propagation immediately generalizes the geometric 
perspective of automatic differentiation to higher-orders.

In this section I first derive the pushforwards of velocities
and the pure forward mode automatic differentiation algorithms 
they define before considering the pullback of covelocities and
the pure reverse mode automatic differentiation algorithms they
define.  Finally I exploit the subtle duality between covelocities
and velocities to derive mixed mode automatic differentiation
algorithms.

\subsection{Pushing Forwards Velocities}

Any smooth map $F: X \rightarrow Y$ induces a map from 
higher-dimensional curves on $X$, $c : \mathbb{R}^{k} \rightarrow X$, 
to higher-order curves on $Y$, $c^{*}: \mathbb{R}^{k} \rightarrow Y$,
through composition, $c^{*} = F \circ c$.  Consequently we can define 
the pushforward of any $(k, R)$-velocity as the $R$-truncated
Taylor expansion of the pushforward curve $c^{*}$ for any curve $c$ 
belonging to the equivalence class of the original velocity.

This pushforward action can also be calculated in coordinates as
a transformation from the coordinates of the initial velocity
over $X$ to the coordinates of the pushforward velocity over $Y$.
Here we consider the pushforwards of first, second, and 
third-order velocities and discuss the differential operators 
that they implicitly implement.

\subsubsection{First-Order Pushforwards}

In local coordinates the pushforward of a first-order velocity 
is given by the first derivative of any pushforward curve,
\begin{align*}
(\delta v_{*})^{l} 
&= \frac{ \partial }{ \partial t } (F \circ c)^{l}
\\
&= (\mathcal{J}_{F})^{l}_{i}(x) \cdot \frac{ \partial c^{i} }{ \partial t }
\\
&= (\mathcal{J}_{F})^{l}_{i}(x) \cdot (\delta v)^{i}.
\end{align*}
This behavior mirrors the transformation properties of tangent 
vector coordinates, which is not a coincidence as tangent 
vectors are identically first-order velocities.

\subsubsection{Second-Order Pushforwards}

Given the projective structure of velocities and their coordinates,
the first-order coordinates of higher-order velocities transform
like the coordinates of first-order velocities derived above.

The transformation of the higher-order coordinates of a $(1, 2)$-velocity
follows from the repeated action of the one temporal derivative,
\begin{align*}
(\delta^{2} v_{*})^{l} 
&= \frac{ \partial^{2} }{ \partial t_{1}^{2} } (F \circ c)^{l}
\\
&= \frac{ \partial }{ \partial t_{1} } \left(
(\mathcal{J}_{F})^{l}_{i}(x) \cdot \frac{ \partial c^{i} }{ \partial t_{1} } \right)
\\
&= 
(\mathcal{J}_{F})^{l}_{ij}(x)
\cdot \frac{ \partial c^{i} }{ \partial t_{1} } \cdot \frac{ \partial c^{j} }{ \partial t_{1} }
+
(\mathcal{J}_{F})^{l}_{i}(x) \cdot \frac{ \partial^{2} c^{i} }{ \partial t_{1}^{2} }
\\
&= 
(\mathcal{J}_{F})^{l}_{ij}(x) \cdot (\delta v)^{i} \cdot (\delta v)^{j}
+
(\mathcal{J}_{F})^{l}_{i}(x) \cdot (\delta^{2} v)^{i}.
\end{align*}
Under a pushforward the second-order coordinates mix with the
first-order coordinates in a transformation that clearly 
distinguishes them from the coordinates of a vector.

The aligned second-order coordinates coordinates of a two-dimensional 
second-order velocity, $(\delta^{2} v)^{i}$ and $(\delta^{2} u)^{i}$, 
transform in the same way as the second-order coordinates of a 
$(1, 2)$-velocity.  The second-order cross term, $(\delta v \delta u)^{i}$
instead mixes the two corresponding first-order coordinates together,
\begin{align*}
(\delta v \delta u_{*})^{l}
&= \frac{ \partial^{2} }{ \partial t_{2} \partial t_{1} } (F \circ c)^{l}
\\
&= \frac{ \partial }{ \partial t_{2} } \left(
(\mathcal{J}_{F})^{l}_{i}(x) \cdot \frac{ \partial c^{i} }{ \partial t_{1} } \right)
\\
&= 
(\mathcal{J}_{F})^{l}_{ij}(x)
\cdot \frac{ \partial c^{i} }{ \partial _{1} } \cdot \frac{ \partial c^{j} }{ \partial t_{2} }
+
(\mathcal{J}_{F})^{l}_{i}(x) \cdot \frac{ \partial^{2} c^{i} }{ \partial t_{1} \partial t_{2} }
\\
&= 
(\mathcal{J}_{F})^{l}_{ij}(x) \cdot (\delta u)^{i} \cdot (\delta v)^{j}
+
(\mathcal{J}_{F})^{l}_{i}(x) \cdot (\delta u \delta v)^{i}.
\end{align*}

While the second-order coordinates all mix with the first-order
coordinates, they do not mix with each other.  This hints at
the rich algebraic structure of higher-order velocities.

\subsubsection{Third-Order Pushforwards}

The transformation of the aligned third-order coordinates of a 
third-order velocity follow from one more application of the temporal 
derivative,
\begin{align*}
(\delta^{3} v_{*})^{l} 
&= \frac{ \partial^{3} }{ \partial t_{1}^{3} } (F \circ c)^{l}
\\
&=
(\mathcal{J}_{F})^{l}_{ijk}(x)
\cdot (\delta v)^{i} \cdot (\delta v)^{j} \cdot (\delta v)^{k}
+
3 (\mathcal{J}_{F})^{l}_{ij}(x) \cdot (\delta v)^{i} \cdot (\delta^{2} v)^{j}
+
(\mathcal{J}_{F})^{l}_{i}(x) \cdot (\delta^{3} v)^{i}.
\end{align*}

Similarly, the maximally-mixed third-order coordinates transforms as
\begin{align*}
(\delta v \delta u \delta w)^{l}_{*} 
&= \frac{ \partial^{3} }{ \partial t_{1} \partial t_{2} \partial t_{3} } (F \circ c)^{l}
\\
&= \quad
(\mathcal{J}_{F})^{l}_{ijk}(x)
\cdot (\delta v)^{i} \cdot (\delta u)^{j} \cdot (\delta w)^{k}
\\
&\quad +
(\mathcal{J}_{F})^{l}_{ij}(x) \cdot 
\left(
  \delta v^{i} \cdot (\delta u \delta w)^{j}
+ \delta u^{i} \cdot (\delta v \delta w)^{j}
+ \delta w^{i} \cdot (\delta v \delta u)^{j}
\right)
\\
&\quad +
(\mathcal{J}_{F})^{l}_{i}(x) \cdot (\delta v \delta u \delta w)^{i}.
\end{align*}
In particular, we can transform these maximally-mixed coordinates 
using only information from three aligned first-order coordinates, 
$(\delta v)^{i}$, $(\delta u)^{i}$, and $(\delta w)^{i}$, the three 
mixed second-order coordinates, $(\delta u \delta w)^{i}$, 
$(\delta v \delta w)^{i}$, and $(\delta v \delta u)^{i}$,  and the 
one mixed third-order coordinates, $(\delta v \delta u \delta w)^{i}$.
Indeed, exploiting the increasingly sparse dependencies of the 
higher-order velocity coordinates on the lower-order coordinates is 
key to propagating only the differential information of interest and 
avoiding unnecessary computation.

\subsubsection{Forward Mode Automatic Differentiation} \label{sec:forward_mode}

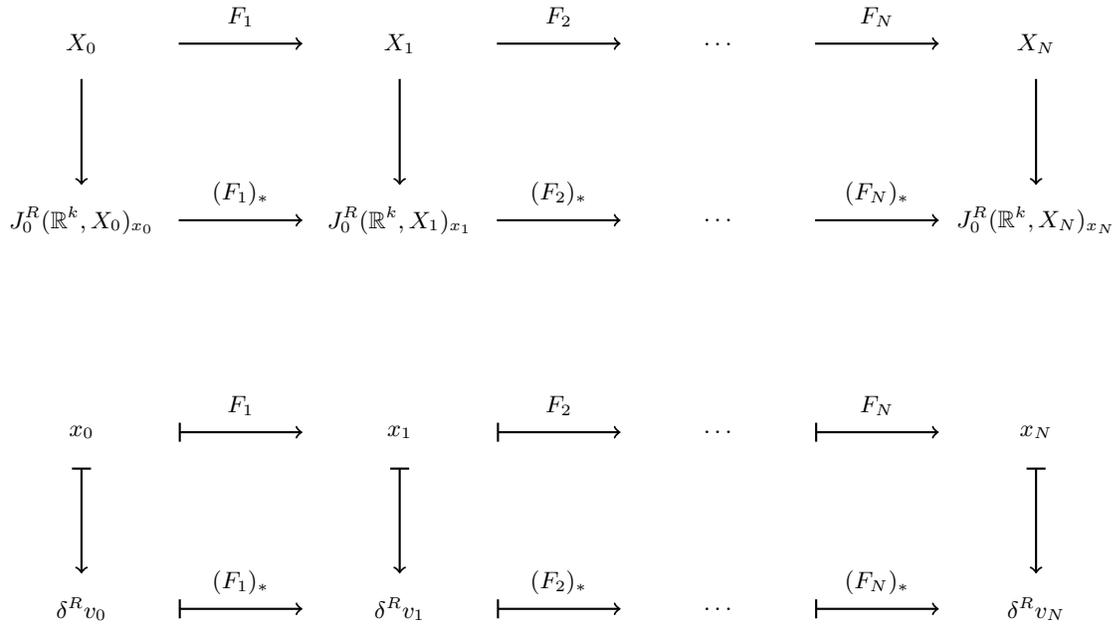
\begin{figure*}
\centering
\begin{tikzpicture}[scale=0.47, thick]

\pgfmathsetmacro{\dx}{1.75}
\pgfmathsetmacro{\dy}{-11}

\node[] at (-13.5, 0) {$X_{0}$};

\node[] at (-9, 0.75) {$F_{1}$};
\draw[->] (-9 - \dx, 0) -- +(2 * \dx, 0);

\node[] at (-4.5, 0) {$X_{1}$};

\node[] at (0, 0.75) {$F_{2}$};
\draw[->] (0 - \dx, 0) -- +(2 * \dx, 0);

\node[] at (4.5, 0) {$\ldots$};

\node[] at (9, 0.75) {$F_{N}$};
\draw[->] (9 - \dx, 0) -- +(2 * \dx, 0);

\node[] at (13.5, 0) {$X_{N}$};

\draw[->] (-13.5, -1) -- +(0, -3);
\draw[->] (-4.5, -1) -- +(0, -3);
\draw[->] (13.5, -1) -- +(0, -3);

\node[] at (-13.5, -5) {$J^{R}_{0}(\mathbb{R}^{k}, X_{0})_{x_{0}}$};

\node[] at (-9, -4.25) {$(F_{1})_{*}$};
\draw[->] (-9 - \dx, -5) -- +(2 * \dx, 0);

\node[] at (-4.5, -5) {$J^{R}_{0} (\mathbb{R}^{k}, X_{1})_{x_{1}}$};

\node[] at (0, -4.25) {$(F_{2})_{*}$};
\draw[->] (0 - \dx, -5) -- +(2 * \dx, 0);

\node[] at (4.5, -5) {$\ldots$};

\node[] at (9, -4.25) {$(F_{N})_{*}$};
\draw[->] (9 - \dx, -5) -- +(2 * \dx, 0);

\node[] at (13.5, -5) {$J^{R}_{0} (\mathbb{R}^{k}, X_{N})_{x_{N}}$};

\node[] at (-13.5, 0 + \dy) {$x_{0}$};

\node[] at (-9, 0.75 + \dy) {$F_{1}$};
\draw[|->] (-9 - \dx, 0 + \dy) -- +(2 * \dx, 0);

\node[] at (-4.5, 0 + \dy) {$x_{1}$};

\node[] at (0, 0.75 + \dy) {$F_{2}$};
\draw[|->] (0 - \dx, 0 + \dy) -- +(2 * \dx, 0);

\node[] at (4.5, 0 + \dy) {$\ldots$};

\node[] at (9, 0.75 + \dy) {$F_{N}$};
\draw[|->] (9 - \dx, 0 + \dy) -- +(2 * \dx, 0);

\node[] at (13.5, 0 + \dy) {$x_{N}$};

\draw[|->] (-13.5, -1 + \dy) -- +(0, -3);
\draw[|->] (-4.5, -1 + \dy) -- +(0, -3);
\draw[|->] (13.5, -1 + \dy) -- +(0, -3);

\node[] at (-13.5, -5 + \dy) {$\delta^{R} v_{0}$};

\node[] at (-9, -4.25 + \dy) {$(F_{1})_{*}$};
\draw[|->] (-9 - \dx, -5 + \dy) -- +(2 * \dx, 0);

\node[] at (-4.5, -5 + \dy) {$\delta^{R} v_{1}$};

\node[] at (0, -4.25 + \dy) {$(F_{2})_{*}$};
\draw[|->] (0 - \dx, -5 + \dy) -- +(2 * \dx, 0);

\node[] at (4.5, -5 + \dy) {$\ldots$};

\node[] at (9, -4.25 + \dy) {$(F_{N})_{*}$};
\draw[|->] (9 - \dx, -5 + \dy) -- +(2 * \dx, 0);

\node[] at (13.5, -5 + \dy) {$\delta^{R} v_{N}$};

\end{tikzpicture}
\caption{
Pushing $\delta^{R} v_{0}$, a $(k, R)$-velocity in 
$J^{R}_{0} (\mathbb{R}^{k}, X_{0})_{x_{0}}$, forwards through 
a function $F$ yields, $\delta^{R} v_{N}$, a $(k, R)$-velocity 
in $J^{R}_{0} (\mathbb{R}^{k}, X_{N})_{x_{N}}$ that captures 
information about the differential structure of $F$ around 
$x_{0}$.  When $F$ is a composite function we can compute this 
pushforward iteratively by pushing the initial velocity forwards 
through each component function.
}
\label{fig:pushforwards} 
\end{figure*}

Pushing forward velocities, or equivalently transforming their
coordinate representations, through a composite function,
\begin{equation*}
F = F_{N} \circ F_{N - 1} \circ \ldots \circ F_{2} \circ F_{1},
\end{equation*}
is readily accomplished by pushing forward an initial velocity 
through each component function iteratively (Figure 
\ref{fig:pushforwards}).  This sequence of pushforwards implicitly 
implements the chain rule for higher-order derivatives and hence 
provides a basis for pure forward mode higher-order automatic 
differentiation.

Consider, for example, a composite function 
$F : \mathbb{R}^{D} \rightarrow \mathbb{R}$.  In this case 
the coordinate representation of a first-order Jacobian array
is given by components of the gradient,
\begin{equation*}
(\mathcal{J}_{F})_{i} (x)
=
\frac{ \partial F }
{ \partial x^{i} }(x),
\end{equation*}
the coordinate representation of the second-order Jacobian array
is given by the components of the Hessian,
\begin{equation*}
(\mathcal{J}_{F})_{ij} (x)
=
\frac{ \partial F }
{ \partial x^{i} \partial x^{j}}(x),
\end{equation*}
and the coordinate representation of the third-order Jacobian
array is given by all of the third-order partial derivatives,
\begin{equation*}
(\mathcal{J}_{F})_{ijk} (x)
=
\frac{ \partial F }
{ \partial x^{i} \partial x^{j} \partial x^{k} }(x).
\end{equation*}

Pushing a $(1, 1)$-velocity forward through $F$ evaluates the
\emph{first-order directional derivative}
\begin{equation*}
(\delta v_{*}) = \frac{ \partial F }{ \partial x^{i} } (x) \cdot (\delta v)^{i},
\end{equation*}
which is just standard first-order forward-mode automatic 
differentiation.  

Similarly, pushing a $(2, 2)$-velocity whose coordinates all vanish except
for $(\delta v)^{i}$ and $(\delta u)^{i}$ evaluates three second-order
directional derivatives,
\begin{align*}
(\delta^{2} v_{*}) &= 
\frac{ \partial^{2} F }{ \partial x^{i} \partial x^{j}} (x)
\cdot (\delta v)^{i} \cdot (\delta v)^{j}
\\
(\delta^{2} u_{*}) &= 
\frac{ \partial^{2} F }{ \partial x^{i} \partial x^{j}} (x)
\cdot (\delta u)^{i} \cdot (\delta u)^{j}
\\
(\delta v \delta u_{*}) &= 
\frac{ \partial^{2} F }{ \partial x^{i} \partial x^{j}} (x)
\cdot (\delta v)^{i} \cdot (\delta u)^{j},
\end{align*}
along with the corresponding first-order directional derivatives,
\begin{align*}
(\delta v_{*}) &= \frac{ \partial F }{ \partial x^{i} } (x) \cdot (\delta v)^{i}
\\
(\delta u_{*}) &= \frac{ \partial F }{ \partial x^{i} } (x) \cdot (\delta u)^{i}.
\end{align*}
Because the second-order coordinates don't mix, in practice we 
need only compute one second-order directional derivative at a time.

Finally, the pushforward of an initial $(3, 3)$-velocity whose 
higher-order coordinates vanish evaluates all of the directional 
derivatives up to third-order, including
\begin{align*}
(\delta v \delta u \delta w_{*}) &= 
\frac{ \partial^{3} F }{ \partial x^{i} \partial x^{j} \partial x^{k}} (x)
\cdot (\delta v)^{i} \cdot (\delta u)^{j} \cdot (\delta w)^{k}
\\
(\delta v \delta u_{*}) &= 
\frac{1}{2} \frac{ \partial^{2} F }{ \partial x^{i} \partial x^{j}} (x)
\cdot (\delta v)^{i} \cdot (\delta u)^{j}
\\
(\delta v \delta w_{*}) &= 
\frac{ \partial^{2} F }{ \partial x^{i} \partial x^{j}} (x)
\cdot (\delta v)^{i} \cdot (\delta w)^{j}
\\
(\delta u \delta w_{*}) &= 
\frac{ \partial^{2} F }{ \partial x^{i} \partial x^{j}} (x)
\cdot (\delta u)^{i} \cdot (\delta w)^{j}
\\
(\delta v_{*}) &= \frac{ \partial F }{ \partial x^{i} } (x) \cdot (\delta v)^{i}
\\
(\delta u_{*}) &= \frac{ \partial F }{ \partial x^{i} } (x) \cdot (\delta u)^{i}
\\
(\delta w_{*}) &= \frac{ \partial F }{ \partial x^{i} } (x) \cdot (\delta w)^{i}.
\end{align*}

The pushforward of one-dimensional velocities provides a geometric
generalization of univariate Taylor series methods \citep{GriewankEtAl:2000}.

\subsection{Pulling Back Covelocities}

Any smooth map $F: X \rightarrow Y$ induces a map from real-valued
functions on $Y$, $g : Y \rightarrow \mathbb{R}$, to real-valued
functions on $X$, $g^{*}: X \rightarrow \mathbb{R}$, through composition, 
$g^{*} = g \circ F$.  Consequently we can define the pullback of any 
$(1, R)$-covelocity as the $R$-truncated Taylor expansion of the 
pullback function $g^{*}$ for any real-valued function $g$ belonging 
to the equivalence class of the original covelocity.  The pullback of 
higher-dimensional covelocities can be derived in the same way, but 
since they decouple into copies of one-dimensional covelocity pullbacks 
we will focus on the transformation of one-dimensional covelocities 
here.

This pullback action can also be calculated in coordinates as
a transformation from the coordinates of the initial covelocity
over $Y$ to the coordinates of the pullback covelocity over $X$.
Here we consider the pullbacks of first, second, and third-order 
covelocities and remark on the differential operators that 
they implement.

\subsubsection{First-Order Pullbacks}

The $1$-truncated Taylor expansion of a real-valued function 
$g : Y \rightarrow \mathbb{R}$ around $F(x) \in Y$ is given by
\begin{equation*}
T_{1} (g(y)) = (\mathcal{J}_{g})_{i} (F(x)) \cdot y^{i}_{\phi}(y)
\end{equation*}
and hence specifies a $(1, 1)$-covelocity in $J^{1}_{y}(Y, \mathbb{R})_{0}$
with the coordinates
\begin{equation*}
(\mathfrak{d} \alpha)_{i} = 
(\mathcal{J}_{g})_{i} (F(x)).
\end{equation*}

At the same time the $1$-truncated Taylor expansion of
the composition $g \circ F$ around $x \in X$ defines the 
polynomial
\begin{equation*}
T_{1} (g \circ F(x'))
=
(\mathcal{J}_{g})_{l} (F(x)) \cdot (\mathcal{J}_{F})^{l}_{i}(x) \cdot x^{i}_{\phi}(x')
\end{equation*}
and hence specifies a $(1, 1)$-covelocity in $J^{1}_{x}(X, \mathbb{R})_{0}$
with coordinates
\begin{equation*}
(\mathfrak{d} \alpha^{*})_{i} = 
(\mathcal{J}_{g})_{l} (F(x)) \cdot (\mathcal{J}_{F})^{l}_{i}(x).
\end{equation*}

Rewriting the coordinates of the compositional covelocity on 
$X$ in terms of the coefficients of the initial covelocity on 
$Y$ gives the transformation
\begin{equation*}
(\mathfrak{d} \alpha^{*})_{i} = 
(\mathcal{J}_{F})^{l}_{i}(x) \cdot (\mathfrak{d} \alpha)_{l}.
\end{equation*}
In words, the coordinates of the pullback covelocity are given 
by a mixture of the coordinates of the initial covelocity 
weighted by the first-order partial derivatives of $F$ in
the local chart.  This transformational behavior mirrors that 
of cotangent vectors, which is none too surprising given that 
cotangent vectors are identically first-order covelocities.

\subsubsection{Second-Order Pullbacks}

Similarly, the $2$-truncated Taylor expansion of a real-valued 
function $g : Y \rightarrow \mathbb{R}$ around $F(x) \in Y$ 
is given by
\begin{align*}
T_{2} (g(y)) 
&= \quad\;\;\; (\mathcal{J}_{g})_{i} (F(x)) \cdot y^{i}_{\phi}(y)
\\
&\quad + \frac{1}{2} (\mathcal{J}_{g})_{ij} (F(x)) 
\cdot y^{i}_{\phi}(y) \cdot y^{j}_{\phi}(y)
\end{align*}
and hence specifies a $(1, 2)$-covelocity in $J^{2}_{y}(Y, \mathbb{R})_{0}$
with the coordinates
\begin{align*}
(\mathfrak{d} \alpha)_{i} 
&= 
(\mathcal{J}_{g})_{i} (F(x))
\\
(\mathfrak{d} \alpha)_{ij} 
&= 
(\mathcal{J}_{g})_{ij} (F(x))
\end{align*}

At the same time the $2$-truncated Taylor expansion of
the composition $g \circ F$ around $x \in X$ defines the 
polynomial
\begin{align*}
T_{2} (g \circ F(x'))
&= \quad\;\;\;
(\mathcal{J}_{g})_{l} (F(x)) \cdot (\mathcal{J}_{F})^{l}_{i}(x) \cdot x^{i}_{\phi}(x')
\\
& \quad +
\frac{1}{2} \left( 
(\mathcal{J}_{g})_{l} (F(x)) \cdot (\mathcal{J}_{F})^{l}_{ij}(x)
+
(\mathcal{J}_{g})_{lm} (F(x)) \cdot (\mathcal{J}_{F})^{l}_{i}(x) \cdot (\mathcal{J}_{F})^{m}_{j}(x)
\right)
\\
& \quad\quad\quad \cdot
x^{i}_{\phi}(x') \cdot x^{j}_{\phi}(x')
\end{align*}
and hence specifies a $(1, 2)$-covelocity in $J^{2}_{x}(X, \mathbb{R})_{0}$
with coordinates
\begin{align*}
(\mathfrak{d} \alpha^{*})_{i} &= 
(\mathcal{J}_{g})_{l} (F(x)) \cdot (\mathcal{J}_{F})^{l}_{i}(x)
\\
(\mathfrak{d} \alpha^{*})_{ij} &= 
(\mathcal{J}_{g})_{l} (F(x)) \cdot (\mathcal{J}_{F})^{l}_{ij}(x)
+
(\mathcal{J}_{g})_{lm} (F(x)) \cdot (\mathcal{J}_{F})^{l}_{i}(x) \cdot (\mathcal{J}_{F})^{m}_{j}(x)
\end{align*}

Rewriting the coordinates of the compositional covelocity on 
$X$ in terms of the coefficients of the initial covelocity on 
$Y$ gives the transformations
\begin{align*}
(\mathfrak{d} \alpha^{*})_{i} &= 
(\mathcal{J}_{F})^{l}_{i}(x) \cdot(\mathfrak{d} \alpha)_{l}
\\
(\mathfrak{d}^{2} \alpha^{*})_{ij} &= 
(\mathcal{J}_{F})^{l}_{ij}(x) \cdot (\mathfrak{d} \alpha)_{l}
+
(\mathcal{J}_{F})^{l}_{i}(x) \cdot 
(\mathcal{J}_{F})^{m}_{j}(x) \cdot 
(\mathfrak{d}^{2} \alpha)_{lm}
\end{align*}

As expected from the projective structure of covelocities,
the first-order coordinates transform in the same way as
the coordinates of a first-order covelocity.  The second-order
coordinates, however, must mix with the first-order coordinates
in order to form the correct transformation.

\subsubsection{Third-Order Pullbacks}

The continuation to third-order follows in turn.  The coordinates of
a $(1, 3)$-covelocity in $J^{3}_{y}(Y, \mathbb{R})_{0}$ define the coordinates
of a pushforward $(1, 3)$-covelocity in $J^{3}_{x}(X, \mathbb{R})_{0}$ as
\begin{align*}
(\mathfrak{d} \alpha^{*})_{i} &= 
(\mathcal{J}_{F})^{l}_{i}(x) \cdot (\mathfrak{d} \alpha)_{l}
\\
(\mathfrak{d}^{2} \alpha^{*})_{ij} &= 
(\mathcal{J}_{F})^{l}_{ij}(x) \cdot (\mathfrak{d} \alpha)_{l}
+
(\mathcal{J}_{F})^{l}_{i}(x) \cdot (\mathcal{J}_{F})^{m}_{j}(x) \cdot (\mathfrak{d}^{2} \alpha)_{lm}
\\
(\mathfrak{d}^{3} \alpha^{*})_{ijk} 
&= \quad 
(\mathcal{J}_{F})^{l}_{ijk}(x) \cdot (\mathfrak{d} \alpha)_{l}
\\ 
& \quad +
\left( 
(\mathcal{J}_{F})^{l}_{i}(x) \cdot (\mathcal{J}_{F})^{m}_{jk}(x) 
+ (\mathcal{J}_{F})^{l}_{j}(x) \cdot (\mathcal{J}_{F})^{m}_{ik}(x) 
+ (\mathcal{J}_{F})^{l}_{k}(x) \cdot (\mathcal{J}_{F})^{m}_{ij}(x) 
\right)
(\mathfrak{d}^{2} \alpha)_{lm}
\\ 
& \quad +
(\mathcal{J}_{F})^{l}_{i}(x) \cdot (\mathcal{J}_{F})^{m}_{j}(x) \cdot (\mathcal{J}_{F})^{q}_{k}(x) 
\cdot (\mathfrak{d}^{3} \alpha)_{lmq}
\end{align*}

The projective structure is again apparent, with the first-order
coordinates transforming as the coordinates of a first-order covelocity
and the second-order coordinates transforming as the coordinates of
a second-order velocity.  The transformation of the third-order
coordinates mixes coordinates across all orders with first, 
second, and third-order partial derivatives.

\subsubsection{Reverse-Mode Automatic Differentiation}

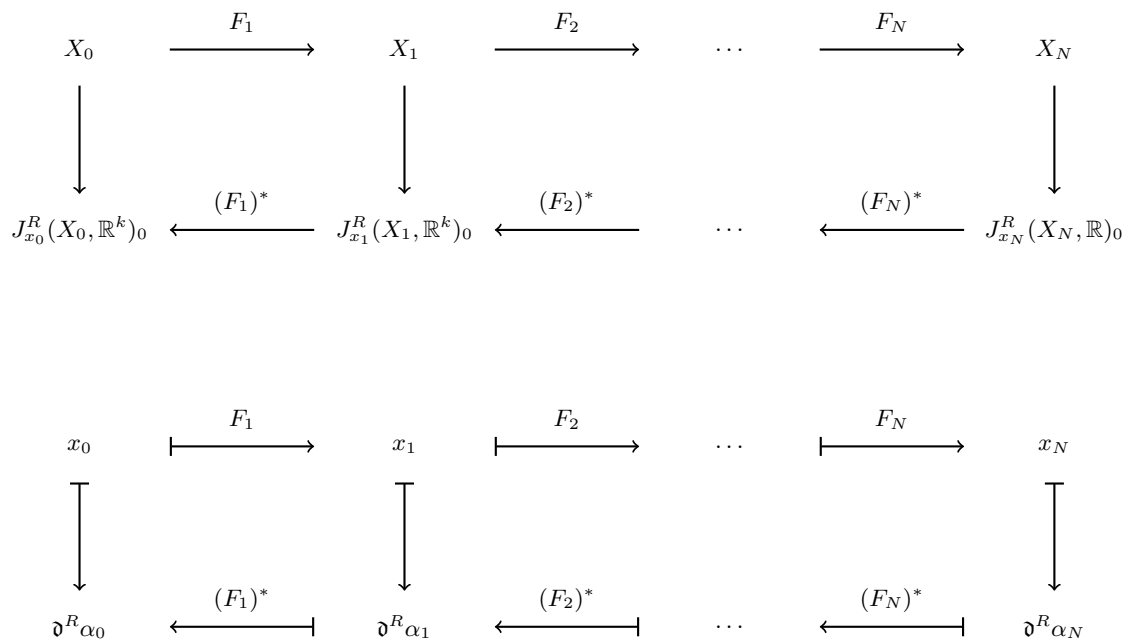
\begin{figure*}
\centering
\begin{tikzpicture}[scale=0.48, thick]

\pgfmathsetmacro{\dx}{2}
\pgfmathsetmacro{\dy}{-11}

\node[] at (-13.5, 0) {$X_{0}$};

\node[] at (-9, 0.75) {$F_{1}$};
\draw[->] (-9 - \dx, 0) -- +(2 * \dx, 0);

\node[] at (-4.5, 0) {$X_{1}$};

\node[] at (0, 0.75) {$F_{2}$};
\draw[->] (0 - \dx, 0) -- +(2 * \dx, 0);

\node[] at (4.5, 0) {$\ldots$};

\node[] at (9, 0.75) {$F_{N}$};
\draw[->] (9 - \dx, 0) -- +(2 * \dx, 0);

\node[] at (13.5, 0) {$X_{N}$};

\draw[->] (-13.5, -1) -- +(0, -3);
\draw[->] (-4.5, -1) -- +(0, -3);
\draw[->] (13.5, -1) -- +(0, -3);

\node[] at (-13.5, -5) {$J^{R}_{x_{0}}(X_{0}, \mathbb{R}^{k})_{0}$};

\node[] at (-9, -4.25) {$(F_{1})^{*}$};
\draw[<-] (-9 - \dx, -5) -- +(2 * \dx, 0);

\node[] at (-4.5, -5) {$J^{R}_{x_{1}} (X_{1}, \mathbb{R}^{k})_{0}$};

\node[] at (0, -4.25) {$(F_{2})^{*}$};
\draw[<-] (0 - \dx, -5) -- +(2 * \dx, 0);

\node[] at (4.5, -5) {$\ldots$};

\node[] at (9, -4.25) {$(F_{N})^{*}$};
\draw[<-] (9 - \dx, -5) -- +(2 * \dx, 0);

\node[] at (13.5, -5) {$J^{R}_{x_{N}} (X_{N}, \mathbb{R})_{0}$};

\node[] at (-13.5, 0 + \dy) {$x_{0}$};

\node[] at (-9, 0.75 + \dy) {$F_{1}$};
\draw[|->] (-9 - \dx, 0 + \dy) -- +(2 * \dx, 0);

\node[] at (-4.5, 0 + \dy) {$x_{1}$};

\node[] at (0, 0.75 + \dy) {$F_{2}$};
\draw[|->] (0 - \dx, 0 + \dy) -- +(2 * \dx, 0);

\node[] at (4.5, 0 + \dy) {$\ldots$};

\node[] at (9, 0.75 + \dy) {$F_{N}$};
\draw[|->] (9 - \dx, 0 + \dy) -- +(2 * \dx, 0);

\node[] at (13.5, 0 + \dy) {$x_{N}$};

\draw[|->] (-13.5, -1 + \dy) -- +(0, -3);
\draw[|->] (-4.5, -1 + \dy) -- +(0, -3);
\draw[|->] (13.5, -1 + \dy) -- +(0, -3);

\node[] at (-13.5, -5 + \dy) {$\mathfrak{d}^{R} \alpha_{0}$};

\node[] at (-9, -4.25 + \dy) {$(F_{1})^{*}$};
\draw[<-|] (-9 - \dx, -5 + \dy) -- +(2 * \dx, 0);

\node[] at (-4.5, -5 + \dy) {$\mathfrak{d}^{R} \alpha_{1}$};

\node[] at (0, -4.25 + \dy) {$(F_{2})^{*}$};
\draw[<-|] (0 - \dx, -5 + \dy) -- +(2 * \dx, 0);

\node[] at (4.5, -5 + \dy) {$\ldots$};

\node[] at (9, -4.25 + \dy) {$(F_{N})^{*}$};
\draw[<-|] (9 - \dx, -5 + \dy) -- +(2 * \dx, 0);

\node[] at (13.5, -5 + \dy) {$\mathfrak{d}^{R} \alpha_{N}$};

\end{tikzpicture}
\caption{
Pulling $\mathfrak{d}^{R} \alpha_{N}$, a $(k, R)$-covelocity in 
$J^{R}_{x_{N}} (X_{N}, \mathbb{R}^{k})_{0}$,
backwards through a function $F$ yields, $\mathfrak{d}^{R} \alpha_{0}$, 
a $(k, R)$-covelocity in $J^{R}_{x_{0}} (X_{0}, \mathbb{R})_{0}$ 
that captures information about the differential structure of $F$
around $x_{0}$.  When $F$ is a composite function we can compute 
this pullback iteratively by pulling the final covelocity
backwards through each component function.
}
\label{fig:pullbacks} 
\end{figure*}

We can construct the pullback of a covelocity along a composite function 
\begin{equation*}
F = F_{N} \circ F_{N - 1} \circ \ldots \circ F_{2} \circ F_{1}
\end{equation*}
by pulling back an final covelocity through each component function
iteratively (Figure \ref{fig:pullbacks}).  This sequence of 
pullbacks implicitly implements the higher-order chain rule, and 
hence provides a basis for pure reverse mode higher-order automatic 
differentiation.

For example, consider a composite function 
$F : \mathbb{R}^{D} \rightarrow \mathbb{R}$, in which case the
coordinate representations of the Jacobian array become the
partial derivatives as in Section \ref{sec:forward_mode}.  In
this case the pullback coordinates at each component function
$F_{n}$ transform as
\begin{align*}
(\mathfrak{d} \alpha^{*})_{i} &= 
\frac{ \partial (F_{n})^{l} }{ \partial x^{i} } (x_{n - 1}) \cdot (\mathfrak{d} \alpha)_{l} 
\\
(\mathfrak{d}^{2} \alpha^{*})_{ij} &= 
\frac{ \partial^{2} (F_{n})^{l} }{ \partial x^{i} \partial x^{j} } (x_{n - 1}) \cdot 
(\mathfrak{d} \alpha)_{l}
+
\frac{ \partial (F_{n})^{l} }{ \partial x^{i} } (x_{n - 1}) \cdot
\frac{ \partial (F_{n})^{m} }{ \partial x^{j} } (x_{n - 1}) \cdot
(\mathfrak{d}^{2} \alpha)_{lm}
\\
(\mathfrak{d}^{3} \alpha^{*})_{ijk} &= 
\quad \frac{\partial^{3} (F_{n})^{l}}{\partial x^{i} x^{j} x^{k}} (x_{n - 1}) 
\cdot (\mathfrak{d} \alpha)_{l} 
\\
& \quad +
\left( \quad\;
\frac{\partial (F_{n})^{l} }{\partial x^{i}}(x_{n - 1}) \cdot 
\frac{\partial^{2} (F_{n})^{m} }{\partial x^{j} x^{k}}(x_{n - 1}) \right.
\\
& \quad\quad\quad +
\frac{\partial (F_{n})^{l} }{\partial x^{j}}(x_{n - 1}) \cdot
\frac{\partial^{2} (F_{n})^{m} }{\partial x^{i} x^{k}}(x_{n - 1})
\\
& \quad\quad\quad + \left.
 \frac{\partial (F_{n})^{l} }{\partial x^{k}}(x_{n - 1}) \cdot
\frac{\partial^{2} (F_{n})^{m} }{\partial x^{i} x^{j}}(x_{n - 1})
\right) 
\cdot (\mathfrak{d}^{2} \alpha)_{lm}
\\
& \quad +
\frac{\partial (F_{n})^{l}}{\partial x^{i}}(x_{n - 1}) \cdot
\frac{\partial (F_{n})^{m}}{\partial x^{j}}(x_{n - 1}) \cdot
\frac{\partial (F_{n})^{q}}{\partial x^{k}}(x_{n - 1}) \cdot 
(\mathfrak{d}^{3} \alpha)_{lmq}.
\end{align*}

Pulling the $(1, 1)$-covelocity with coordinate 
$(\mathfrak{d} \alpha)_{1} = 1$ back through $F$ yields the 
components of the gradient evaluated at the input,
\begin{equation*}
(\mathfrak{d} \alpha^{*})_{i} = \frac{ \partial F }{ \partial x^{i} } (x),
\end{equation*}
which is just standard first-order reverse-mode automatic differentiation.

Pulling a $(1, 2)$-covelocity with first-order coefficients 
$(\mathfrak{d} \alpha)_{1} = 1$ and second-order coefficients 
$(\mathfrak{d}^{2} \alpha)_{11} = 0$ back through $F$ yields not only the 
components of the gradient but also the components of the Hessian
evaluated at the input,
\begin{align*}
(\mathfrak{d} \alpha^{*})_{i} &= \frac{ \partial F }{ \partial x^{i} } (x)
\\
(\mathfrak{d}^{2} \alpha^{*})_{ij} &= 
\frac{ \partial^{2} F }{ \partial x^{i} \partial x^{j}} (x).
\end{align*}

Pulling a $(1, 3)$-covelocity with first-order coefficients 
$(\mathfrak{d} \alpha)_{1} = 1$ and vanishing higher-order coefficients 
$(\mathfrak{d}^{2} \alpha)_{11} = (\mathfrak{d}^{3} \alpha)_{111} = 0$ 
back through $F$ yields also the third-order partial derivatives of
the composite function,
\begin{align*}
(\mathfrak{d} \alpha^{*})_{i} &= \frac{ \partial F }{ \partial x^{i} } (x)
\\
(\mathfrak{d}^{2} \alpha^{*})_{ij} &= 
\frac{ \partial^{2} F }{ \partial x^{i} \partial x^{j}} (x).
\\
(\mathfrak{d}^{3} \alpha^{*})_{ijk} &= 
\frac{ \partial^{2} F }{ \partial x^{i} \partial x^{j} \partial x^{k}} (x).
\end{align*}
In general the pullback of a $R$-th order covelocity whose higher-order
coefficients all vanish will yield all of the partial derivatives up to order 
$R$ evaluated at $x \in X$,
\begin{equation*}
\frac{ \partial F^{j} }{ \partial x^{i} } (x), \, 
\frac{ \partial^{2} F^{j} }{ \partial x^{i} \partial x^{j} } (x), \, 
\ldots, \,
\frac{ \partial^{R} F^{j} }{ \partial x^{i_{1}} \ldots \partial x^{i_{R}}} (x).
\end{equation*}

The pullback of a $(1, R)$-covelocity yields an explicit version of
the recursive updates introduced in \cite{Neidinger:1992}.

\subsection{Interleaving Pushforwards and Pullbacks} \label{sec:mixed_mode}

Pushing forwards velocities and pulling back covelocities do not
define all of the natural differential operators that exist at a
given order.  For example, neither evaluates the Hessian-vector 
product of a many-to-one function directly.  We could run $D$ 
forward mode sweeps to compute each element of the product one-by-one,
or we could run a single, memory-intensive second-order reverse mode 
sweep to compute the full Hessian and only then contract it against 
the desired vector.  Fortunately we can exploit the subtle duality 
of velocities and covelocities to derive explicit calculations for 
the gradients of the differential operators implemented with
forward mode methods, such as the Hessian-vector product as the
gradient of a first-order directional derivative.

At first-order velocities and covelocities manifest the duality 
familiar from tangent vectors and cotangent vectors: the space of
linear transformations from $(k, 1)$-covelocities to the real numbers
is isomorphic to the space of $(k, 1)$-velocities, and the space of 
linear transformations from $(k, 1)$-velocities to the real numbers is
isomorphic to the space of $(k, 1)$-covelocities.  In other words,
any $(k, 1)$-velocity serves as a map that takes any $(k, 1)$-covelocity 
to a real number and vice-versa.  

As we proceed to higher-orders this duality isn't quite as clean.
In general the space of linear transformations from covelocities to 
real numbers is not contained within any single velocity space but 
instead spans a mixture of velocity spaces up to the given dimension
and order.  Consequently constructing explicit duals to a given velocity 
or covelocity requires care.  Fortunately we can always verify a valid 
dual pair by demonstrating that the action of a dual on a pulled back 
covelocity is equal to the action of the pushed forward dual on the 
initial covelocity (Figure \ref{fig:dual_consistency}).

\begin{figure*}
\centering
\begin{tikzpicture}[scale=0.35, thick]

\draw[|->, color=gray80] (-4.5, -1) -- +(0, -3);

\draw[|->, color=gray80] (-5.5, -1) -- +(0, -8);
\fill[color=white] (-7, -6) rectangle(-3, -4);

\draw[|->, color=gray80] (5.5, -1) -- +(0, -3);

\draw[|->, color=gray80] (4.5, -1) -- +(0, -8);
\fill[color=white] (3, -6) rectangle(7, -4);

\node[] at (-10, 0) {$\ldots$};

\node[] at (-5, 0) {$x_{n - 1}$};

\node[] at (0, 0.75) {$F_{n}$};
\draw[|->] (-2, 0) -- +(4, 0);

\node[] at (5, 0) {$x_{n}$};

\node[] at (10, 0) {$\ldots$};

\node[] at (-10, -5) {$\ldots$};

\node[] at (-5, -5) {$v_{n - 1}$};

\node[] at (0, -4.25) {$(F_{n})_{*}$};
\draw[|->] (-2, -5) -- +(4, 0);

\node[] at (5, -5) {$v_{n}$};

\node[] at (10, -5) {$\ldots$};

\node[] at (-10, -10) {$\ldots$};

\node[] at (-5, -10) {$\mathfrak{d}^{R} \alpha_{n - 1}$};

\node[] at (0, -9.25) {$(F_{n})^{*}$};
\draw[<-|] (-2, -10) -- +(4, 0);

\node[] at (5, -10) {$\mathfrak{d}^{R} \alpha_{n}$};

\node[] at (10, -10) {$\ldots$};

\end{tikzpicture}
\caption{
The dual space of a given covelocity space consists of the linear
maps from the covelocity space to the real numbers.  For example,
$\delta^{R} v_{n - 1}( \mathfrak{d}^{R} \alpha_{n} ) = r \in \mathbb{R}$.
This duality is consist with respect to pushforwards and pullbacks
-- $\delta^{R} v_{n} ( \mathfrak{d}^{R} \alpha_{n - 1} ) = r$ as 
well.  This consistency is useful for verifying that we've identified
an appropriate dual for any given covelocity.
}
\label{fig:dual_consistency} 
\end{figure*}

Once we've identified valid covelocity duals we can then define 
contractions with well-defined pullbacks by projecting the duals 
to lower orders.  These contractions then propagate through composite 
functions, carrying the the differential information required to
evaluate the gradients of directional derivatives with them.

\subsubsection{Second-Order Contraction Pullbacks} \label{sec:second_order_contraction}

Consider a $(1, 2)$-covelocity, 
$\mathfrak{d}^{2} \alpha \in J^{2} (X, \mathbb{R})_{0}$ with coordinates 
$( (\mathfrak{d} \alpha)_{l}, (\mathfrak{d}^{2} \alpha)_{lm})$.
In this case there are multiple duals, but let's take the $(2, 2)$-velocity,
$\delta u \delta v \in J^{2}_{0}(\mathbb{R}^{2}, X)$ with non-vanishing 
coordinates $( (\delta v)_{i}, (\delta u)_{i}, (\delta v \delta u)_{i} )$.

In coordinates the action of this dual on the pullback of the covelocity 
is given by
\begin{align*}
\delta u \delta v (\mathfrak{d}^{2} \alpha_{*})
&=
(\delta v)^{i} \cdot (\delta u)^{j} \cdot (\mathfrak{d}^{2} \alpha^{*})_{ij} 
+ (\delta v \delta u)^{i} \cdot (\mathfrak{d} \alpha^{*})_{i}
\\
&= \quad
(\mathcal{J}_{F})^{l}_{i}(x) \cdot (\mathcal{J}_{F})^{m}_{j}(x) \cdot
(\delta v)^{i} \cdot (\delta u)^{j} \cdot (\mathfrak{d}^{2} \alpha)_{lm} 
\\
& \quad +
(\mathcal{J}_{F})^{l}_{ij}(x) \cdot
(\delta v)^{i} \cdot (\delta u)^{j} \cdot (\mathfrak{d} \alpha)_{l} 
+ (\mathcal{J}_{F})^{l}_{i}(x) \cdot (\delta v \delta u)^{i} \cdot
(\mathfrak{d} \alpha)_{l} 
\\
&=
\Big[ (\mathcal{J}_{F})^{l}_{i}(x) \cdot (\delta v)^{i} \Big]
\Big[ (\mathcal{J}_{F})^{m}_{j}(x) \cdot (\delta u)^{j} \Big]
(\mathfrak{d}^{2} \alpha)_{lm} 
\\
& \quad +
\Big[ (\mathcal{J}_{F})^{l}_{ij}(x) \cdot
(\delta v)^{i} (\delta u)^{j} 
+ (\mathcal{J}_{F})^{l}_{i}(x) \cdot (\delta v \delta u)^{i} \Big] \cdot
(\mathfrak{d} \alpha)_{l} 
\\
&=
(\delta v^{*})^{l} \cdot (\delta u^{*})^{m} \cdot (\mathfrak{d}^{2} \alpha)_{lm} 
+(\delta v \delta u^{*})^{l} \cdot  (\mathfrak{d} \alpha)_{l} 
\\
&= (\delta u \delta v^{*})(\mathfrak{d}^{2} \alpha),
\end{align*}
which is just the action of the pushforward dual on the initial covelocity,
as expected from the required consistency.

In coordinates the projection of the dual to a $(2, 1)$-velocity is given 
by
\begin{equation*}
\pi^{2}_{1} ((\delta v)^{i}, (\delta u)^{i}, (\delta v \delta u)^{i}) 
= (\delta v^{i}, \delta u^{i}).
\end{equation*}
which we can further project to the two $(1, 1)$-velocities with
coordinates $(\delta v)^{i}$ and $(\delta u)^{i}$, respectively.

Either projection can then be applied to the $(1, 2)$-covelocity
which in coordinates gives
\begin{align*}
(\delta v)^{i} \cdot (\mathfrak{d}^{2} \alpha^{*})_{ij} + (\mathfrak{d} \alpha^{*})_{j}
&= \quad
(\mathcal{J}_{F})^{l}_{i}(x) \cdot (\mathcal{J}_{F})^{m}_{j}(x) \cdot 
(\delta v)^{i} \cdot (\mathfrak{d}^{2} \alpha)_{lm}
\\
& \quad +
(\mathcal{J}_{F})^{l}_{ij}(x) \cdot (\delta v)^{i} \cdot (\mathfrak{d} \alpha)_{l} 
+
(\mathcal{J}_{F})^{l}_{j}(x) \cdot (\mathfrak{d} \alpha)_{l}
\\
&= \quad
(\mathcal{J}_{F})^{l}_{j}(x) \cdot 
\Big[ (\mathcal{J}_{F})^{m}_{i}(x) \cdot (\delta v)^{i} \Big] \cdot
(\mathfrak{d}^{2} \alpha)_{lm} 
\\
& \quad +
(\mathcal{J}_{F})^{l}_{ij}(x) \cdot (\delta v)^{i} \cdot
(\mathfrak{d} \alpha)_{l} 
+
(\mathcal{J}_{F})^{l}_{j}(x) \cdot
(\mathfrak{d} \alpha)_{l} 
\\
&=
(\mathcal{J}_{F})^{l}_{j}(x) \cdot (\delta v_{*})^{m} \cdot (\mathfrak{d}^{2} \alpha)_{lm}
+
(\mathcal{J}_{F})^{l}_{ij}(x) \cdot (\delta v)^{i} \cdot (\mathfrak{d} \alpha)_{l}
+
(\mathcal{J}_{F})^{l}_{j}(x) \cdot (\mathfrak{d} \alpha)_{l} .
\end{align*}
Defining the contraction 
$(\mathfrak{d} \beta)_{l} = (\delta v_{*})^{m} (\mathfrak{d}^{2} \alpha)_{lm} $
this simplifies to
\begin{equation*}
(\mathfrak{d} \beta^{*})_{j} + (\mathfrak{d} \alpha^{*})_{j}
=
(\mathcal{J}_{F})^{l}_{j}(x) \cdot (\mathfrak{d} \beta)_{l}
+
(\mathcal{J}_{F})^{l}_{ij}(x) \cdot (\delta v)^{i} \cdot (\mathfrak{d} \alpha)_{l}
+
(\mathcal{J}_{F})^{l}_{j}(x) \cdot (\mathfrak{d} \alpha)_{l},
\end{equation*}
which decouples into the sequential updates: first
\begin{equation*}
(\mathfrak{d} \alpha^{*})_{j} = (\mathcal{J}_{F})^{l}_{j}(x) \cdot (\mathfrak{d} \alpha)_{l},
\end{equation*}
and then 
\begin{equation*}
(\mathfrak{d} \beta^{*})_{j}
=
(\mathcal{J}_{F})^{l}_{j}(x) \cdot (\mathfrak{d} \beta)_{l}
+
(\mathcal{J}_{F})^{l}_{ij}(x) \cdot (\delta v)^{i} \cdot (\mathfrak{d} \alpha)_{l}.
\end{equation*}
In other words, provided that we've already pushed $\delta v$ forwards 
and pulled $\mathfrak{d} \alpha$ back, the contraction $\mathfrak{d} \beta$ 
features a well-defined pull back that we can use to propagate new 
differential information through a composite function.

Given the function
\begin{equation*}
F = F_{N} \circ F_{N - 1} \circ \ldots \circ F_{2} \circ F_{1},
\end{equation*}
we can then push the $(1, 1)$-velocity $\delta v$ forwards through
each component function via
\begin{equation*}
(\delta v_{*})^{l} 
= 
(\mathcal{J}_{F_{n}})^{l}_{i}(x_{n - 1}) \cdot (\delta v)^{i},
\end{equation*}
and then pull the $(1, 1)$-covelocity $\mathfrak{d} \alpha$
backwards through the component functions via
\begin{equation*}
(\mathfrak{d} \alpha^{*})_{i} 
= 
(\mathcal{J}_{F_{n}})^{l}_{i}(x_{n - 1}) \cdot (\mathfrak{d} \alpha)_{l}.
\end{equation*}
As we're pulling back $\mathfrak{d} \alpha$ we can use the intermediate
pushforwards of $\delta v$ to pull back $\mathfrak{d} \beta$ as well via
\begin{equation*}
(\mathfrak{d} \beta^{*})_{i} 
=
(\mathcal{J}_{F_{n}})^{l}_{i}(x_{n - 1}) \cdot (\mathfrak{d} \beta)_{l}
+
(\mathcal{J}_{F_{n}})^{l}_{ij}(x_{n - 1}) \cdot 
(\delta v)^{j} \cdot (\mathfrak{d} \alpha)_{l} .
\end{equation*}
This sequence implements mixed mode higher-order automatic differentiation 
that evaluates gradients of directional derivatives.

For example, take the composite function 
$F : \mathbb{R}^{D} \rightarrow \mathbb{R}$.  Pushing forwards $\delta v$ 
through $F$ gives a pushforward velocity at each component function and then 
ultimately the first-order directional derivative,
\begin{equation*}
(\delta v_{*}) = \frac{ \partial F }{ \partial x^{i} } (x) \cdot (\delta v)^{i}.
\end{equation*}
Pulling $(\mathfrak{d} \alpha)^{l} = 1$ and $(\mathfrak{d} \beta)^{i} = 0$ 
back through $F$, then yields the components of the gradient,
\begin{equation*}
(\mathfrak{d} \alpha^{*})_{i} = \frac{ \partial F }{ \partial x^{i} } (x)
\end{equation*}
and the components of the gradient of the first-order directional 
derivative
\begin{equation*}
(\mathfrak{d} \beta_{*})_{i} = 
\frac{ \partial }{ \partial x^{i} } \left(
\frac{ \partial F }{ \partial x^{j}} (x) \cdot (\delta v)^{j} \right)
=
\frac{ \partial^{2} F }{ \partial x^{i} \partial x^{j}} (x) \cdot
(\delta v)^{j}
\end{equation*}
in a single reverse sweep.

As we will see below, this mixed mode method is particularly
useful because, unlike higher-order reverse mode methods, it 
can be implemented using only information local to each
subexpression in a computer program.  Indeed these methods
reduce to explicit versions of \emph{higher-order adjoint methods}
\citep{GriewankEtAl:2008}.  There $\mathfrak{d} \beta$ is called
a \emph{second-order adjoint}, but because its transformation 
mixes pushforwards and pullbacks the it is perhaps more 
accurately denoted a \emph{conditional} second-order adjoint.

\subsubsection{Third-Order Contraction Pullbacks}

Similarly, consider a $(1, 3)$-covelocity, 
$\mathfrak{d}^{3} \alpha \in J^{3}_{x} (X, \mathbb{R})_{0}$
with coordinates 
$( (\mathfrak{d} \alpha)_{l}, (\mathfrak{d}^{2} \alpha)_{lm}, (\mathfrak{d}^{3} \alpha)_{lmq})$.
One possible dual is the $(3, 3)$-velocity,
$\delta u \delta v \delta w \in J^{3}_{0}(\mathbb{R}^{3}, X)_{x}$ with non-vanishing 
coordinates 
\begin{equation*}
( (\delta v)_{i}, (\delta u)_{i}, (\delta w)_{i}, (\delta v \delta u)_{i},
(\delta v \delta w)_{i}, (\delta u \delta w)_{i}, (\delta v \delta u \delta w)_{i}).
\end{equation*}

In coordinates the action of this dual on the pullback of a given covelocity 
is given by
\begin{align*}
\delta u \delta v \delta w (\mathfrak{d}^{3} \alpha_{*})
&= \quad
(\delta v)^{i} \cdot (\delta u)^{j} \cdot (\delta w)^{j} \cdot 
(\mathfrak{d}^{3} \alpha^{*})_{ij} 
\\
& \quad +
\Big[ 
(\delta v)^{i} \cdot (\delta u \delta w)^{j} 
+ (\delta u)^{i} \cdot (\delta v \delta w)^{j} 
+ (\delta w)^{i} \cdot (\delta v \delta u)^{j} 
\Big] \cdot 
(\mathfrak{d}^{2} \alpha^{*})_{ij} 
\\
& \quad +
(\delta v \delta u \delta w)^{i} \cdot (\mathfrak{d} \alpha^{*})_{i}.
\end{align*}

In coordinates the projection of the dual to a $(3, 2)$-velocity is given 
by
\begin{align*}
\pi^{3}_{2} ((\delta v)^{i}, (\delta u)^{i}, (\delta w)^{i},
(\delta v \delta u)^{i}, & (\delta v \delta w)^{i}, (\delta u \delta w)^{i},
(\delta v \delta u \delta w)^{i} ) 
\\
&= 
((\delta v)^{i}, (\delta u)^{i}, (\delta w)^{i}, 
(\delta v \delta u)^{i}, (\delta v \delta w)^{i}, (\delta u \delta w)^{i}).
\end{align*}
which we can further project to three $(2, 2)$-velocities with the 
coordinates
\begin{align*}
((\delta v)^{i}, (\delta u)^{i}, (\delta v \delta u)^{i})
\\
((\delta v)^{i}, (\delta w)^{i}, (\delta v \delta w)^{i})
\\
((\delta u)^{i}, (\delta w)^{i}, (\delta u \delta w)^{i}).
\end{align*}

Applying the first to a $(1, 3)$-covelocity gives the coordinate
action
\begin{align*}
(\delta v)^{i} \cdot (\delta u)^{j} \cdot & (\mathfrak{d}^{3} \alpha^{*})_{ijk}
+ (\delta v \delta u)^{i} \cdot (\mathfrak{d}^{2} \alpha^{*})_{ik} 
+ (\mathfrak{d} \alpha^{*})_{k}
\\
&= \quad 
(\mathcal{J}_{F})^{l}_{ijk}(x) \cdot 
(\delta v)^{i} \cdot (\delta u)^{j} \cdot  (\mathfrak{d} \alpha)_{l}
\\ 
& \quad +
\left( 
(\mathcal{J}_{F})^{l}_{i}(x) \cdot (\mathcal{J}_{F})^{m}_{jk}(x) 
+ (\mathcal{J}_{F})^{l}_{j}(x) \cdot (\mathcal{J}_{F})^{m}_{ik}(x) 
+ (\mathcal{J}_{F})^{l}_{k}(x) \cdot (\mathcal{J}_{F})^{m}_{ij}(x) 
\right)
\\
& \quad\quad \cdot
(\delta v)^{i} \cdot (\delta u)^{j} \cdot (\mathfrak{d}^{2} \alpha)_{lm}
\\ 
& \quad +
(\mathcal{J}_{F})^{l}_{i}(x) \cdot (\mathcal{J}_{F})^{m}_{j}(x) \cdot (\mathcal{J}_{F})^{q}_{k}(x) \cdot
(\delta v)^{i} \cdot (\delta u)^{j} \cdot (\mathfrak{d}^{3} \alpha)_{lmq}
\\
& \quad +
(\mathcal{J}_{F})^{l}_{i}(x) \cdot (\mathcal{J}_{F})^{m}_{k}(x) \cdot 
(\delta v \delta u)^{i} \cdot (\mathfrak{d}^{2} \alpha)_{lm} 
+
(\mathcal{J}_{F})^{l}_{ik}(x) \cdot (\delta v \delta u)^{i} \cdot (\mathfrak{d} \alpha)_{l} 
\\
& \quad +
(\mathcal{J}_{F})^{l}_{k}(x) \cdot (\mathfrak{d} \alpha)_{l} 
\\
&= \quad 
(\mathcal{J}_{F})^{l}_{ijk}(x) \cdot 
(\delta v)^{i} \cdot (\delta u)^{j} \cdot (\mathfrak{d} \alpha)_{l}
\\ 
& \quad +
(\mathcal{J}_{F})^{m}_{jk}(x) \cdot
\Big[ (\mathcal{J}_{F})^{l}_{i}(x) (\delta v)^{i} \Big] \cdot
(\delta u)^{j} \cdot (\mathfrak{d}^{2} \alpha)_{lm}
\\
& \quad + 
(\mathcal{J}_{F})^{m}_{ik}(x) \cdot
\Big[ (\mathcal{J}_{F})^{l}_{j}(x) (\delta u)^{j} \Big] \cdot
(\delta v)^{i} \cdot (\mathfrak{d}^{2} \alpha)_{lm}
\\
& \quad +
(\mathcal{J}_{F})^{l}_{ik}(x) \cdot (\delta v \delta u)^{i} \cdot (\mathfrak{d} \alpha)_{l} 
\\
& \quad +
(\mathcal{J}_{F})^{l}_{k}(x) \cdot
\Big[ 
(\mathcal{J}_{F})^{m}_{i}(x) \cdot (\delta v \delta u)^{i}
+
(\mathcal{J}_{F})^{m}_{ij}(x) \cdot (\delta v)^{i} \cdot (\delta u)^{j} 
\Big] \cdot
(\mathfrak{d}^{2} \alpha)_{lm}
\\ 
& \quad +
(\mathcal{J}_{F})^{n}_{k}(x) \cdot
\Big[ (\mathcal{J}_{F})^{l}_{i}(x) \cdot (\delta v)^{i} \Big] \cdot
\Big[ (\mathcal{J}_{F})^{m}_{j}(x) \cdot (\delta u)^{j} \Big] \cdot
(\mathfrak{d}^{3} \alpha)_{lmn}
\\
& \quad +
(\mathcal{J}_{F})^{l}_{k}(x) \cdot (\mathfrak{d} \alpha)_{l}
\\
&= \quad 
(\mathcal{J}_{F})^{l}_{ijk}(x) (\delta v)^{i} \cdot (\delta u)^{j} \cdot (\mathfrak{d} \alpha)_{l}
\\ 
& \quad +
(\mathcal{J}_{F})^{m}_{jk}(x) \cdot 
(\delta v^{*})^{l} \cdot (\delta u)^{j} \cdot (\mathfrak{d}^{2} \alpha)_{lm}
+ 
(\mathcal{J}_{F})^{m}_{ik}(x) \cdot 
(\delta u^{*})^{l} \cdot (\delta v)^{i} \cdot (\mathfrak{d}^{2} \alpha)_{lm}
\\
& \quad +
(\mathcal{J}_{F})^{l}_{ik}(x) \cdot (\delta v \delta u)^{i} \cdot (\mathfrak{d} \alpha)_{l}
\\
& \quad +
(\mathcal{J}_{F})^{l}_{k}(x) \cdot 
(\delta v \delta u^{*})^{m} \cdot (\mathfrak{d}^{2} \alpha)_{lm}
\\ 
& \quad +
(\mathcal{J}_{F})^{n}_{k}(x) \cdot
(\delta v^{*})^{l} \cdot (\delta u^{*})^{m} \cdot
 (\mathfrak{d}^{3} \alpha)_{lmn}
\\
& \quad +
(\mathcal{J}_{F})^{l}_{k}(x) \cdot (\mathfrak{d} \alpha)_{l}
\\
&= \quad 
(\mathcal{J}_{F})^{l}_{ijk}(x) \cdot (\delta v)^{i} \cdot (\delta u)^{j} \cdot (\mathfrak{d} \alpha)_{l}
\\ 
& \quad +
(\mathcal{J}_{F})^{m}_{jk}(x) \cdot (\delta u)^{j} \cdot 
\Big[ (\delta v^{*})^{l} (\mathfrak{d}^{2} \alpha)_{lm} \Big]
+ 
(\mathcal{J}_{F})^{m}_{ik}(x) \cdot (\delta v)^{i} \cdot 
\Big[ (\delta u^{*})^{l} (\mathfrak{d}^{2} \alpha)_{lm}  \Big]
\\
& \quad +
(\mathcal{J}_{F})^{l}_{ik}(x) \cdot (\delta v \delta u)^{i} \cdot (\mathfrak{d} \alpha)_{l} 
\\
& \quad +
(\mathcal{J}_{F})^{l}_{k}(x) 
\Big[ 
(\delta v^{*})^{l} \cdot (\delta u^{*})^{m} \cdot (\mathfrak{d}^{3} \alpha)_{lmn} 
+ (\delta v \delta u^{*})^{m} \cdot (\mathfrak{d}^{2} \alpha)_{lm} 
\Big]
\\
& \quad +
(\mathcal{J}_{F})^{l}_{k}(x) \cdot (\mathfrak{d} \alpha)_{l} .
\end{align*}

Defining the partial contractions
\begin{align*}
(\mathfrak{d} \beta)_{l} 
&= (\delta v^{*})^{m} \cdot (\mathfrak{d}^{2} \alpha)_{lm}
\\
(\mathfrak{d} \gamma)_{l} 
&= (\delta u^{*})^{m} \cdot (\mathfrak{d}^{2} \alpha)_{lm}
\\
(\mathfrak{d} \epsilon)_{l} 
&= 
(\delta v^{*})^{l} \cdot (\delta u^{*})^{m} \cdot (\mathfrak{d}^{3} \alpha)_{lmn} 
+ (\delta v \delta u^{*})^{l} \cdot (\mathfrak{d}^{2} \alpha)_{ln}
\end{align*}
this simplifies to
\begin{align*}
(\mathfrak{d} \epsilon^{*})_{k} + (\mathfrak{d} \alpha^{*})_{k}
&=
(\mathcal{J}_{F})^{l}_{ijk}(x) \cdot (\delta v)^{i} \cdot (\delta u)^{j} \cdot (\mathfrak{d} \alpha)_{l}
\\ 
& \quad +
(\mathcal{J}_{F})^{m}_{jk}(x) \cdot (\delta u)^{j} \cdot (\mathfrak{d} \beta)_{m}
+ 
(\mathcal{J}_{F})^{m}_{ik}(x) \cdot (\delta v)^{i} \cdot (\mathfrak{d} \gamma)_{m}
+
(\mathcal{J}_{F})^{l}_{ik}(x) \cdot (\delta v \delta u)^{i} \cdot (\mathfrak{d} \alpha)_{l} 
\\
& \quad +
(\mathcal{J}_{F})^{l}_{k}(x) \cdot (\mathfrak{d} \epsilon)_{l}
\\
& \quad +
(\mathcal{J}_{F})^{l}_{k}(x) \cdot (\mathfrak{d} \alpha)_{l},
\end{align*}
which decouples into three sequential updates: first 
\begin{equation*}
(\mathfrak{d} \alpha^{*})_{j} = (\mathcal{J}_{F})^{l}_{j}(x) \cdot (\mathfrak{d} \alpha)_{l},
\end{equation*}
then two updates that follow from Section \ref{sec:second_order_contraction},
\begin{align*}
(\mathfrak{d} \beta^{*})_{i} 
&= 
(\mathcal{J}_{F})^{l}_{i}(x) \cdot \beta_{l}
+
(\mathcal{J}_{F})^{l}_{ij}(x) \cdot (\delta v)^{j} \cdot (\mathfrak{d} \alpha)_{l}
\\
(\mathfrak{d} \gamma^{*})_{i} 
&= 
(\mathcal{J}_{F})^{l}_{i}(x) \cdot \gamma_{l}
+
(\mathcal{J}_{F})^{l}_{ij}(x) \cdot (\delta u)^{j} \cdot (\mathfrak{d} \alpha)_{l},
\end{align*}
and finally
\begin{align*}
(\mathfrak{d} \epsilon^{*})_{k}
&=
(\mathcal{J}_{F})^{l}_{k}(x) \cdot (\mathfrak{d} \epsilon)_{l}
\\
& \quad +
(\mathcal{J}_{F})^{l}_{ijk}(x) (\delta v)^{i} \cdot (\delta u)^{j} \cdot (\mathfrak{d} \alpha)_{l} 
\\ 
& \quad +
(\mathcal{J}_{F})^{m}_{jk}(x) \cdot (\delta u)^{j} \cdot (\mathfrak{d} \beta)_{m}
+ 
(\mathcal{J}_{F})^{m}_{ik}(x) \cdot (\delta v)^{i} \cdot (\mathfrak{d} \gamma)_{m}
+
(\mathcal{J}_{F})^{l}_{ik}(x) \cdot (\delta v \delta u)^{i} \cdot (\mathfrak{d} \alpha)_{l} .
\end{align*}
In other words, provided that we've already pushed $\delta v$, $\delta u$,
and $\delta v \delta u$ forwards and pulled $\mathfrak{d} \alpha$, $\mathfrak{d} \beta$,
and $\mathfrak{d} \gamma$ back the contraction $\mathfrak{d} \epsilon$ admits a 
well-defined pull back that we can use to propagate new differential information 
through a composite function.

Given the function
\begin{equation*}
F = F_{N} \circ F_{N - 1} \circ \ldots \circ F_{2} \circ F_{1},
\end{equation*}
we can then push the velocities forwards through each component function via
\begin{align*}
(\delta v_{*})^{l} 
&= 
(\mathcal{J}_{F_{n}})^{l}_{i}(x_{n - 1}) \cdot (\delta v)^{i}
\\
(\delta u_{*})^{l} 
&= 
(\mathcal{J}_{F_{n}})^{l}_{i}(x_{n - 1}) \cdot (\delta u)^{i}
\\
(\delta v \delta u_{*})^{l} 
&= 
(\mathcal{J}_{F_{n}})^{l}_{i}(x_{n - 1}) \cdot (\delta v \delta u)^{i}
+ (\mathcal{J}_{F_{n}})^{l}_{ij}(x_{n - 1}) \cdot (\delta v)^{i} \cdot (\delta v)^{j} 
\end{align*}
and then pull the $(1, 1)$-covelocity $\mathfrak{d} \alpha$
backwards through the component functions via
\begin{equation*}
(\mathfrak{d} \alpha^{*})_{i} 
= 
(\mathcal{J}_{F_{n}})^{l}_{i}(x_{n - 1}) \cdot (\mathfrak{d} \alpha)_{l}.
\end{equation*}
As we're pulling back $\mathfrak{d} \alpha$ we can use the intermediate
pushforwards of the velocities to pull back the contractions as well,
\begin{align*}
(\mathfrak{d} \beta^{*})_{i} 
&=
(\mathcal{J}_{F_{n}})^{l}_{i}(x_{n - 1}) \cdot (\mathfrak{d} \beta)_{l}
+
(\mathcal{J}_{F_{n}})^{l}_{ij}(x_{n - 1}) \cdot (\delta v)^{j} \cdot (\mathfrak{d} \alpha)_{l} 
\\
(\mathfrak{d} \gamma^{*})_{i} 
&=
(\mathcal{J}_{F_{n}})^{l}_{i}(x_{n - 1}) \cdot (\mathfrak{d} \gamma)_{l}
+
(\mathcal{J}_{F_{n}})^{l}_{ij}(x_{n - 1}) \cdot (\delta u)^{j} \cdot (\mathfrak{d} \alpha)_{l} 
\\
(\mathfrak{d} \epsilon^{*})_{k}
&= \quad
(\mathcal{J}_{F_{n}})^{l}_{k}(x_{n - 1}) \cdot (\mathfrak{d} \epsilon)_{l}
\\
& \quad +
(\mathcal{J}_{F_{n}})^{l}_{ijk}(x_{n - 1}) \cdot (\delta v)^{i} (\delta u)^{j} \cdot (\mathfrak{d} \alpha)_{l} 
\\ 
& \quad +
(\mathcal{J}_{F_{n}})^{m}_{jk}(x_{n - 1}) \cdot (\delta u)^{j} \cdot (\mathfrak{d} \beta)_{m}
\\
& \quad +
(\mathcal{J}_{F_{n}})^{m}_{ik}(x_{n - 1}) \cdot (\delta v)^{i} \cdot (\mathfrak{d} \gamma)_{m}
\\
& \quad +
(\mathcal{J}_{F_{n}})^{l}_{ik}(x_{n - 1}) \cdot (\delta v \delta u)^{i} \cdot (\mathfrak{d} \alpha)_{l}.
\end{align*}
This sequence implements mixed mode higher-order automatic differentiation 
that evaluates the gradient of second-order directional derivatives.

For example, take the many-to-one composite function 
$F : \mathbb{R}^{D} \rightarrow \mathbb{R}$.  Pushing forwards
$\delta v$, $\delta u$, and $\delta v \delta u = 0$ through $F$ defines 
a pushforward velocity at each component function and then ultimately 
the first and second-order directional derivatives,
\begin{align*}
(\delta v_{*}) 
&= 
\frac{ \partial F }{ \partial x^{i} } (x) \cdot (\delta v)^{i}
\\
(\delta u_{*}) 
&= 
\frac{ \partial F }{ \partial x^{i} } (x) \cdot (\delta u)^{i}
\\
(\delta v \delta u_{*}) 
&= 
\frac{ \partial^{2} F }{ \partial x^{i} \partial x^{j} } (x) \cdot 
(\delta v)^{i} \cdot (\delta v)^{j}.
\end{align*}
Pulling $(\mathfrak{d} \alpha)^{l}  = 1$ and 
$(\mathfrak{d} \beta)^{l} = (\mathfrak{d} \gamma)^{l} = (\mathfrak{d} \epsilon)^{l} = 0$ 
back through $F$ yields the components of the gradient,
\begin{equation*}
(\mathfrak{d} \alpha^{*})_{i} = \frac{ \partial F }{ \partial x^{i} } (x)
\end{equation*}
the components of the two possible first-order directional
derivative gradients,
\begin{align*}
(\mathfrak{d} \beta_{*})_{i} 
&= 
\frac{ \partial^{2} F }{ \partial x^{i} \partial x^{j}} (x) \cdot
(\delta v)^{j}
\\
(\mathfrak{d} \gamma_{*})_{i} 
&= 
\frac{ \partial^{2} F }{ \partial x^{i} \partial x^{j}} (x) \cdot
(\delta u)^{j},
\end{align*}
and then finally the gradient of the second-order directional
derivative,
\begin{equation*}
(\mathfrak{d} \epsilon_{*})_{i} 
= 
\frac{ \partial }{ \partial x^{i} } \left( 
\frac{ \partial^{2} F }{ \partial x^{j} \partial x^{k}} (x) 
\cdot (\delta v)^{j} \cdot (\delta u)^{k} \right)
= 
\frac{ \partial^{3} F }{ \partial x^{i} \partial x^{j} \partial x^{k}} (x) 
\cdot (\delta v)^{j} \cdot (\delta u)^{k}.
\end{equation*}

\section{Practical Implementations of Automatic Differentiation}

The automatic differentiation methods defined by the geometric
transformations introduced in Section \ref{sec:higher_order} are 
straightforward to implement given a composite function.  Decomposing 
a computer program into the necessary component functions in practice, 
however, is far from trivial.  In order to define truly automatic 
differentiation we need a means of automatically specifying appropriate
component functions, or finding away around that requirement.

In this section I review what is needed to construct valid component 
functions from the subexpressions in a computer program.  Given the
existence of such a decomposition I present explicit algorithms that 
implement many of the differential operators introduced in Section 
\ref{sec:higher_order}.  I will then consider the operators for
which the component functions can be further simplified into \emph{local
functions} for which this procedure is significantly easier and
more in line with existing automatic differentiation algorithms.

\subsection{Building Composite Functions}

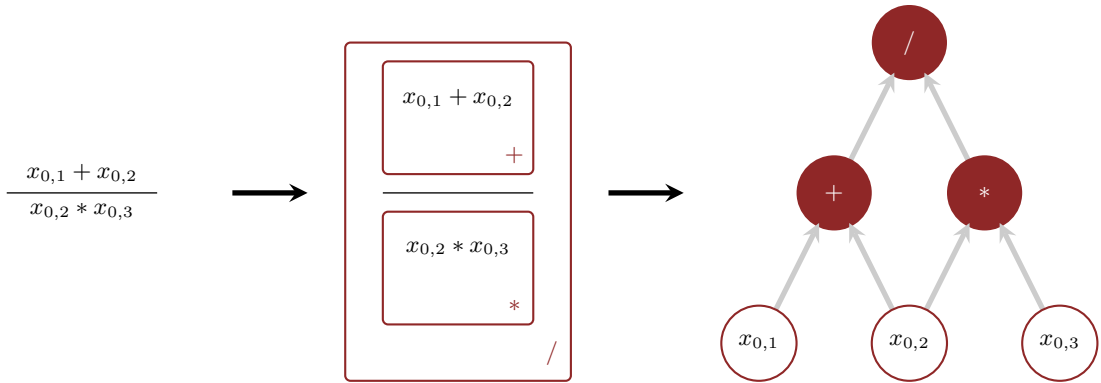
\begin{figure*}
\centering
\begin{tikzpicture}[scale=0.5, thick]

\node at (-18, 4.5) { $x_{0, 1} + x_{0, 2}$ };
\node at (-18, 3.5) { $x_{0, 2} * x_{0, 3}$ };
\draw[-, line width=0.5] (-20, 4) -- (-16, 4);

\draw[->, >=stealth, line width=2] (-14, 4) -- (-12, 4);

\node at (-8, 6.5) { $x_{0, 1} + x_{0, 2}$ };
\draw[rounded corners=2pt, color=dark] (-10, 4.5) rectangle +(4, 3);
\node[color=dark] at (-6.5, 5) { $+$ };

\node at (-8, 2.5) { $x_{0, 2} * x_{0, 3}$ };
\draw[rounded corners=2pt, color=dark] (-10, 0.5) rectangle +(4, 3);
\node[color=dark] at (-6.5, 1) { $*$ };

\draw[-, line width=0.5] (-10, 4) -- (-6, 4);

\draw[rounded corners=2pt, color=dark] (-11, -1) rectangle +(6, 9);
\node[color=dark] at (-5.5, -0.25) { $/$ };

\draw[->, >=stealth, line width=2] (-4, 4) -- (-2, 4);

\fill[color=dark, text=white] (4, 8) circle (1)
node[] { $/$ };

\draw[<->, >=stealth, color=gray80, line width=2] (2, 4) -- +(+1.6, 3.2);

\fill[color=dark, text=white] (2, 4) circle (1)
node[] { $+$ };

\draw[<->, >=stealth, color=gray80, line width=2] (6, 4) -- +(-1.6, 3.2);

\fill[color=dark, text=white] (6, 4) circle (1)
node[] { $*$ };

\draw[<->, >=stealth, color=gray80, line width=2] (0, 0) -- +(+1.6, 3.2);

\filldraw[draw=dark, fill=white] (0, 0) circle (1)
node[] { $x_{0, 1}$ };

\draw[<->, >=stealth, color=gray80, line width=2] (4, 0) -- +(-1.6, 3.2);
\draw[<->, >=stealth, color=gray80, line width=2] (4, 0) -- +(+1.6, 3.2);

\filldraw[draw=dark, fill=white] (4, 0) circle (1)
node[] { $x_{0, 2}$ };

\draw[<->, >=stealth, color=gray80, line width=2] (8, 0) -- +(-1.6, 3.2);

\filldraw[draw=dark, fill=white] (8, 0) circle (1)
node[] { $x_{0, 3}$ };

\end{tikzpicture}
\caption{
The subexpressions of a computer program that computes a function
between manifolds define a directed acyclic graph known as 
an \textit{expression graph}, where each internal node defines a 
subexpression in the program and the leaf nodes define input 
variables.  A program implementing the function 
$y = (x_{0, 1} + x_{0, 2}) / (x_{0, 2} * x_{0, 3} )$, for example,
generates a graph with three subexpression nodes and the three 
input leaves, $\left\{ x_{0, 1}, x_{0, 2}, x_{0, 3} \right\}$.
}
\label{fig:expr_graph} 
\end{figure*}

Consider a computer program that implements a function between two
manifolds, $F : X \rightarrow Y$ by computing the output $F(x) \in Y$ 
for any input $x \in X$.  The first step towards turning such a 
program into a composite function is to represent it as an 
\textit{expression graph}, with internal nodes designating each 
subexpression in the program, leaf nodes designating the input 
variables, and edges designating functional dependencies (Figure 
\ref{fig:expr_graph}).

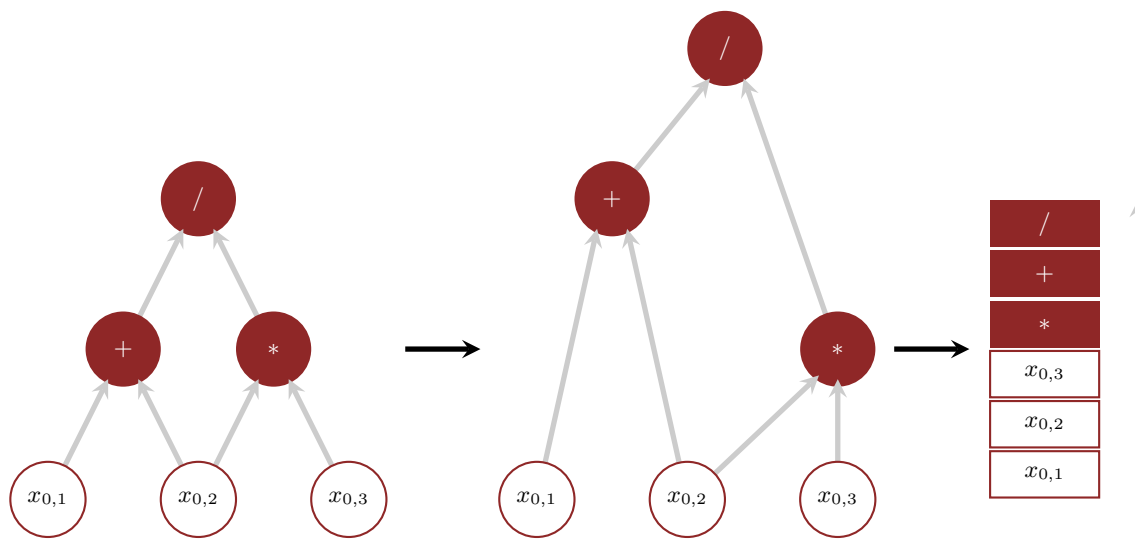
\begin{figure*}
\centering
\begin{tikzpicture}[scale=0.5, thick]

\fill[color=dark, text=white] (4, 8) circle (1)
node[] { $/$ };

\draw[->, >=stealth, color=gray80, line width=2] (2, 4) -- +(+1.6, 3.2);

\fill[color=dark, text=white] (2, 4) circle (1)
node[] { $+$ };

\draw[->, >=stealth, color=gray80, line width=2] (6, 4) -- +(-1.6, 3.2);

\fill[color=dark, text=white] (6, 4) circle (1)
node[] { $*$ };

\draw[->, >=stealth, color=gray80, line width=2] (0, 0) -- +(+1.6, 3.2);

\filldraw[draw=dark, fill=white] (0, 0) circle (1)
node[] { $x_{0, 1}$ };

\draw[->, >=stealth, color=gray80, line width=2] (4, 0) -- +(-1.6, 3.2);
\draw[->, >=stealth, color=gray80, line width=2] (4, 0) -- +(+1.6, 3.2);

\filldraw[draw=dark, fill=white] (4, 0) circle (1)
node[] { $x_{0, 2}$ };

\draw[->, >=stealth, color=gray80, line width=2] (8, 0) -- +(-1.6, 3.2);

\filldraw[draw=dark, fill=white] (8, 0) circle (1)
node[] { $x_{0, 3}$ };

\draw[->, >=stealth, line width=2] (9.5, 4) -- +(2, 0);

\pgfmathsetmacro{\dxx}{10}

\fill[color=dark, text=white] (8 + \dxx, 12) circle (1)
node[] { $/$ };

\draw[<->, >=stealth, color=gray80, line width=2] (5 + \dxx, 8) -- +(2.6, 3.2);
\fill[color=dark, text=white] (5 + \dxx, 8) circle (1)
node[] { $+$ };

\draw[<->, >=stealth, color=gray80, line width=2] (11 + \dxx, 4) -- +(-2.5, 7.2);
\fill[color=dark, text=white] (11 + \dxx, 4) circle (1)
node[] { $*$ };

\draw[<->, >=stealth, color=gray80, line width=2] (3 + \dxx, 0) -- +(1.6, 7.2);

\filldraw[draw=dark, fill=white] (3 + \dxx, 0) circle (1)
node[] { $x_{0, 1}$ };

\draw[<->, >=stealth, color=gray80, line width=2] (7 + \dxx, 0) -- +(-1.6, 7.2);
\draw[<->, >=stealth, color=gray80, line width=2] (7 + \dxx, 0) -- +(3.5, 3.3);

\filldraw[draw=dark, fill=white] (7 + \dxx, 0) circle (1)
node[] { $x_{0, 2}$ };

\draw[<->, >=stealth, color=gray80, line width=2] (11 + \dxx, 0) -- +(0, 3.2);

\filldraw[draw=dark, fill=white] (11 + \dxx, 0) circle (1)
node[] { $x_{0, 3}$ };

\draw[->, >=stealth, line width=2] (22.5, 4) -- +(2, 0);

\pgfmathsetmacro{\dxxx}{25}
\pgfmathsetmacro{\ddx}{0.05}

\filldraw[draw=dark, fill=white] (\dxxx + \ddx, 0.000 + \ddx) rectangle +(3 - 2 * \ddx, 1.333 - 2 * \ddx)
node[midway] { $x_{0, 1}$ };

\filldraw[draw=dark, fill=white] (\dxxx + \ddx, 1.333 + \ddx) rectangle +(3 - 2 * \ddx, 1.334 - 2 * \ddx)
node[midway] { $x_{0, 2}$ };

\filldraw[draw=dark, fill=white] (\dxxx + \ddx, 2.667 + \ddx) rectangle +(3 - 2 * \ddx, 1.333 - 2 * \ddx)
node[midway] { $x_{0, 3}$ };

\fill[color=dark, text=white] (\dxxx + \ddx, 4.000 + \ddx) rectangle +(3 - 2 * \ddx, 1.333 - 2 * \ddx)
node[midway] { $*$ };

\fill[color=dark, text=white] (\dxxx + \ddx, 5.333 + \ddx) rectangle +(3 - 2 * \ddx, 1.333 - 2 * \ddx)
node[midway] { $+$ };

\fill[color=dark, text=white] (\dxxx + \ddx, 6.667 + \ddx) rectangle +(3 - 2 * \ddx, 1.333 - 2 * \ddx)
node[midway] { $/$ };

\draw[->, >=stealth, color=gray80, line width=2] (\dxxx + 4, 0) -- (\dxxx + 4, 8);

\end{tikzpicture}
\caption{
A topological sort of an expression graph yields a non-unique
ordering of the subexpressions, often called a \emph{stack} or 
a \emph{tape} in the context of automatic differentiation.  The 
ordering guarantees that when sweeping across the stack the 
subexpression defined at each node will not be processed until 
all of the subexpressions on which it depends have been processed.
}
\label{fig:topological_sort} 
\end{figure*}

A \emph{topological sort} of the nodes in the expression graph 
yields a non-unique ordering such that each ndoes follows its 
parents (Figure \ref{fig:topological_sort}).  This sorting ensures 
that if we sweep along the sorted nodes then any subexpression 
will be considered only once all of the subexpressions on which 
it depends have already been considered.

With this ordering the subexpressions define a sequence of functions
that \emph{almost} compose together to yield the target function.
The immediate limitation is that the subexpressions within each node 
can depend on \emph{any} of the previous subexpression whereas 
composition utilizes only the output of the function immediately 
proceeding each function.  

\begin{figure*}
\centering
\begin{tikzpicture}[scale=0.5, thick]

\pgfmathsetmacro{\dxx}{-14}

\fill[color=dark, text=white] (8 + \dxx, 12) circle (1)
node[] { $/$ };

\draw[<->, >=stealth, color=gray80, line width=2] (5 + \dxx, 8) -- +(2.6, 3.2);
\fill[color=dark, text=white] (5 + \dxx, 8) circle (1)
node[] { $+$ };

\draw[<->, >=stealth, color=gray80, line width=2] (11 + \dxx, 4) -- +(-2.5, 7.2);
\fill[color=dark, text=white] (11 + \dxx, 4) circle (1)
node[] { $*$ };

\draw[<->, >=stealth, color=gray80, line width=2] (3 + \dxx, 0) -- +(1.6, 7.2);

\filldraw[draw=dark, fill=white] (3 + \dxx, 0) circle (1)
node[] { $x_{0, 1}$ };

\draw[<->, >=stealth, color=gray80, line width=2] (7 + \dxx, 0) -- +(-1.6, 7.2);
\draw[<->, >=stealth, color=gray80, line width=2] (7 + \dxx, 0) -- +(3.5, 3.3);

\filldraw[draw=dark, fill=white] (7 + \dxx, 0) circle (1)
node[] { $x_{0, 2}$ };

\draw[<->, >=stealth, color=gray80, line width=2] (11 + \dxx, 0) -- +(0, 3.2);

\filldraw[draw=dark, fill=white] (11 + \dxx, 0) circle (1)
node[] { $x_{0, 3}$ };

\draw[->, >=stealth, line width=2] (-1, 4) -- (1, 4);

\fill [rounded corners=2pt, color=gray95] (1.5, 10.5) rectangle (12.5, 13.5);
\fill [rounded corners=2pt, color=gray95] (1.5, 6.5) rectangle (12.5, 9.5);
\fill [rounded corners=2pt, color=gray95] (1.5, 2.5) rectangle (12.5, 5.5);

\fill[color=dark, text=white] (8, 12) circle (1)
node[] { $/$ };

\draw[<->, >=stealth, color=gray80, line width=2] (11, 8) -- +(-2.6, 3.2);
\fill[color=dark, text=white] (11, 8) circle (1)
node[] { $I$ };

\draw[<->, >=stealth, color=gray80, line width=2] (5, 8) -- +(2.6, 3.2);
\fill[color=dark, text=white] (5, 8) circle (1)
node[] { $+$ };

\draw[<->, >=stealth, color=gray80, line width=2] (3, 4) -- +(1.6, 3.2);
\fill[color=dark, text=white] (3, 4) circle (1)
node[] { $I$ };

\draw[<->, >=stealth, color=gray80, line width=2] (7, 4) -- +(-1.6, 3.2);
\fill[color=dark, text=white] (7, 4) circle (1)
node[] { $I$ };

\draw[<->, >=stealth, color=gray80, line width=2] (11, 4) -- +(0, 3.2);
\fill[color=dark, text=white] (11, 4) circle (1)
node[] { $*$ };

\draw[<->, >=stealth, color=gray80, line width=2] (3, 0) -- +(0, 3.2);

\filldraw[draw=dark, fill=white] (3, 0) circle (1)
node[] { $x_{0, 1}$ };

\draw[<->, >=stealth, color=gray80, line width=2] (7, 0) -- +(0, 3.2);
\draw[<->, >=stealth, color=gray80, line width=2] (7, 0) -- +(3.5, 3.3);

\filldraw[draw=dark, fill=white] (7, 0) circle (1)
node[] { $x_{0, 2}$ };

\draw[<->, >=stealth, color=gray80, line width=2] (11, 0) -- +(0, 3.2);

\filldraw[draw=dark, fill=white] (11, 0) circle (1)
node[] { $x_{0, 3}$ };

\end{tikzpicture}
\caption{
In order to define valid component functions, each subexpression in a
topologically-ordered expression graph must be complemented with identity 
maps that carry forward intermediate variables needed by future 
subexpressions.
}
\label{fig:encapsulated_expressions} 
\end{figure*}
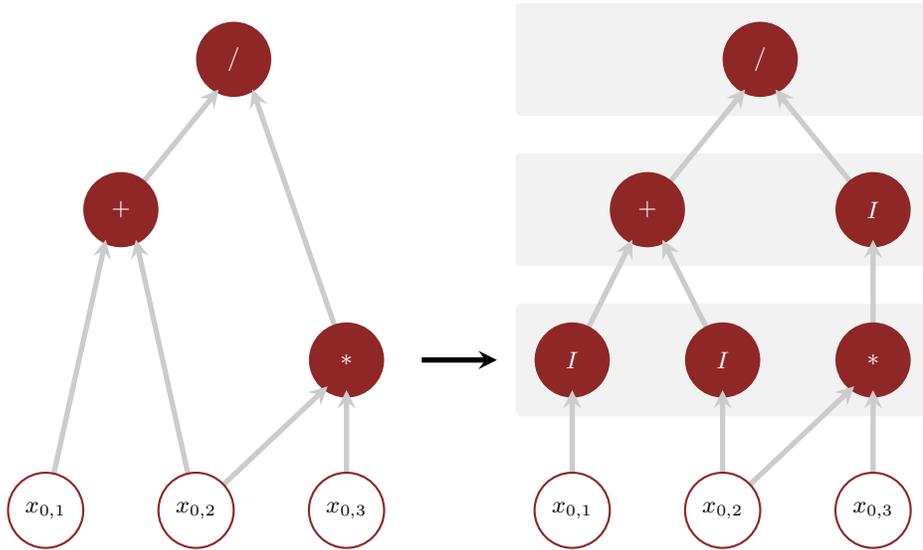

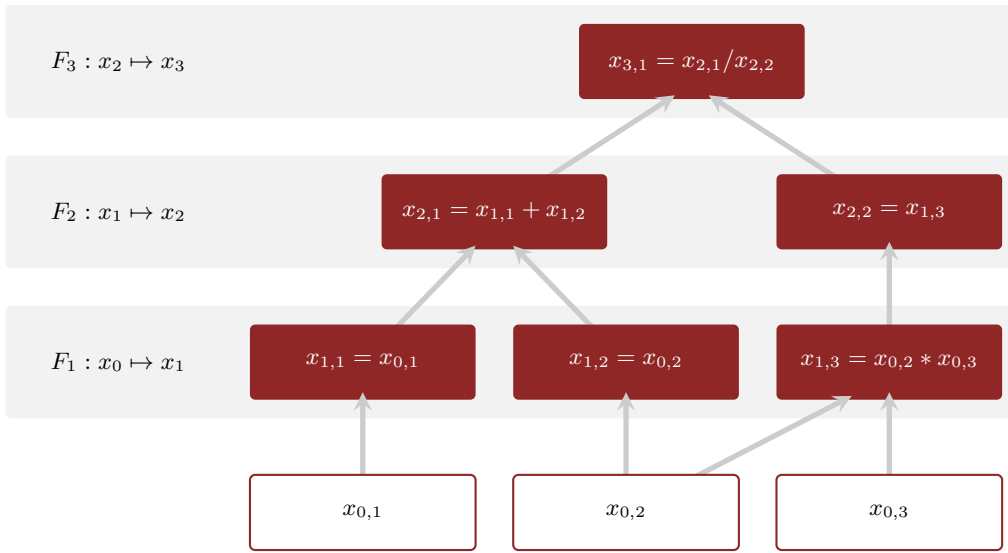
\begin{figure*}
\centering
\begin{tikzpicture}[scale=0.5, thick]

\fill [rounded corners=2pt, color=gray95] (-6.5, 10.5) rectangle (20.5, 13.5);
\node at (-3.5, 12) { $F_{3} : x_{2} \mapsto x_{3}$ };

\fill [rounded corners=2pt, color=gray95] (-6.5, 6.5) rectangle (20.5, 9.5);
\node at (-3.5, 8) { $F_{2} : x_{1} \mapsto x_{2}$ };

\fill [rounded corners=2pt, color=gray95] (-6.5, 2.5) rectangle (20.5, 5.5);
\node at (-3.5, 4) { $F_{1} : x_{0} \mapsto x_{1}$ };

\fill[rounded corners=2, color=dark] (8.75, 11) rectangle +(6, 2);
\node[text=white] at (11.75, 12) { $x_{3, 1} = x_{2, 1} / x_{2, 2}$ };

\draw[<->, >=stealth, color=gray80, line width=2] (17, 8) -- +(-4.8, 3.1);
\fill[rounded corners=2, color=dark] (14, 7) rectangle +(6, 2);
\node[text=white] at (17, 8) { $x_{2, 2} = x_{1, 3}$ };

\draw[<->, >=stealth, color=gray80, line width=2] (6.5, 8) -- +(4.8, 3.1);
\fill[rounded corners=2, color=dark] (3.5, 7) rectangle +(6, 2);
\node[text=white] at (6.5, 8) { $x_{2, 1} = x_{1, 1} + x_{1, 2}$ };

\draw[<->, >=stealth, color=gray80, line width=2] (3, 4) -- +(3, 3.1);
\fill[rounded corners=2, color=dark] (0, 3) rectangle +(6, 2);
\node[text=white] at (3, 4) { $x_{1, 1} = x_{0, 1}$ };

\draw[<->, >=stealth, color=gray80, line width=2] (10, 4) -- +(-3, 3.1);
\fill[rounded corners=2, color=dark] (7, 3) rectangle +(6, 2);
\node[text=white] at (10, 4) { $x_{1, 2} = x_{0, 2}$ };

\draw[<->, >=stealth, color=gray80, line width=2] (17, 4) -- +(0, 3.2);

\fill[rounded corners=2, color=dark] (14, 3) rectangle +(6, 2);
\node[text=white] at (17, 4) { $x_{1, 3} = x_{0, 2} * x_{0, 3}$ };

\draw[<->, >=stealth, color=gray80, line width=2] (3, 0) -- +(0, 3.2);

\filldraw[rounded corners=2, draw=dark, fill=white] (0, -1) rectangle +(6, 2);
\node[] at (3, 0) { $x_{0, 1}$ };

\draw[<->, >=stealth, color=gray80, line width=2] (10, 0) -- +(0, 3.2);
\draw[<->, >=stealth, color=gray80, line width=2] (10, 0) -- +(6, 3.1);

\filldraw[rounded corners=2, draw=dark, fill=white] (7, -1) rectangle +(6, 2);
\node[] at (10, 0) { $x_{0, 2}$ };

\draw[<->, >=stealth, color=gray80, line width=2] (17, 0) -- +(0, 3.2);

\filldraw[rounded corners=2, draw=dark, fill=white] (14, -1) rectangle +(6, 2);
\node[] at (17, 0) { $x_{0, 3}$ };

\end{tikzpicture}
\caption{
Each subexpression and accompanying identity maps define component functions
that can be composed together to yield the function implemented by the
initial computer program, here $F = F_{3} \circ F_{2} \circ F_{1}$.
}
\label{fig:component_functions} 
\end{figure*}

In order to define valid component functions we need to propagate
any intermediate variables that are used by future subexpressions
along with the output of a given subexpression.  This can be
implemented by complementing each subexpression with identify 
functions that map any necessary variables forward (Figure 
\ref{fig:encapsulated_expressions}).  This supplement then yields 
a sequence of component functions that depend only on the 
immediately proceeding component function and hence can be 
composed together (Figure \ref{fig:component_functions}).

Once we've decomposed the subexpressions in a computer program
into component functions we can implement any of the automatic
differentiation methods defined by geometric transformations,
which become sweeps across the ordered component functions.  
This translation of subexpressions into component functions,
however, is absent in most treatments of automatic differentiation.
As we will see below, it ends up being unnecessary for methods
that utilize only one-dimensional arrays, such as those defined 
in forward mode and mixed mode automatic differentiation, as well
as first-order reverse mode automatic differentiation.  The
translation is critical, however, for higher-order reverse mode 
automatic differentiation methods.

\subsection{General Implementations}

Following convention in the automatic differentiation literature
let's consider computer programs that define many-to-one 
real-valued functions, $F : \mathbb{R}^{D} \rightarrow \mathbb{R}$.
Further let's presume the existence of an algorithm that can convert 
the subexpressions in the computer program into the component
functions
\begin{equation*}
F = F_{N} \circ F_{N - 1} \circ \ldots \circ F_{2} \circ F_{1},
\end{equation*}
where 
\begin{alignat*}{6}
F_{n} :\; &\mathbb{R}^{D_{n - 1}}& &\rightarrow& \; &\mathbb{R}^{D_{n}}&
\\
&x_{n - 1}& &\mapsto& &x_{n}&
\end{alignat*}
with $D_{0} = D$ and $D_{N} = 1$

In this circumstance the intermediate Jacobians arrays become 
arrays of the partial derivatives of the intermediate, real-valued 
functions,
\begin{equation*}
(\mathcal{J}_{F_{n}})^{l}_{i_{1} \ldots i_{K}}
=
\frac{ \partial^{K} (F_{n})^{l} }
{ \partial x_{i_{1}} \ldots \partial x_{i_{K}} } (x_{n - 1}).
\end{equation*}

Each geometric object that we push forward or pull back across 
the sequence of composite function implements a unique differential 
operator.  Moreover, these methods evaluate not only the desired 
differential operator but also various \emph{adjunct} operators
at the same time which may be of use in of themselves.  Here cost 
is quantified relative to a single function evaluation. 

\subsubsection{Forward Mode Methods}

In order to implement first order methods the $n$-th component 
function and its output value, $x_{n}$ must be complemented with
a $(1,1)$-velocity represented with the coordinates $(\delta v_{n})^{i}$.

\begin{tcolorbox}[colback=white,colframe=gray90, coltitle=black,boxrule=3pt,
fonttitle=\bfseries,title=First-Order Directional Derivative]

\begin{description}

\item[Form:] $\displaystyle \mathbf{\delta v}^{T} \cdot \mathbf{g} 
		               = \sum_{i = 1}^{N} (v)_{i} \frac{ \partial F }{ \partial x_{i} } (x) $
		               
\item[Algorithm:] \hfill
\begin{algorithmic}[0]
\State Construct composite functions, $(F_{1}, \ldots, F_{N})$.
\State $x_{0} \gets x$, $(\delta v_{0})^{i} \gets v^{i}$.
\State
\For{$1 \le n \le N$}
  \State $x_{n} \gets F_{n}(x_{n - 1})$
  \State $\displaystyle (\delta v_{n})^{l} \gets \sum_{j = 1}^{D_{n - 1}} 
         \frac{ \partial (F_{n})^{l} }{ \partial x_{i} }(x_{n - 1}) \cdot (\delta v_{n - 1})^{i}$
\EndFor
\State
\Return $(\delta v_{N})^{1}$
\end{algorithmic}

\item[Cost:] $\mathcal{O} \! \left( 1 \right)$
\item[Adjuncts:] $ f $
\end{description}
	
\end{tcolorbox}

For second order methods the $n$-th component function needs
to be complemented with only a three of the coordinates of a 
$(2, 2)$-velocity: $(\delta v_{n})^{i}$, $(\delta u_{n})^{i}$, 
and $(\delta v \delta u_{n})^{i}$.  The additional coordinates
$(\delta^{2} v_{n})^{i}$ and $(\delta^{2} v_{n})^{i}$ are
unnecessary.

\begin{tcolorbox}[colback=white,colframe=gray90, coltitle=black,boxrule=3pt,
fonttitle=\bfseries,title=Second-Order Directional Derivative]

\begin{description}

\item[Form:] $\displaystyle \mathbf{\delta v}^{T} \cdot \mathbf{H} \cdot \mathbf{\delta u} 
		        = \sum_{i = 1}^{N} \sum_{j = 1}^{N} (v)_{i} (u)_{j} 
		          \frac{ \partial^{2} F }{ \partial x_{i} \partial x_{j} } (x)$
		          
\item[Algorithm:] \hfill
\begin{algorithmic}[0]
\State Construct composite functions, $(F_{1}, \ldots, F_{N})$.
\State $x_{0} \gets $, $(\delta v_{0})^{i} \gets v^{i}$, $(\delta u_{0})^{i} \gets u^{i}$ 
\State $(\delta v \delta u_{0})^{i} \gets 0$.
\State
\For{$1 \le n \le N$}
  \State $x_{n} \gets F_{n}(x_{n - 1})$
  \State $\displaystyle (\delta v_{n})^{l} \gets \sum_{j = 1}^{D_{n - 1}} 
         \frac{ \partial (F_{n})^{l} }{ \partial x_{i} }(x_{n - 1}) \cdot (\delta v_{n - 1})^{i}$
  \State $\displaystyle (\delta u_{n})^{l} \gets \sum_{j = 1}^{D_{n - 1}} 
         \frac{ \partial (F_{n})^{l} }{ \partial x_{i} }(x_{n - 1}) \cdot (\delta u_{n - 1})^{i}$
         
  \State $\displaystyle (\delta v \delta u_{n})^{l} \gets
         \hspace{4mm} \sum_{i = 1}^{D_{n - 1}} 
         \frac{ \partial (F_{n})^{l} }{ \partial x_{i} }(x_{n - 1}) \cdot (\delta u \delta v_{n - 1})^{i}$
  \State $\hspace{20mm} \displaystyle + \sum_{i = 1}^{D_{n - 1}} \sum_{j = 1}^{D_{n - 1}} 
         \frac{ \partial^{2} (F_{n})^{l} }{ \partial x_{i} \partial x_{j} }(x_{n - 1}) 
         \cdot (\delta u_{n - 1})^{i} \cdot (\delta v_{n - 1})^{j}$
\EndFor
\State
\Return $(\delta v \delta u_{N})^{1}$
\end{algorithmic}

\item[Cost:] $\mathcal{O} \! \left( 1 \right)$
\item[Adjuncts:] $ f, \, \mathbf{v}^{T} \cdot \mathbf{g} = \delta v_{N}, \, 
                         \mathbf{u}^{T} \cdot \mathbf{g} = \delta u_{N}$
\end{description}
	
\end{tcolorbox}

A third-order directional derivative requires complementing 
the $n$-th component function with only 7 of the sixteen
coordinates of a $(3,3)$-velocity: $(\delta v_{n})^{i}$, 
$(\delta u_{n})^{i}$, $(\delta w_{n})^{i}$, $(\delta v \delta u_{n})^{i}$, 
$(\delta v \delta w_{n})^{i}$, $(\delta u \delta w_{n})^{i}$ and 
$(\delta v \delta u \delta w_{n})^{i}$.

\begin{tcolorbox}[colback=white,colframe=gray90, coltitle=black,boxrule=3pt,
fonttitle=\bfseries,title=Third-Order Directional Derivative]

\begin{description}

\item[Form:] $\displaystyle \mathbf{\delta v}^{T} \cdot \mathbf{H} \cdot \mathbf{\delta u} 
		        = \sum_{i=1}^{N} \sum_{j=1}^{N} \sum_{k=1}^{N} (v)_{i} (u)_{j} (w)_{k} 
		          \frac{ \partial^{3} F }{ \partial x_{i} \partial x_{j} \partial x_{k} } (x)$
		          
\item[Algorithm:] \hfill
\begin{algorithmic}[0]
\State Construct composite functions, $(F_{1}, \ldots, F_{N})$.
\State $x_{0} \gets x$, $(\delta v_{0})^{i} \gets v^{i}$, $(\delta u_{0})^{i} \gets u^{i}$, 
       $(\delta w_{0})^{i} \gets w^{i}$
\State $(\delta v \delta u_{0})^{i} \gets 0$, $(\delta v \delta w_{0})^{i} \gets 0$, 
       $(\delta u \delta w_{0})^{i} \gets 0$, $(\delta v \delta u \delta w_{0})^{i} \gets 0$.
\State
\For{$1 \le n \le N$}
  \State $x_{n} \gets F_{n}(x_{n - 1})$
  \State $(\delta v_{n})^{l} \gets \sum_{j = 1}^{D_{n - 1}} 
         \frac{ \partial (F_{n})^{l} }{ \partial x_{i} }(x_{n - 1}) \cdot (\delta v_{n - 1})^{i}$
  \State $(\delta u_{n})^{l} \gets \sum_{j = 1}^{D_{n - 1}} 
         \frac{ \partial (F_{n})^{l} }{ \partial x_{i} }(x_{n - 1}) \cdot (\delta u_{n - 1})^{i}$
  \State $ (\delta w_{n})^{l} \gets \sum_{j = 1}^{D_{n - 1}} 
         \frac{ \partial (F_{n})^{l} }{ \partial x_{i} }(x_{n - 1}) \cdot (\delta w_{n - 1})^{i}$
         
  \State $(\delta v \delta u_{n})^{l} \gets 
         \hspace{4mm} \sum_{i = 1}^{D_{n - 1}} 
         \frac{ \partial (F_{n})^{l} }{ \partial x_{i} }(x_{n - 1}) \cdot (\delta v \delta u_{n - 1})^{i}$
  \State $\hspace{20mm} + \sum_{i = 1}^{D_{n - 1}} \sum_{j = 1}^{D_{n - 1}} 
         \frac{ \partial^{2} (F_{n})^{l} }{ \partial x_{i} \partial x_{j} }(x_{n - 1}) 
         \cdot (\delta v_{n - 1})^{i} \cdot (\delta u_{n - 1})^{j}$
         
  \State $(\delta v \delta w_{n})^{l} \gets 
         \hspace{4mm} \sum_{i = 1}^{D_{n - 1}} 
         \frac{ \partial (F_{n})^{l} }{ \partial x_{i} }(x_{n - 1}) \cdot (\delta v \delta w_{n - 1})^{i}$
  \State $\hspace{20mm} + \sum_{i = 1}^{D_{n - 1}} \sum_{j = 1}^{D_{n - 1}} 
         \frac{ \partial^{2} (F_{n})^{l} }{ \partial x_{i} \partial x_{j} }(x_{n - 1}) 
         \cdot (\delta v_{n - 1})^{i} \cdot (\delta w_{n - 1})^{j}$
         
  \State $(\delta u \delta w_{n})^{l} \gets 
         \hspace{4mm} \sum_{i = 1}^{D_{n - 1}} 
         \frac{ \partial (F_{n})^{l} }{ \partial x_{i} }(x_{n - 1}) \cdot (\delta u \delta w_{n - 1})^{i}$
  \State $\hspace{20mm} + \sum_{i = 1}^{D_{n - 1}} \sum_{j = 1}^{D_{n - 1}} 
         \frac{ \partial^{2} (F_{n})^{l} }{ \partial x_{i} \partial x_{j} }(x_{n - 1}) 
         \cdot (\delta u_{n - 1})^{i} \cdot (\delta w_{n - 1})^{j}$
         
  \State $(\delta v \delta u \delta w_{n})^{l} \gets 
         \hspace{4mm} \sum_{i = 1}^{D_{n - 1}} 
         \frac{ \partial (F_{n})^{l} }{ \partial x_{i} }(x_{n - 1}) 
           \cdot (\delta v \delta u \delta w_{n - 1})^{i}$
  \State $\hspace{25mm} + \sum_{i = 1}^{D_{n - 1}} \sum_{j = 1}^{D_{n - 1}} 
         \frac{ \partial^{2} (F_{n})^{l} }{ \partial x_{i} \partial x_{j} }(x_{n - 1}) $
  \State $\hspace{35mm} \cdot \Big( \quad (\delta v_{n - 1})^{i} \cdot (\delta u \delta w_{n - 1})^{j}$
  \State $\hspace{39mm} + (\delta u_{n - 1})^{i} \cdot (\delta v \delta w_{n - 1})^{j}$
  \State $\hspace{39mm} + (\delta w_{n - 1})^{i} \cdot (\delta v \delta u_{n - 1})^{j} \Big)$
  \State $\hspace{25mm} + \sum_{i = 1}^{D_{n - 1}} \sum_{j = 1}^{D_{n - 1}} \sum_{k = 1}^{D_{n - 1}} 
         \frac{ \partial^{3} (F_{n})^{l} }{ \partial x_{i} \partial x_{j} \partial x_{k} }(x_{n - 1}) $
  \State $\hspace{35mm} \cdot (\delta v_{n - 1})^{i} \cdot (\delta u_{n - 1})^{j} \cdot (\delta w_{n - 1})^{k}$
\EndFor
\State
\Return $(\delta v \delta u \delta w_{N})^{1}$
\end{algorithmic}

\item[Cost:] $\mathcal{O} \! \left( 1 \right)$
\item[Adjuncts:] $ f, \, \mathbf{v}^{T} \cdot \mathbf{g} = \delta v_{N}, \, 
    \mathbf{u}^{T} \cdot \mathbf{g} = \delta u_{N}, \mathbf{w}^{T} \cdot \mathbf{g} = \delta w_{N}, \
    \mathbf{v}^{T} \cdot \mathbf{H} \cdot \mathbf{u} = \delta v \delta u_{N}, \
    \mathbf{v}^{T} \cdot \mathbf{H} \cdot \mathbf{w} = \delta v \delta w_{N}, \
    \mathbf{u}^{T} \cdot \mathbf{H} \cdot \mathbf{w} = \delta u \delta w_{N}$
\end{description}
	
\end{tcolorbox}

\subsubsection{Reverse Mode Methods}

In order to implement a first-order reverse mode method we
must complement the $n$-th component function with the 
coordinates of a $(1, 1)$-covelocity, $(\mathfrak{d} \alpha_{n})_{l}$. 

\begin{tcolorbox}[colback=white,colframe=gray90, coltitle=black,boxrule=3pt,
fonttitle=\bfseries,title=Gradient]
	
\begin{description}
\item[Form:] $\displaystyle g_{i} = \frac{ \partial F }{ \partial x_{i} } (x)$
\item[Algorithm:] \hfill

\begin{algorithmic}[0]
\State Construct composite functions, $(F_{1}, \ldots, F_{N})$
\State $x_{0} \gets x$, $(\mathfrak{d} \alpha_{0})_{l} \gets 0$
\State 
\For{$1 \le n \le N$}
  \State $x_{n} \gets F_{n}(x_{n - 1})$
  \State $(\mathfrak{d} \alpha_{n})_{l} \gets 0$
\EndFor
\State $(\mathfrak{d} \alpha_{N})_{1} \gets 1$
\For{$N \ge n \ge 1$}
  \State $\displaystyle (\mathfrak{d} \alpha_{n - 1})_{i} \gets \sum_{l = 1}^{D_{n}} 
         \frac{ \partial (F_{n})^{l} }{ \partial x_{i} }(x_{n - 1}) \cdot (\mathfrak{d} \alpha_{n})_{l}$
\EndFor
\State
\Return $\{ (\mathfrak{d} \alpha_{0})_{1}, \ldots, (\mathfrak{d} \alpha_{0})_{N} \}$
\end{algorithmic}

\item[Cost:] $\mathcal{O} \! \left( 1 \right)$
\item[Adjuncts:] $ f $
\end{description}
	
\end{tcolorbox}

For second-order methods we need all of the coordinates of
a $(1, 2)$-covelocity: $(\mathfrak{d} \alpha_{n})_{l}$ and 
$(\mathfrak{d}^{2} \alpha_{n})_{lm}$. 

\begin{tcolorbox}[colback=white,colframe=gray90, coltitle=black,boxrule=3pt,
fonttitle=\bfseries,title=Hessian]
	
\begin{description}
\item[Form:] $\displaystyle H_{ij} = \frac{ \partial^{2} F }{ \partial x_{i} \partial x_{j} } (x)$

\item[Algorithm:] \hfill
\begin{algorithmic}[0]
\State Construct composite functions, $(F_{1}, \ldots, F_{N})$
\State $x_{0} \gets x$, $(\mathfrak{d} \alpha_{0})_{l} \gets 0$, $(\mathfrak{d}^{2} \alpha_{0})_{lm} \gets 0$,
\State
\For{$n \in \left\{ 1, \ldots, N \right\}$}
  \State $x_{n} \gets F_{n}(x_{n - 1})$
  \State $(\mathfrak{d} \alpha_{n})_{l} \gets 0$, $(\mathfrak{d}^{2} \alpha_{n})_{lm} \gets 0$
\EndFor
\State $(\mathfrak{d} \alpha_{N})_{l} \gets 1$
\For{$N \ge n \ge 1$}
  \State $\displaystyle (\mathfrak{d} \alpha_{n - 1})_{i} \gets \sum_{l = 1}^{D_{n}} 
         \frac{ \partial (F_{n})^{l} }{ \partial x_{i} }(x_{n - 1}) \cdot (\mathfrak{d} \alpha_{n})_{l}$
  \State $\displaystyle \displaystyle (\mathfrak{d}^{2} \alpha_{n - 1})_{ij} \gets
         \hspace{4mm} \sum_{l = 1}^{D_{n}} 
         \frac{ \partial^{2} (F_{n})^{l} }{ \partial x_{i} \partial x_{j} }(x_{n - 1}) 
         \cdot (\mathfrak{d} \alpha_{n - 1})_{l}$
  \State $\hspace{23mm} \displaystyle + \sum_{l = 1}^{D_{n}} \sum_{m = 1}^{D_{n}} 
         \frac{ \partial (F_{n})^{l} }{ \partial x_{i} }(x_{n - 1}) \cdot
         \frac{ \partial (F_{n})^{m} }{ \partial x_{j} }(x_{n - 1}) \cdot
         (\mathfrak{d}^{2} \alpha_{n - 1})_{lm}$
\EndFor
\State
\Return $\{ (\mathfrak{d}^{2} \alpha_{0})_{11}, \ldots, (\mathfrak{d}^{2} \alpha_{0})_{NN} \}$
\end{algorithmic}

\item[Cost:] $\mathcal{O} \! \left( 1 \right)$
\item[Adjuncts:] $ f, \, 
  \mathbf{g} = \{ (\mathfrak{d} \alpha_{0})_{1}, \ldots, (\mathfrak{d} \alpha_{0})_{N} \}$
\end{description}

\end{tcolorbox}

Third-order methods we need all of the coordinates of a 
$(1, 3)$-covelocity: $(\mathfrak{d} \alpha_{n})_{l}$,
$(\mathfrak{d}^{2} \alpha_{n})_{lm}$, and 
$(\mathfrak{d}^{3} \alpha_{n})_{lmq}$

\begin{tcolorbox}[colback=white,colframe=gray90, coltitle=black,boxrule=3pt,
fonttitle=\bfseries,title=Third-Order Partial Derivative Array]
	
\begin{description}
\item[Form:] $\displaystyle \frac{ \partial^{2} F }{ \partial x_{i} \partial x_{j} \partial x_{k} } (x)$

\item[Algorithm:] \hfill
\begin{algorithmic}[0]
\State Construct composite functions, $(F_{1}, \ldots, F_{N})$
\State $x_{0} \gets 0$, $(\mathfrak{d} \alpha_{0})_{l} \gets 0$,
       $(\mathfrak{d}^{2} \alpha_{0})_{lm} \gets 0$, $(\mathfrak{d}^{3} \alpha_{0})_{lmq} \gets 0$,
\State
\For{$1 \le n \le N $}
  \State $x_{n} \gets F_{n}(x_{n - 1})$
  \State $(\mathfrak{d} \alpha_{n})_{l} \gets 0$,
         $(\mathfrak{d}^{2} \alpha_{n})_{lm} \gets 0$,
         $(\mathfrak{d}^{3} \alpha_{n})_{lmq} \gets 0$
\EndFor
\State $(\mathfrak{d} \alpha_{N})_{l} \gets 1$,
\For{$N \ge n \ge 1$}
  \State $\displaystyle (\mathfrak{d} \alpha_{n - 1})_{i} \gets \sum_{l = 1}^{D_{n}} 
         \frac{ \partial (F_{n})^{l} }{ \partial x_{i} }(x_{n - 1}) \cdot (\mathfrak{d} \alpha_{n})_{l}$
  \State $\displaystyle \displaystyle (\mathfrak{d}^{2} \alpha_{n - 1})_{ij} \gets
         \hspace{4mm} \sum_{l = 1}^{D_{n}} 
         \frac{ \partial^{2} (F_{n})^{l} }{ \partial x_{i} \partial x_{j} }(x_{n - 1}) 
         \cdot (\mathfrak{d} \alpha_{n - 1})_{l}$
  \State $\hspace{23mm} \displaystyle + \sum_{l = 1}^{D_{n}} \sum_{m = 1}^{D_{n}} 
         \frac{ \partial (F_{n})^{l} }{ \partial x_{i} }(x_{n - 1}) \cdot
         \frac{ \partial (F_{n})^{m} }{ \partial x_{j} }(x_{n - 1}) \cdot
         (\mathfrak{d}^{2} \alpha_{n - 1})_{lm}$
  \State $\displaystyle \displaystyle (\mathfrak{d}^{3} \alpha_{n - 1})_{ijk} \gets 
         \hspace{5mm} \sum_{l = 1}^{D_{n}} 
         \frac{ \partial^{3} (F_{n})^{l} }{ \partial x_{i} \partial x_{j} \partial x_{k} }(x_{n - 1}) 
         \cdot (\mathfrak{d} \alpha_{n - 1})_{l}$
  \State $\hspace{25mm} \displaystyle + \sum_{l = 1}^{D_{n}} \sum_{m = 1}^{D_{n}} 
         \Big( \hspace{2mm}
             \frac{ \partial (F_{n})^{l} }{ \partial x_{i} }(x_{n - 1}) \cdot
             \frac{ \partial^{2} (F_{n})^{m} }{ \partial x_{j} \partial x_{k} }(x_{n - 1})$
  \State $\hspace{45mm} \displaystyle    
           + \frac{ \partial (F_{n})^{l} }{ \partial x_{j} }(x_{n - 1}) \cdot
             \frac{ \partial^{2} (F_{n})^{m} }{ \partial x_{i} \partial x_{k} }(x_{n - 1})$
  \State $\hspace{45mm} \displaystyle
           + \frac{ \partial (F_{n})^{l} }{ \partial x_{k} }(x_{n - 1}) \cdot
             \frac{ \partial^{2} (F_{n})^{m} }{ \partial x_{i} \partial x_{j} }(x_{n - 1})
         \Big) \cdot
         (\mathfrak{d}^{2} \alpha_{n - 1})_{lm}$
  \State $\hspace{23mm} \displaystyle + \sum_{l = 1}^{D_{n}} \sum_{m = 1}^{D_{n}} \sum_{q = 1}^{D_{n}} 
         \frac{ \partial (F_{n})^{l} }{ \partial x_{i} }(x_{n - 1}) \cdot
         \frac{ \partial (F_{n})^{m} }{ \partial x_{j} }(x_{n - 1}) \cdot
         \frac{ \partial (F_{n})^{q} }{ \partial x_{k} }(x_{n - 1}) \cdot$
  \State $\hspace{45mm} \cdot (\mathfrak{d}^{3} \alpha_{n - 1})_{lmq}$
\EndFor
\State
\Return $\{ (\mathfrak{d}^{3} \alpha_{0})_{111}, \ldots, (\mathfrak{d}^{3} \alpha_{0})_{NNN} \}$
\end{algorithmic}

\item[Cost:] $\mathcal{O} \! \left( 1 \right)$
\item[Adjuncts:] $ f, \,
  \mathbf{g} = \{ (\mathfrak{d} \alpha_{0})_{1}, \ldots, (\mathfrak{d} \alpha_{0})_{N} \}, \,
  \mathbf{H} = \{ (\mathfrak{d}^{2} \alpha_{0})_{11}, \ldots, (\mathfrak{d}^{2} \alpha_{0})_{NN} \}$
\end{description}
	
\end{tcolorbox}

\subsubsection{Mixed Mode Methods}

Mixed mode methods complement each component function with some of
the coordinates of velocities, covelocities, and well-defined 
contractions between the two.  A second-order mixed method requires
the coordinates of a $(1,1)$-velocity, $(\delta v_{n})^{i}$, the 
coordinates of a $(1,1)$-covelocity, $(\mathfrak{d} \alpha_{n})_{l}$,
and the coordinates of the contraction $(\mathfrak{d} \beta_{n})_{l}$. 

\begin{tcolorbox}[colback=white,colframe=gray90, coltitle=black,boxrule=3pt,
fonttitle=\bfseries,title=Gradient of First-Order Directional Derivative]
	
\begin{description}
		
\item[Form:] $\displaystyle
              \frac{ \partial }{ \partial x_{i} }
              \left( \sum_{j = 1}^{N} v^{j} \frac{ \partial F }{ \partial x_{j} }(x) \right)
              =
		      \sum_{j = 1}^{N} v^{j} \frac{ \partial^{2} F }{ \partial x_{i} \partial x_{j} }(x) 
		      = \mathbf{H} \cdot \mathbf{v} $
		
\item[Algorithm:] \hfill \\
\begin{algorithmic}[0]
\State Construct composite functions, $(F_{1}, \ldots, F_{N})$
\State $x_{0} \gets x$, $(\delta v_{0})^{i} \gets v^{i}$, 
       $(\mathfrak{d} \alpha_{0})_{l} \gets 0$, $(\mathfrak{d} \beta_{0})_{l} \gets 0$,
\State
\For{$1 \le n \le N$}
  \State $ x_{s} \gets f_{s}(x_{I(s)})$
  \State $\displaystyle (\delta v_{n})^{l} \gets \sum_{j = 1}^{D_{n - 1}} 
         \frac{ \partial (F_{n})^{l} }{ \partial x_{i} }(x_{n - 1}) \cdot (\delta v_{n - 1})^{i}$
  \State $(\mathfrak{d} \alpha_{n})_{l} \gets 0$, $(\mathfrak{d} \beta_{n})_{l} \gets 0$
\EndFor
\State $(\mathfrak{d} \alpha_{N})_{l} \gets 1$
\For{$N \ge n \ge 1$}
  \State $\displaystyle (\mathfrak{d} \alpha_{n - 1})_{i} \gets \sum_{l = 1}^{D_{n}} 
         \frac{ \partial (F_{n})^{l} }{ \partial x_{i} }(x_{n - 1}) \cdot (\mathfrak{d} \alpha_{n})_{l}$  
    \State $\displaystyle (\mathfrak{d} \beta_{n - 1})_{i} \gets \hspace{3mm} \sum_{l = 1}^{D_{n}} 
         \frac{ \partial (F_{n})^{l} }{ \partial x_{i} }(x_{n - 1}) \cdot (\mathfrak{d} \beta_{n})_{l}$  
    \State $\hspace{18mm} \displaystyle + \sum_{l = 1}^{D_{n}} \sum_{j = 1}^{D_{n - 1}} 
         \frac{ \partial^{2} (F_{n})^{l} }{ \partial x_{i} \partial x_{j} }(x_{n - 1})
         \cdot (\mathfrak{d} \alpha_{n})_{l} \cdot (\delta v_{n - 1})^{j}$
\EndFor
\State
\Return $\{ (\mathfrak{d} \beta_{0})_{1}, \ldots, (\mathfrak{d} \beta_{0})_{N} \}$
\end{algorithmic}

\item[Cost:] $\mathcal{O} \! \left( 1 \right)$
\item[Adjuncts:] $ f, \,
                   \mathbf{v}^{T} \cdot \mathbf{g} = \delta v_{N}, \,
                   \mathbf{g} = \{ (\mathfrak{d} \alpha_{0})_{1}, \ldots, (\mathfrak{d} \alpha_{0})_{N} \}$

\end{description}
	
\end{tcolorbox}

We can compute the $i$-th column of the full Hessian array by taking
the input vector to be a vectors whose elements all vanish except for 
the $i$-th element, $\mathbf{v}_{n} = \delta^{i}_{n}$,   Consequently 
we can compute the full Hessian array using only local information with 
$N$ executions of the Hessian-vector product algorithm, one for each 
column.

A third-order mixed method requires three of the coordinates of a 
$(2, 2)$-velocity, $(\delta v_{n})^{i}$, $(\delta u_{n})^{i}$, and
$(\delta v \delta u_{n})^{i}$, the coordinates of a $(1,1)$-covelocity, 
$(\mathfrak{d} \alpha_{n})_{l}$, and the coordinates of the three 
contractions, $(\mathfrak{d} \beta_{n})_{l}$, $(\mathfrak{d} \gamma_{n})_{l}$,
and $(\mathfrak{d} \epsilon_{n})_{l}$. 

\begin{tcolorbox}[colback=white,colframe=gray90, coltitle=black,boxrule=3pt,
fonttitle=\bfseries,title=Gradient of a Second-Order Directional Derivative]
	
\begin{description}
		
\item[Form:] $\displaystyle 
  \frac{ \partial }{ \partial x_{i} }
  \left( \sum_{j = 1}^{N} \sum_{k = 1}^{N} v^{j} u^{k} 
  \frac{ \partial^{2} F }{ \partial x_{j}  \partial x_{k} }(x) \right)
  =
  \sum_{j = 1}^{N} \sum_{k = 1}^{N} v^{i} u^{j} 
  \frac{ \partial^{3} F }{ \partial x_{i}  \partial x_{j}  \partial x_{k} }(x) $

\item[Algorithm:] \hfill \\
\begin{algorithmic}[0]
\State Construct composite functions, $(F_{1}, \ldots, F_{N})$
\State $x_{0} \gets x$, $(\delta v_{0})^{i} \gets v^{i}$, $(\delta u_{0})^{i} \gets u^{i}$, 
       $(\delta v \delta u_{0})^{i} \gets 0$
\State $(\mathfrak{d} \alpha_{0})_{l} \gets 0$, $(\mathfrak{d} \beta_{0})_{l} \gets 0$,
       $(\mathfrak{d} \gamma_{0})_{l} \gets 0$, $(\mathfrak{d} \epsilon_{0})_{l} \gets 0$,
\State
\For{$1 \le n \le N$}
  \State $ x_{s} \gets f_{s}(x_{I(s)})$
  \State $(\delta v_{n})^{l} \gets \sum_{j = 1}^{D_{n - 1}} 
         \frac{ \partial (F_{n})^{l} }{ \partial x_{i} }(x_{n - 1}) \cdot (\delta v_{n - 1})^{i}$
  \State $(\delta u_{n})^{l} \gets \sum_{j = 1}^{D_{n - 1}} 
         \frac{ \partial (F_{n})^{l} }{ \partial x_{i} }(x_{n - 1}) \cdot (\delta u_{n - 1})^{i}$
         
  \State $(\delta v \delta u_{n})^{l} \gets
         \hspace{4mm} \sum_{i = 1}^{D_{n - 1}} 
         \frac{ \partial (F_{n})^{l} }{ \partial x_{i} }(x_{n - 1}) \cdot (\delta u \delta v_{n - 1})^{i}$
  \State $\hspace{20mm} + \sum_{i = 1}^{D_{n - 1}} \sum_{j = 1}^{D_{n - 1}} 
         \frac{ \partial^{2} (F_{n})^{l} }{ \partial x_{i} \partial x_{j} }(x_{n - 1}) 
         \cdot (\delta u_{n - 1})^{i} \cdot (\delta v_{n - 1})^{j}$
         
  \State $(\mathfrak{d} \alpha_{n})_{l} \gets 0$, $(\mathfrak{d} \beta_{n})_{l} \gets 0$,
         $(\mathfrak{d} \gamma_{n})_{l} \gets 0$, $(\mathfrak{d} \epsilon_{n})_{l} \gets 0$
\EndFor

\State $(\mathfrak{d} \alpha_{N})_{l} \gets 1$

\For{$N \ge n \ge 1$}
  \State $(\mathfrak{d} \alpha_{n - 1})_{i} \gets \sum_{l = 1}^{D_{n}} 
         \frac{ \partial (F_{n})^{l} }{ \partial x_{i} }(x_{n - 1}) \cdot (\mathfrak{d} \alpha_{n})_{l}$  

    \State $(\mathfrak{d} \beta_{n - 1})_{i} \gets \hspace{3mm} \sum_{l = 1}^{D_{n}} 
         \frac{ \partial (F_{n})^{l} }{ \partial x_{i} }(x_{n - 1}) \cdot (\mathfrak{d} \beta_{n})_{l}$  
    \State $\hspace{18mm} + \sum_{l = 1}^{D_{n}} \sum_{j = 1}^{D_{n - 1}}
         \frac{ \partial^{2} (F_{n})^{l} }{ \partial x_{i} \partial x_{j} }(x_{n - 1})
         \cdot (\mathfrak{d} \alpha_{n})_{l} \cdot (\delta v_{n - 1})^{j}$
         
    \State $(\mathfrak{d} \gamma_{n - 1})_{i} \gets \hspace{3mm} \sum_{l = 1}^{D_{n}} 
         \frac{ \partial (F_{n})^{l} }{ \partial x_{i} }(x_{n - 1}) \cdot (\mathfrak{d} \gamma_{n})_{l}$  
    \State $\hspace{18mm} + \sum_{l = 1}^{D_{n}} \sum_{j = 1}^{D_{n - 1}} 
         \frac{ \partial^{2} (F_{n})^{l} }{ \partial x_{i} \partial x_{j} }(x_{n - 1})
         \cdot (\mathfrak{d} \alpha_{n})_{l} \cdot (\delta u_{n - 1})^{j}$

    \State $(\mathfrak{d} \epsilon_{n - 1})_{i} \gets \hspace{3mm} \sum_{l = 1}^{D_{n}} 
         \frac{ \partial (F_{n})^{l} }{ \partial x_{i} }(x_{n - 1}) \cdot (\mathfrak{d} \epsilon_{n})_{l}$  
    \State $\hspace{18mm} + \sum_{l = 1}^{D_{n}} \sum_{j = 1}^{D_{n - 1}} 
         \frac{ \partial^{2} (F_{n})^{l} }{ \partial x_{i} \partial x_{j} }(x_{n - 1})
         \cdot \Big( \hspace{3mm} (\mathfrak{d} \alpha_{n})_{l} \cdot (\delta v \delta u_{n - 1})^{j}$
    \State $\hspace{71mm} + (\mathfrak{d} \beta_{n})_{l} \cdot (\delta u_{n - 1})^{j}$
    \State $\hspace{71mm} + (\mathfrak{d} \gamma_{n})_{l} \cdot (\delta v_{n - 1})^{j} \Big)$
    \State $\hspace{18mm} + \sum_{l = 1}^{D_{n}} \sum_{j = 1}^{D_{n - 1}} \sum_{k = 1}^{D_{n - 1}} 
         \frac{ \partial^{3} (F_{n})^{l} }{ \partial x_{i} \partial x_{j} \partial x_{k} }(x_{n - 1})
         \cdot (\mathfrak{d} \alpha_{n})_{l} \cdot (\delta v_{n - 1})^{j} \cdot (\delta u_{n - 1})^{k}$
\EndFor
\State
\Return $\{ \mathsf{\epsilon}_{1}, \ldots, \mathsf{\epsilon}_{N} \}$
\end{algorithmic}
		
\item[Cost:] $\mathcal{O} \! \left( n \right)$
\item[Adjuncts:] $ f, \, \mathbf{v}^{T} \mathbf{g} = \delta v_{N}, \, 
  \mathbf{u}^{T} \mathbf{g} = \delta u_{N}, \,
  \mathbf{v}^{T} H \mathbf{u} = \delta v \delta u_{N}, \\ 
  \mathbf{g} = \{ (\mathfrak{d} \alpha_{0})_{1}, \ldots, (\mathfrak{d} \alpha_{0})_{N} \}, \,
  \mathbf{H} \mathbf{v} = \{ (\mathfrak{d} \beta_{0})_{1}, \ldots, (\mathfrak{d} \beta_{0})_{N} \}, \\
  \mathbf{H} \mathbf{u} = \{ (\mathfrak{d} \gamma_{0})_{1}, \ldots, (\mathfrak{d} \gamma_{0})_{N} \}$

\end{description}
	
\end{tcolorbox}

The gradient of a second-order directional derivative is rarely of 
immediate use, but it can be used repeatedly to build up the gradient 
of the trace of the product of a matrix times the Hessian,
\begin{equation*}
\frac{\partial}{\partial x_{i} } \mathrm{Tr} \! \left[ \mathbf{M} \, \mathbf{H} \right]
= \sum_{j = 1}^{N} \sum_{k = 1}^{N} M^{k}_{j} 
\frac{ \partial^{3} f }{ \partial x_{i} \partial x_{j}  \partial x_{k} }
\end{equation*}
which commonly arises when taking the derivative of the determinant of
the Hessian.  To see this rewrite the above as
\begin{equation*}
\frac{\partial}{\partial x_{i} } \mathrm{Tr} \! \left[ \mathbf{M} \, \mathbf{H} \right]
= \sum_{k = 1}^{N} \left( \sum_{j = 1}^{N} M^{k}_{j} \delta^{j}_{k}
\frac{ \partial^{3} f }{ \partial x_{i} \partial x_{j}  \partial x_{k} } \right).
\end{equation*}
In words, we set $\mathbf{v}$ to the $k$-th column of the matrix $\mathbf{M}$ 
and $u^{n} = \delta^{n}_{k}$ and execute the gradient of the second-order
directional derivative algorithm and then repeat for each $k$.  This
repeated execution will also yield the full gradient and Hessian as
adjuncts.

\subsection{Local Implementations}

Unfortunately these general implementations leave much to be desired in
practice because of their dependence on explicit component functions.
The construction of these component functions requires understanding
the dependencies of each subexpression in the computer program which 
itself requires a careful static analysis of the expression graph.  
Not only is this static analysis subtle to implement well it also 
obstructs the use of these methods for computer programs that define 
dynamic expression graphs, in particular those utilizing control flow 
statements.

Conveniently, many of the geometric transformations we have introduced
decouple into transformations \emph{local} to each subexpression,
rendering explicit component functions unnecessary.  The corresponding
automatic differentiation methods then become significantly easier to
implement.

Consider, the pushforward of a $(1, 1)$-velocity that implicitly 
computes a first-order directional derivative.  The coordinates of 
the intermediate velocities between each component function are given 
by one-dimensional arrays that naturally separate across the 
corresponding subexpressions, the component $(\delta v_{n})^{i}$
being assigned to the $i$-th subexpression in the component
function.

Moreover these separated coordinates transform completely 
independently of one another, each subexpression component
aggregating contributions from each of its inputs,
\begin{equation*}
(\delta v_{n})
= 
\sum_{i = 1}^{I_{n}} (\mathcal{J}_{f_{l}})_{l} (x_{n}) \cdot (\delta v_{i})^{l}.
\end{equation*}  
In particular the action of the auxiliary identity maps 
becomes trivial,
\begin{equation*}
(\mathcal{J}_{I})_{i} (x_{n}) = 0
\end{equation*}
and hence the auxiliary maps can be completely ignored.  

Consequently the pushforward of the initial velocity can be 
implemented with calculations entirely local to each subexpression
and the subexpressions on which it depends.  Without having 
to propagate information along the auxiliary identity maps the 
resulting transformations not only are much simpler to implement
but also consume significantly less memory and computing power.  
Moveover these local updates admit the implementation of these 
methods through well-known techniques like local source code 
transformation and operator overloading that are the basis for 
most automatic differentiation tools.

This simplification arises for \emph{any} method transforming
an object that admits coordinates formed from a direct product
of one-dimensional arrays.  This includes all forward mode methods, 
the first-order reverse mode method, and the mixed mode methods 
we derived in Section \ref{sec:mixed_mode}.  The important 
exception are the higher-order reverse mode methods that pull 
back covelocities whose coordinates require higher-order arrays.  
Intuitively, these higher-order arrays provide the space needed 
to propagate the \emph{mixed derivatives} that are needed to build 
up higher-order Jacobians arrays in a single sweep.  Local 
implementations of the same functionality, on the other hand, 
requires multiple sweeps through the topologically sorted 
expression graph.

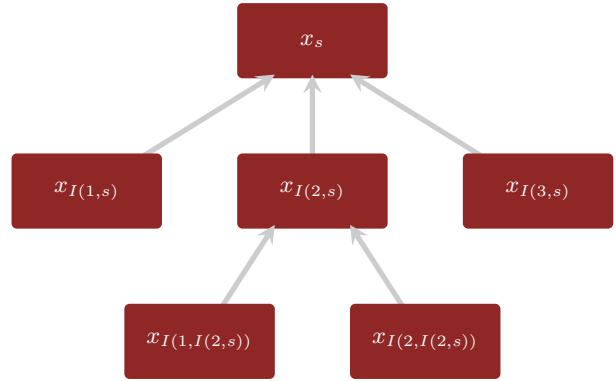
\begin{figure*}
\centering
\begin{tikzpicture}[scale=0.5, thick]

\fill[rounded corners=2, color=dark] (2, 3) rectangle +(4, 2);
\node[text=white] at (4, 4) { $x_{s}$ };

\draw[->, >=stealth, color=gray80, line width=2] (-2, 0) -- +(5, 3.1);

\fill[rounded corners=2, color=dark] (-4, -1) rectangle +(4, 2);
\node[text=white] at (-2, 0) { $x_{I(1, s)}$ };

\draw[->, >=stealth, color=gray80, line width=2] (4, 0) -- +(0, 3.1);

\fill[rounded corners=2, color=dark] (2, -1) rectangle +(4, 2);
\node[text=white] at (4, 0) { $x_{I(2, s)}$ };

\draw[->, >=stealth, color=gray80, line width=2] (10, 0) -- +(-5, 3.1);

\fill[rounded corners=2, color=dark] (8, -1) rectangle +(4, 2);
\node[text=white] at (10, 0) { $x_{I(3, s)}$ };

\draw[->, >=stealth, color=gray80, line width=2] (1, -4) -- +(+2, 3.1);

\fill[rounded corners=2, color=dark] (-1, -5) rectangle +(4, 2);
\node[text=white] at (1, -4) { $x_{I(1, I(2, s))}$ };

\draw[->, >=stealth, color=gray80, line width=2] (7, -4) -- +(-2, 3.1);

\fill[rounded corners=2, color=dark] (5, -5) rectangle +(4, 2);
\node[text=white] at (7, -4) { $x_{I(2, I(2, s))}$ };

\end{tikzpicture}
\caption{
The $s$-th subexpression in the expression stack features $N_{I}(s)$
inputs with $I(i, s)$ denoting the global index of the $i$-th of those
inputs.  This notation allows us to identify parent and child nodes
from any given node.
}
\label{fig:local_indices} 
\end{figure*}

In the following sections I present explicit algorithms that
utilize only local information.  Let the function 
\verb|construct_stack| assign global indices 
$s_{1}, \ldots, s_{S}$ to each of the $S = N + N_{\text{subexp}}$ 
nodes in the expression graph beginning with the $N$ input 
nodes and the $N_{\text{subexp}}$ nodes for the topologically 
sorted subexpressions.  Additionally assume that all subexpressions 
are many-to-one with the $s$-th subexpression featuring $N_{I}(s)$ 
scalar inputs such that the it is of the form 
$F_{s} : \mathbb{R}^{N_{I}(s)} \rightarrow \mathbb{R}$.  
Finally let $x_{I(s)}$ denote the set of input values to the 
$s$-th node in the stack and let $I(i, s)$ define the the global 
index of the $i$-th input to the $s$-th node in the stack
(Figure \ref{fig:local_indices}).

\subsubsection{Forward Mode Methods}

At first order the $s$-th node in the stack contains the output 
value, $x_{s}$, and the $(1,1)$-velocity $\delta v_{s}$.  Because 
we have limited consideration to many-to-one functions both will 
be one-dimensional.  To simplify the notation I will denote the 
single coordinate that defines each velocity as
\begin{align*}
\mathsf{v}_{s} = (\delta v_{s})^{1}.
\end{align*}

\begin{tcolorbox}[colback=white,colframe=gray90, coltitle=black,boxrule=3pt,
fonttitle=\bfseries,title=First-Order Directional Derivative]

\begin{description}

\item[Form:] $\displaystyle \mathbf{v}^{T} \cdot \mathbf{g} 
		               = \sum_{i = 1}^{N} v^{i} \frac{ \partial f }{ \partial x^{i} } (x) $
\item[Algorithm:] \hfill
\begin{algorithmic}[0]
\State \verb|construct_stack|
\State
\For{$1 \le n \le N$}
  \State $\mathsf{v}_{n} \gets v^{n}$
\EndFor
\State
\For{$N + 1 \le s \le S$}
  \State $ x_{s} \gets f_{s}(x_{I(s)})$
  \State $\mathsf{v}_{s} \gets 0$
  \For{$1 \le i \le I(s)$}
  \State $\displaystyle \mathsf{v}_{s} \gets \mathsf{v}_{s} + 
         \frac{ \partial f_{s} }{ \partial x^{i} }(x_{I(s)}) \cdot \mathsf{v}_{I(i, s)}$
  \EndFor
\EndFor
\State
\Return $\mathsf{v}_{S}$
\end{algorithmic}

\item[Cost:] $\mathcal{O} \! \left( 1 \right)$ Sweeps
\item[Adjuncts:] $ f $
\end{description}
	
\end{tcolorbox}

At second-order the $s$-th node in expression graph will contain 
the scalar output value, $x_{s}$, and three of the components of
a $(2,2)$-velocity, $\delta v_{s}$, $\delta u_{s}$, and 
$\delta v \delta u_{s}$.  Because we have limited consideration 
to many-to-one functions all will be one-dimensional and simplify
the notation I will denote the single coordinates as
\begin{align*}
\mathsf{v}_{s} &= (\delta v_{s})^{1}
\\
\mathsf{u}_{s} &= (\delta u_{s})^{1}
\\
\mathsf{vu}_{s} &= (\delta v \delta u_{s})^{1}.
\end{align*}

\begin{tcolorbox}[colback=white,colframe=gray90, coltitle=black,boxrule=3pt,
fonttitle=\bfseries,title=Second-Order Directional Derivative]

\begin{description}

\item[Form:] $\displaystyle \mathbf{v}^{T} \cdot \mathbf{H} \cdot \mathbf{u} 
		        = \sum_{i=1}^{N} \sum_{j=1}^{N} v^{i} u^{j} 
		          \frac{ \partial^{2} f }{ \partial x^{i} \partial x^{j} } (x)$
		          
\item[Algorithm:] \hfill
\begin{algorithmic}[0]
\State \verb|construct_stack|
\State
\For{$1 \le n \le N$}
  \State $\mathsf{v}_{n} \gets v^{n}$
  \State $\mathsf{u}_{n} \gets u^{n}$
  \State $\mathsf{vu}_{n} \gets 0$
\EndFor
\State
\For{$N + 1 \le s \le S$}
  \State $x_{s} = f_{n}(x_{I(s)})$
  \State $\mathsf{v}_{s} \gets 0$
  \State $\mathsf{u}_{s} \gets 0$ 
  \State $\mathsf{vu}_{s} \gets 0$
  \For{$1 \le i \le I(s)$}
  \State $\displaystyle \mathsf{v}_{s} \gets \mathsf{v}_{s} + 
         \frac{ \partial f_{s} }{ \partial x^{i} }(x_{i}) \cdot \mathsf{v}_{I(i, s)}$
  \State $\displaystyle \mathsf{u}_{n} \gets \mathsf{u}_{n} + 
         \frac{ \partial f_{s} }{ \partial x^{i} }(x_{I(s)}) \cdot \mathsf{u}_{I(i, s)}$
  \State $\displaystyle \mathsf{vu}_{s} \gets \mathsf{vu}_{s}
             + \frac{ \partial f_{s} }{ \partial x^{i} }(x_{I(s)}) \cdot \mathsf{vu}_{I(i, s)}$
  \For{$1 \le j \le I(s)$}
    \State $\displaystyle \mathsf{vu}_{s} \gets \mathsf{vu}_{s}
             + \frac{ \partial^{2} f_{s} }{ \partial x^{i} \partial x^{j} } 
                      \cdot \mathsf{v}_{I(i, s)} \cdot \mathsf{u}_{I(j, s)}$
    \EndFor
  \EndFor
\EndFor
\State
\Return $\mathsf{vu}_{S}$
\end{algorithmic}

\item[Cost:] $\mathcal{O} \! \left( 1 \right)$
\item[Adjuncts:] $ f, \, \mathbf{v}^{T} \cdot \mathbf{g} = \mathsf{v}_{S}, \, 
                   \mathbf{u}^{T} \cdot \mathbf{g} = \mathsf{u}_{S}$
\end{description}
	
\end{tcolorbox}

To compute a third-order directional derivative the $s$-th node in 
expression graph needs to contain the scalar output value, $x_{s}$, 
and seven of the components of a $(3,3)$-velocity, $\delta v_{s}$, 
$\delta u_{s}$, $\delta w_{s}$, $\delta v \delta u_{s}$, 
$\delta v \delta w_{s}$, $\delta u \delta w_{s}$ and 
$\delta v \delta u \delta w_{s}$.  Because we have limited 
consideration to many-to-one functions all will be one-dimensional 
and simplify the notation I will denote the single coordinates as
\begin{align*}
\mathsf{v}_{s} &= (\delta v_{s})^{1}
\\
\mathsf{u}_{s} &= (\delta u_{s})^{1}
\\
\mathsf{w}_{s} &= (\delta w_{s})^{1}
\\
\mathsf{vu}_{s} &= (\delta v \delta u_{s})^{1}
\\
\mathsf{vw}_{s} &= (\delta v \delta w_{s})^{1}
\\
\mathsf{uw}_{s} &= (\delta u \delta w_{s})^{1}
\\
\mathsf{vuw}_{s} &= (\delta v \delta u \delta w_{s})^{1}
\end{align*}

\begin{tcolorbox}[colback=white,colframe=gray90, coltitle=black,boxrule=3pt,
fonttitle=\bfseries,title=Third-Order Directional Derivative]

\begin{description}

\item[Form:] $\displaystyle \sum_{i=1}^{N} \sum_{j=1}^{N} \sum_{k=1}^{N} v^{i} u^{j} w^{k} 
		          \frac{ \partial^{3} f }{ \partial x^{i} \partial x^{j} \partial x^{k} } (x)$
		          
\item[Algorithm:] \hfill
\begin{algorithmic}[0]
\State \verb|construct_stack|
\State
\For{$1 \le n \le N$}
  \State $\mathsf{v}_{n} \gets v^{n}$, $\mathsf{u}_{n} \gets u^{n}$, $\mathsf{w}_{n} \gets w^{n}$
  \State $\mathsf{vu}_{n} \gets 0$, $\mathsf{vw}_{n} \gets 0$, $\mathsf{uw}_{n} \gets 0$, 
         $\mathsf{vuw}_{n} \gets 0$
\EndFor
\State
\For{$N + 1 \le s \le S$}
  \State $x_{s} = f_{n}(x_{I(s)})$
  \State $\mathsf{v}_{s} \gets 0$, $\mathsf{u}_{s} \gets 0$, $\mathsf{w}_{s} \gets 0$ 
  \State $\mathsf{vu}_{s} \gets 0$, $\mathsf{vw}_{s} \gets 0$, $\mathsf{uw}_{s} \gets 0$,
         $\mathsf{vuw}_{s} \gets 0$
  \For{$1 \le i \le I(s)$}
  \State $\displaystyle \mathsf{v}_{s} \gets \mathsf{v}_{s} + 
         \frac{ \partial f_{s} }{ \partial x^{i} }(x_{i}) \cdot \mathsf{v}_{I(i, s)}$
  \State $\displaystyle \mathsf{u}_{n} \gets \mathsf{u}_{n} + 
         \frac{ \partial f_{s} }{ \partial x^{i} }(x_{I(s)}) \cdot \mathsf{u}_{I(i, s)}$
  \State $\displaystyle \mathsf{w}_{n} \gets \mathsf{w}_{n} + 
         \frac{ \partial f_{s} }{ \partial x^{i} }(x_{I(s)}) \cdot \mathsf{w}_{I(i, s)}$
  \State $\displaystyle \mathsf{vu}_{s} \gets \mathsf{vu}_{s}
             + \frac{ \partial f_{s} }{ \partial x^{i} }(x_{I(s)}) \cdot \mathsf{vu}_{I(i, s)}$
  \State $\displaystyle \mathsf{vw}_{s} \gets \mathsf{vw}_{s}
             + \frac{ \partial f_{s} }{ \partial x^{i} }(x_{I(s)}) \cdot \mathsf{vw}_{I(i, s)}$
  \State $\displaystyle \mathsf{uw}_{s} \gets \mathsf{uw}_{s}
             + \frac{ \partial f_{s} }{ \partial x^{i} }(x_{I(s)}) \cdot \mathsf{uw}_{I(i, s)}$
  \State $\displaystyle \mathsf{vuw}_{s} \gets \mathsf{vuw}_{s}
             + \frac{ \partial f_{s} }{ \partial x^{i} }(x_{I(s)}) \cdot \mathsf{vuw}_{I(i, s)}$
  \For{$1 \le j \le I(s)$}
    \State $\displaystyle \mathsf{vu}_{s} \gets \mathsf{vu}_{s}
             + \frac{ \partial^{2} f_{s} }{ \partial x^{i} \partial x^{j} } 
                      \cdot \mathsf{v}_{I(i, s)} \cdot \mathsf{u}_{I(j, s)}$
    \State $\displaystyle \mathsf{vw}_{s} \gets \mathsf{vw}_{s}
             + \frac{ \partial^{2} f_{s} }{ \partial x^{i} \partial x^{j} } 
                      \cdot \mathsf{v}_{I(i, s)} \cdot \mathsf{w}_{I(j, s)}$
    \State $\displaystyle \mathsf{uw}_{s} \gets \mathsf{uw}_{s}
             + \frac{ \partial^{2} f_{s} }{ \partial x^{i} \partial x^{j} } 
                      \cdot \mathsf{u}_{I(i, s)} \cdot \mathsf{w}_{I(j, s)}$
    \State $\displaystyle \mathsf{vuw}_{s} \gets \mathsf{vuw}_{s}
             + \frac{ \partial^{2} f_{s} }{ \partial x^{i} \partial x^{j} } 
                      \cdot ( \mathsf{v}_{I(i, s)} \cdot \mathsf{uw}_{I(j, s)}
                             + \mathsf{u}_{I(i, s)} \cdot \mathsf{vw}_{I(j, s)}
                             + \mathsf{w}_{I(i, s)} \cdot \mathsf{vu}_{I(j, s)} )$ 
      \For{$1 \le k \le I(s)$}
        \State $\displaystyle \mathsf{vuw}_{s} \gets \mathsf{vuw}_{s}
             + \frac{ \partial^{3} f_{s} }{ \partial x^{i} \partial x^{j} \partial x^{k} } 
                      \cdot \mathsf{v}_{I(i, s)} \cdot \mathsf{u}_{I(j, s)} \cdot \mathsf{w}_{I(k, s)}$
      \EndFor
    \EndFor
  \EndFor
\EndFor
\State
\Return $\mathsf{vuw}_{S}$
\end{algorithmic}

\item[Cost:] $\mathcal{O} \! \left( 1 \right)$
\item[Adjuncts:] $ f, \, \mathbf{v}^{T} \cdot \mathbf{g} = \mathsf{v}_{S}, \, 
                   \mathbf{u}^{T} \cdot \mathbf{g} = \mathsf{u}_{S}, \,
                   \mathbf{w}^{T} \cdot \mathbf{g} = \mathsf{w}_{S}, \
                   \mathbf{v}^{T} \cdot \mathbf{H} \cdot \mathbf{u} = \mathsf{vu}_{S}, \
                   \mathbf{v}^{T} \cdot \mathbf{H} \cdot \mathbf{w} = \mathsf{vw}_{S}, \
                   \mathbf{u}^{T} \cdot \mathbf{H} \cdot \mathbf{w} = \mathsf{uw}_{S}$
\end{description}
	
\end{tcolorbox}

\subsubsection{Reverse Mode Methods}

At first order the $s$-th node in the stack contains the output 
value, $x_{s}$, and the $(1,1)$-covelocity $\mathfrak{d} \alpha_{s}$.  
Because we have limited consideration to many-to-one functions both 
will be one-dimensional.  To simplify the notation I will denote 
the single coordinate that defines each covelocity as
\begin{align*}
\mathsf{a}_{s} = (\mathfrak{d} \alpha_{s})^{1}.
\end{align*}

\begin{tcolorbox}[colback=white,colframe=gray90, coltitle=black,boxrule=3pt,
fonttitle=\bfseries,title=Gradient]
	
\begin{description}

\item[Form:] $\displaystyle g_{i} = \frac{ \partial f }{ \partial x_{i} }(x) $

\item[Algorithm:] \hfill \\
\begin{algorithmic}[0]
\State \verb|construct_stack|
\State
\For{$1 \le n \le N$}
  \State $ \mathsf{a}_{n} \gets 0$
\EndFor
\State
\For{$N + 1 \le s \le S$}
  \State $ x_{s} \gets f_{s}(x_{I(s)})$
  \State $\mathsf{a}_{s} \gets 0$
\EndFor
\State $\mathsf{a}_{S} \gets 1$
\For{$S \ge n > N + 1 $}
  \For{$1 \le i \le I(s)$}
  \State $\displaystyle \mathsf{a}_{I(i, s)} \gets \mathsf{a}_{I(i, s)} + 
         \frac{ \partial f_{s} }{ \partial x^{i} }(x_{I(n)}) \cdot \mathsf{a}_{s}$
  \EndFor
\EndFor
\State
\Return $\{ \mathsf{a}_{1}, \ldots, \mathsf{a}_{N} \}$
\end{algorithmic}

\item[Cost:] $\mathcal{O} \! \left( 1 \right)$
\item[Adjuncts:] $ f $

\end{description}
	
\end{tcolorbox}

\subsubsection{Mixed Mode Methods}

In order to compute a second-order mixed mode calculation the $s$-th node 
in the stack will need to contain the output value, $x_{s}$, the 
$(1,1)$-velocity $\delta v_{s}$, the $(1,1)$-covelocity $\mathfrak{d} \alpha_{s}$,
and the contraction $\mathfrak{d} \beta_{s}$.  Because we have limited 
consideration to many-to-one functions all will be one-dimensional.  
To simplify the notation I will denote the single coordinate that 
defines each object as
\begin{align*}
\mathsf{v}_{s} &= (\delta v_{s})^{1}
\\
\mathsf{a}_{s} &= (\mathfrak{d} \alpha_{s})^{1}
\\
\mathsf{b}_{s} &= (\mathfrak{d} \beta_{s})^{1}.
\end{align*}

\begin{tcolorbox}[colback=white,colframe=gray90, coltitle=black,boxrule=3pt,
fonttitle=\bfseries,title=Gradient of First-Order Directional Derivative]
	
\begin{description}
		
\item[Form:] $\displaystyle
              \frac{ \partial }{ \partial x_{i} }
              \left( \sum_{j = 1}^{N} v^{j} \frac{ \partial f }{ \partial x_{j} }(x) \right)
              =
		      \sum_{j = 1}^{N} v^{j} \frac{ \partial^{2} f }{ \partial x_{i} \partial x_{j} }(x) 
		      = \mathbf{H} \cdot \mathbf{v} $
		
\item[Algorithm:] \hfill \\
\begin{algorithmic}[0]
\State \verb|construct_stack|
\State
\For{$1 \le n \le N$}
  \State $\mathsf{v}_{n} \gets v^{s}$
  \State $\mathsf{a}_{n} \gets 0$
  \State $\mathsf{b}_{n} \gets 0$
\EndFor
\State
\For{$N + 1 \le s \le S$}
  \State $ x_{s} \gets f_{s}(x_{I(s)})$
  \For{$1 \le i \le I(s)$}
    \State $\displaystyle \mathsf{v}_{s} \gets \mathsf{v}_{s} + 
           \frac{ \partial f_{s} }{ \partial x^{i} }(x_{i}) \cdot \mathsf{v}_{I(i, s)}$
  \EndFor
  \State $\mathsf{a}_{s} \gets 0$
  \State $\mathsf{b}_{s} \gets 0$
\EndFor
\State $\mathsf{a}_{S} \gets 1$
\For{$S \ge n > N + 1 $}
  \For{$1 \le i \le I(s)$}
    \State $\displaystyle \mathsf{a}_{I(i, s)} \gets \mathsf{a}_{I(i, s)} + 
           \frac{ \partial f_{s} }{ \partial x^{i} }(x_{I(n)}) \cdot \mathsf{a}_{s}$
    \State $\displaystyle \mathsf{b}_{I(i, s)} \gets \mathsf{b}_{I(i, s)} + 
           \frac{ \partial f_{s} }{ \partial x^{i} }(x_{I(n)}) \cdot \mathsf{b}_{s}$
    \For{$1 \le j \le I(s)$}
      \State $\displaystyle \mathsf{b}_{I(i, s)} \gets \mathsf{b}_{I(i, s)}
             + \frac{ \partial^{2} f_{s} }{ \partial x^{i} \partial x^{j} } 
                      \cdot \mathsf{v}_{I(j, s)} \cdot \mathsf{a}_{s}$
    \EndFor
  \EndFor
\EndFor
\State
\Return $\{ \mathsf{e}_{1}, \ldots, \mathsf{e}_{N} \}$
\end{algorithmic}

\item[Cost:] $\mathcal{O} \! \left( 1 \right)$
\item[Adjuncts:] $ f, \, \mathbf{v}^{T} \mathbf{g} = \mathsf{v}_{S}, \, 
                         \mathbf{g} = \{ \mathsf{a}_{1}, \ldots, \mathsf{a}_{N} \}$

\end{description}
	
\end{tcolorbox}

In order to compute a third-order mixed mode calculation the $s$-th node 
in the stack will need to contain the output value, $x_{s}$, three
components of a $(2,2)$-velocity, $\delta v_{s}$, $\delta u_{s}$,
and $\delta v \delta u_{s}$, the $(1,1)$-covelocity $\mathfrak{d} \alpha_{s}$,
and the three contractions, $\mathfrak{d} \beta_{s}$, $\mathfrak{d} \gamma_{s}$,
and $\mathfrak{d} \epsilon_{s}$.  Because we have limited 
consideration to many-to-one functions all will be one-dimensional.  
To simplify the notation I will denote the single coordinate that 
defines each object as
\begin{align*}
\mathsf{v}_{s} &= (\delta v_{s})^{1}
\\
\mathsf{u}_{s} &= (\delta u_{s})^{1}
\\
\mathsf{vu}_{s} &= (\delta v \delta u_{s})^{1}
\\
\mathsf{a}_{s} &= (\mathfrak{d} \alpha_{s})^{1}
\\
\mathsf{b}_{s} &= (\mathfrak{d} \beta_{s})^{1}
\\
\mathsf{g}_{s} &= (\mathfrak{d} \gamma_{s})^{1}
\\
\mathsf{e}_{s} &= (\mathfrak{d} \epsilon_{s})^{1}.
\end{align*}

\begin{tcolorbox}[colback=white,colframe=gray90, coltitle=black,boxrule=3pt,
fonttitle=\bfseries,title=Gradient of Second-Order Directional Derivative]
	
\begin{description}
		
\item[Form:] $\displaystyle 
  \frac{ \partial }{ \partial x_{i} }
  \left( \sum_{j = 1}^{N} \sum_{k = 1}^{N} v^{j} u^{k} 
  \frac{ \partial^{2} f }{ \partial x_{j}  \partial x_{k} }(x) \right)
  =
  \sum_{j = 1}^{N} \sum_{k = 1}^{N} v^{i} u^{j} 
  \frac{ \partial^{3} f }{ \partial x_{i}  \partial x_{j}  \partial x_{k} }(x) $

\item[Algorithm:] \hfill \\
\begin{algorithmic}[0]
\State \verb|construct_stack|
\State
\For{$1 \le n \le N$}
  \State $\mathsf{v}_{n} \gets v^{n}$, $\mathsf{u}_{n} \gets u^{n}$, $\mathsf{vu}_{n} \gets 0$
  \State $\mathsf{a}_{n} \gets 0$, $\mathsf{b}_{n} \gets 0$, $\mathsf{g}_{n} \gets 0$, $\mathsf{e}_{n} \gets 0$
\EndFor
\State
\For{$N + 1 \le s \le S$}
  \State $ x_{s} \gets f_{s}(x_{I(s)})$
  \For{$1 \le i \le I(s)$}
    \State $\displaystyle \mathsf{v}_{s} \gets \mathsf{v}_{s} + 
           \frac{ \partial f_{s} }{ \partial x^{i} }(x_{i}) \cdot \mathsf{v}_{I(i, s)}$
    \State $\displaystyle \mathsf{u}_{s} \gets \mathsf{u}_{s} + 
           \frac{ \partial f_{s} }{ \partial x^{i} }(x_{i}) \cdot \mathsf{u}_{I(i, s)}$
    \State $\displaystyle \mathsf{vu}_{s} \gets \mathsf{vu}_{s} + 
           \frac{ \partial f_{s} }{ \partial x^{i} }(x_{i}) \cdot \mathsf{vu}_{I(i, s)}$
    \For{$1 \le j \le I(s)$}
      \State $\displaystyle \mathsf{vu}_{s} \gets \mathsf{vu}_{s} + 
             \frac{ \partial f_{s} }{ \partial x^{i} \partial x^{j} }(x_{i}) 
             \cdot \mathsf{v}_{I(i, s)} \cdot \mathsf{u}_{I(i, s)}$
    \EndFor
  \EndFor
  \State $\mathsf{a}_{s} \gets 0$, $\mathsf{b}_{s} \gets 0$, $\mathsf{g}_{s} \gets 0$, $\mathsf{e}_{s} \gets 0$
\EndFor
\State $\mathsf{a}_{S} \gets 1$
\For{$S \ge n > N + 1 $}
  \For{$1 \le i \le I(s)$}
    \State $\displaystyle \mathsf{a}_{I(i, s)} \gets \mathsf{a}_{I(i, s)} + 
           \frac{ \partial f_{s} }{ \partial x^{i} }(x_{I(n)}) \cdot \mathsf{a}_{s}$
    \State $\displaystyle \mathsf{b}_{I(i, s)} \gets \mathsf{b}_{I(i, s)} + 
           \frac{ \partial f_{s} }{ \partial x^{i} }(x_{I(n)}) \cdot \mathsf{b}_{s}$
    \State $\displaystyle \mathsf{g}_{I(i, s)} \gets \mathsf{g}_{I(i, s)} + 
           \frac{ \partial f_{s} }{ \partial x^{i} }(x_{I(n)}) \cdot \mathsf{g}_{s}$
    \State $\displaystyle \mathsf{e}_{I(i, s)} \gets \mathsf{e}_{I(i, s)} + 
           \frac{ \partial f_{s} }{ \partial x^{i} }(x_{I(n)}) \cdot \mathsf{e}_{s}$
    \For{$1 \le j \le I(s)$}
      \State $\displaystyle \mathsf{b}_{I(i, s)} \gets \mathsf{b}_{I(i, s)}
             + \frac{ \partial^{2} f_{s} }{ \partial x^{i} \partial x^{j} } 
                      \cdot \mathsf{v}_{I(j, s)} \cdot \mathsf{a}_{s}$
      \State $\displaystyle \mathsf{g}_{I(j, s)} \gets \mathsf{g}_{I(i, s)}
             + \frac{ \partial^{2} f_{s} }{ \partial x^{i} \partial x^{j} } 
                      \cdot \mathsf{u}_{I(j, s)} \cdot \mathsf{a}_{s}$
      \State $\displaystyle \mathsf{e}_{I(j, s)} \gets \mathsf{e}_{I(i, s)}
             + \frac{ \partial^{2} f_{s} }{ \partial x^{i} \partial x^{j} } 
                      \cdot ( \mathsf{v}_{I(j, s)} \cdot \mathsf{g}_{s}
                             + \mathsf{u}_{I(j, s)} \cdot \mathsf{b}_{s}
                             +\mathsf{vu}_{I(j, s)} \cdot \mathsf{a}_{s})$
      \For{$1 \le k \le I(s)$}
        \State $\displaystyle \mathsf{e}_{I(i, s)} \gets \mathsf{e}_{I(i, s)}
             + \frac{ \partial^{3} f_{s} }{ \partial x^{i} \partial x^{j} \partial x^{k} } 
                      \cdot \mathsf{v}_{I(j, s)} \cdot \mathsf{u}_{I(k, s)} \cdot \mathsf{a}_{s}$
      \EndFor
    \EndFor
  \EndFor
\EndFor
\State
\Return $\{ \mathsf{b}_{1}, \ldots, \mathsf{b}_{N} \}$
\end{algorithmic}
		
\item[Cost:] $\mathcal{O} \! \left( n \right)$
\item[Adjuncts:] $ f, \, \mathbf{v}^{T} \mathbf{g} = \mathsf{v}_{S}, \, 
                         \mathbf{u}^{T} \mathbf{g} = \mathsf{u}_{S}, \,
                         \mathbf{v}^{T} H \mathbf{u} = \mathsf{vu}_{S}, \, 
                         \mathbf{g} = \{ \mathsf{a}_{1}, \ldots, \mathsf{a}_{N} \}, \
                         \mathbf{H} \mathbf{v} = \{ \mathsf{b}_{1}, \ldots, \mathsf{b}_{N} \}, \
                         \mathbf{H} \mathbf{u} = \{ \mathsf{g}_{1}, \ldots, \mathsf{g}_{N} \}$

\end{description}
	
\end{tcolorbox}

\section{Conclusion}

By placing the methods of automatic differentiation within the
context of differential geometry we can develop a comprehensive 
theory that identifies well-defined differential operators and
the procedures needed for their implementation that immediately
generalized beyond first-order methods.

The main difference between the methods derived here and their 
counterparts in the automatic differentiation literature is 
the explicitness of the updates.  Most higher-order automatic 
differentiation methods apply first-order methods recursively,
either across the entire program to locally to each node in
the expression graph, coupling automatic differentiation of 
programs that define first-order partial derivatives with the 
updates that propagate the higher-order derivatives through 
the expression stack.  The explicit algorithms presented here 
cleanly separate the higher-order partial derivatives from 
the updates, facilitating algorithms that implement the 
partial derivatives analytically for improved performance.

In particular these explicit algorithms highlight a critical 
tradeoff often taken for granted in first-order methods
that utilize reverse sweeps through the expression graph.
Computations common to each subexpression and its partial 
derivatives must either be computed twice, once for the 
forward sweep and once for the reverse sweep, or cached in 
memory.  The latter introduces an increased memory burden, 
potentially limiting memory locality and hence practical 
performance.  Because the higher-order partial derivatives 
of complex functions typically share many expensive 
computations, the possibility of efficiently caching 
relevant intermediate values is a critical engineering
challenge to the performance of higher-order methods.

Finally, the implementation advantages of the local mixed mode
methods must be balanced against their additional cost.  For
example, for sufficiently high-dimensional target functions 
the pure reverse mode algorithm for computing higher-order
partial derivatives arrays becomes significantly faster than 
the multi-sweep mixed mode equivalents.  To push the 
performance of calculations like the Hessian it may become 
necessary to deal with the more general implementations and 
develop tools for the efficient expression graph analysis 
necessary to automatically construct explicit component functions.

\section{Acknowledgements}

I thank Bob Carpenter, Dan Simpson, Matt Johnson, Gordon 
Pusch, John Armstrong, Damiano Brigo, and Charles Margossian
for pivotal discussions about higher-order automatic 
differentiation and jets, as well as Charles Margossian 
and Bob Carpenter for helpful feedback on this manuscript.

\bibliography{higher_order_autodiff}
\bibliographystyle{imsart-nameyear}

\end{document}